\documentclass[useAMS,usenatbib]{mnras}
\usepackage[T1]{fontenc}

\usepackage{graphicx}
\usepackage{subcaption}
\usepackage{float}
\usepackage{amsmath}
\usepackage{amssymb}
\usepackage{bm}
\newcommand{\eagle}{{\sc eagle}}
\newcommand{\tng}{{\sc tng}}

\newcommand{\ha}{H$\alpha$}
\renewcommand{\lim}{{\sc lim}}
\newcommand{\sfr}{{\sc sfr}}
\newcommand{\hod}{{\sc hod}}
\renewcommand{\lq}{{`}}
\renewcommand{\rq}{{'}}
\newcommand{\illus}{IllustrisTNG}
\newcommand{\arepo}{{\sc arepo}}
\newcommand{\agn}{{\sc agn}}
\newcommand{\elg}{{\sc elg}}

\newcommand{\desi}{{\sc desi}}
\newcommand{\ssfr}{sSFR}

\newcommand{\iqr}{{\sc iqr}}
\newcommand{\rsd}{{\sc rsd}}

\newcommand{\lm}{$L$ - $M$} 
 
\usepackage{braket}

\usepackage{siunitx}

\newcommand{\mhalounit}[1]{\qty[print-unity-mantissa = false]{e#1}{M_\odot h^{-1}}}
\newcommand{\massunit}{\unit{M_\odot h^{-1}}}
\newcommand{\sfrunit}{\unit{M_\odot yr^{-1}}}

\newcommand{\kunit}{\unit{h\ Mpc^{-1}}}
\newcommand{\lenunit}{\unit{Mpc\ h^{-1}}}

\newcommand{\pkunit}{\unit{h^{-3}\ Mpc^3 }}

\newcommand{\mvir}{$M_{\mathrm{vir}}$}

\newcommand{\rvir}{$r_{\mathrm{vir}}$}

\newcommand{\vmax}{$\tilde{c}$}
\newcommand{\msat}{$M_\mathrm{sat}$}

\newcommand{\rev}[1]{\textcolor{brown}{#1}}

\newcommand{\citepaperone}{Paper I}

\usepackage{comment}

\usepackage{xr}
\usepackage{subfiles}

\usepackage[capitalise]{cleveref}
\crefformat{equation}{Eq. #2#1#3}
\crefrangeformat{equation}{Eqs. #3#1#4-#5#2#6}

\crefrangeformat{figure}{Figs. #3#1#4-#5#2#6}

\usepackage{scalerel}
\usepackage{tikz}
\usetikzlibrary{svg.path}
\definecolor{orcidlogocol}{HTML}{A6CE39}
\tikzset{
  orcidlogo/.pic={
    \fill[orcidlogocol] svg{M256,128c0,70.7-57.3,128-128,128C57.3,256,0,198.7,0,128C0,57.3,57.3,0,128,0C198.7,0,256,57.3,256,128z};
    \fill[white] svg{M86.3,186.2H70.9V79.1h15.4v48.4V186.2z}
                 svg{M108.9,79.1h41.6c39.6,0,57,28.3,57,53.6c0,27.5-21.5,53.6-56.8,53.6h-41.8V79.1z M124.3,172.4h24.5c34.9,0,42.9-26.5,42.9-39.7c0-21.5-13.7-39.7-43.7-39.7h-23.7V172.4z}
                 svg{M88.7,56.8c0,5.5-4.5,10.1-10.1,10.1c-5.6,0-10.1-4.6-10.1-10.1c0-5.6,4.5-10.1,10.1-10.1C84.2,46.7,88.7,51.3,88.7,56.8z};
  }
}

\newcommand\orcidicon[1]{\href{https://orcid.org/#1}{\mbox{\scalerel*{
\begin{tikzpicture}[yscale=-1,transform shape]
\pic{orcidlogo};
\end{tikzpicture}
}{|}}}}

\title[Secondary bias in line-intensity mapping]{Scatter in the star formation rate–halo mass relation: secondary bias and its impact on line-intensity mapping}
\author[R. L. Jun et al.]{Rui Lan Jun$^{1 \orcidicon{0009-0006-7907-1283}}$, Tom Theuns$^{2\, \orcidicon{0000-0002-3790-9520}}$, Kana Moriwaki$^{1,3 \, \orcidicon{0000-0003-3349-4070}}$, Sownak Bose$^{2\, \orcidicon{0000-0002-0974-5266}}$\\
$^{1}$ Department of Physics, Graduate School of Science, The University of Tokyo, 7-3-1 Hongo, Bunkyo, Tokyo 133-0033, Japan\\
$^{2}$ Institute for Computational Cosmology, Durham University, South Road, Durham DH1 3LE, UK\\
$^{3}$ Research Center for the Early Universe, Graduate School of Science, The University of Tokyo, 7-3-1 Hongo, Bunkyo, Tokyo 113-0033, Japan}
\date{Accepted XXX. Received YYY; in original form ZZZ}
\pubyear{2024}

\begin{document}

\maketitle

\begin{abstract}

We use the \illus{} cosmological hydrodynamical simulations to study the impact of secondary bias -- specifically, the correlation between star formation rate (\sfr{}) and halo bias at fixed halo mass -- on the line-intensity mapping (\lim{}) power spectrum. In \lim{}, the galaxy contributions are flux-weighted, and therefore depend on the luminosity of emission line.
We show that the (ensemble-averaged) large-scale two-halo term of the power spectrum depends only on the mean luminosity–halo mass relation if the scatter is uncorrelated with halo bias. However, when luminosity correlates with halo bias at fixed mass, this assumption breaks down. 
For many emission lines (e.g. \ha{}), luminosity is strongly correlated with \sfr{}, making the \sfr{}-weighted power spectrum important to study. 
In \illus{}, secondary bias increases the two-halo term of the \sfr{}-weighted power spectrum by 5 per cent at $z \sim 1.5$ compared to a model with random scatter. We also find that \sfr{}s of central and satellite galaxies are correlated, enhancing the one-halo term -- which depends on the distribution of \sfr{} inside the halo -- by 10
per cent relative to random pairings. To mitigate secondary bias in the two-halo term, we identify halo concentration (for haloes with mass $\log M_h \lesssim 12$) and satellite mass (for $\log M_h \gtrsim 12$) as effective secondary parameters. These results highlight the need to account for secondary bias when building mock catalogues and interpreting \lim{} observations.

\end{abstract}


\section{Introduction}

The first large map of the Universe showed that galaxies 
sampled on scales of up to $\sim 200~\unit{Mpc}$ around us are not randomly sprinkled throughout space but rather display a \lq web-like\rq\ pattern of sheets and filaments of galaxies delineating voids in which galaxies are rarer \citep{Geller89}. This result of the \lq Center for Astrophysics\rq\ (CfA) Redshift Survey has been confirmed by subsequent surveys, such as the Two-degree Field Galaxy Redshift Survey (2dFGRS, \citealt{Colless_2001}) and the Sloan Digital Sky Surveys (SDSS, \citealt{Blanton_2017}), which sample much larger scales of several $\unit{Gpc}$.

The large-scale structure is thought to originate from small, Gaussian density perturbations in the early Universe, imprinted by inflation, which grew through gravitational instability at a rate determined by the cosmological parameters (see e.g. \citealt{LSS_2006} for a review). The details of this large-scale structure and its evolution thus provide a means to study the Universe’s initial conditions and to infer key cosmological parameters that govern its evolution.

Current galaxy surveys, such as Euclid \citep{Ballardini24} and \desi{} \citep{Desi_2016, Desi25} are constructing higher resolution, larger volume maps of the Universe. These enable increasingly accurate measurements of the statistical properties of the large-scale structure and offer deeper insights into the role of dark matter and dark energy in structure formation. Measuring large scales is important because they are less affected by non-linear growth, making them more straightforward to connect to the initial conditions. The smaller scales are interesting because they provide insight into how galaxies evolve. However, obtaining a large map down to faint luminosities is very challenging.

Line-intensity mapping (\lim) is an emerging observational technique that complements current galaxy surveys by efficiently mapping similarly large or potentially even larger volumes. \lim{} surveys (e.g. SPHEREx; \citealt{SPHEREx_2018}, CONCERTO; \citealt{Concerto_2020}) measure the combined flux detected from all sources in a voxel (3D equivalent of pixel) that spans two spatial dimensions and one spectral (wavelength) dimension. 
The wavelength range can be chosen to correspond to a given emission line emitted by galaxies in a small range of redshifts, for example, hydrogen's $n=3\to 2$ H$\alpha$ recombination line centred at redshift $z\sim 1$. The same \lim{} survey can then be used to study H$\alpha$ emission at a different redshift, or from a different line at the same redshift, by simply choosing another wavelength range. The great advantage of \lim{} is that the spectrograph is very efficient, and that the survey strategy does not require us to first identify galaxies. In addition, the redshift and spatial resolution can be optimised for the particular experimental strategy. By detecting total flux, \lim{} also captures emission from sources that may be too faint to detect individually. 
A more in-depth discussion of \lim{} can be found in the review by \citet{Kovetz_2017}, while \citet{Schaan+21-galaxy} compare the \lim{} strategy to that of galaxy surveys.

The survey data need to be compared to models to extract cosmological information.
Hydrodynamical simulations attempt to model the complex processes that determine how galaxies evolve as their host halo grows in mass (see e.g. \citealt{Vogelsberger20, Crain23} for reviews). Unfortunately, these simulations are, at present, too expensive computationally to be able to make a mock Universe on the observed scale with sufficient resolution to resolve the details of galaxy formation. As a result, dark matter-only simulations are often used as an alternative, with haloes populated with galaxies using a statistical model such as the halo occupation distribution ({\sc hod}) method of \cite{Berlind02} or the subhalo abundance matching ({\sc sham}) technique introduced by \cite{Vale04}.

The models can also be used to study the statistical significance of inferences by generating many realisations of the dark matter density field for a given choice of cosmological parameters. It is crucial that these models populate haloes with the correct statistics to avoid biases. For example, the properties of a galaxy may depend on parameters other than just the mass of its host halo.\footnote{Such dependencies have been inferred from the clustering of galaxies in the {\sc sdss} and {\sc boss} surveys \citep[e.g.][]{Vakili2019,Walsh2019,Yuan2021,Salcedo2022,Zhai2023}.} If such dependencies are not captured in models, it can hinder a fair comparison between a cosmological model and the observed Universe.

In mocks for \lim{}, it is typically the star formation rate (\sfr{}) that needs to be modelled accurately, since the luminosity -- and thus the flux -- of many of the strong emission lines of galaxies depends on the galaxy’s \sfr{}. This is especially important for the \lim{} power spectrum, where galaxies contribute weighted by their flux.
A common method is to use an \sfr-halo mass relation to assign \sfr s to haloes according to their mass \citep[e.g.][]{Fonseca_2017,Silva_2018} and then relate \sfr{} to line luminosity using some relation that depends on the type of emission line measured in the \lim{} survey.
However, while halo mass is certainly a useful predictor of \sfr{}, hydrodynamical simulations show that there is large scatter around the mean \sfr-halo mass relation.

\citet{Li16} explore how the standard deviation of the scatter between \sfr{} and halo mass affects the \lim{} power spectrum of galaxies weighted by the luminosity of the CO line (see their figure 5). They conclude that random scatter only affects the power spectrum if the standard deviation is \lq large\rq. In contrast, \citet{Liu_2024} find that bursty star formation affects the [\ion{C}{II}] \lim{} power spectrum in a way that cannot be explained by mass-independent scatter. 

Another important consideration is whether the scatter is random. If it is not, how does the bias in the scatter affect the comparison between different models or between models and data? 
While shot noise increases in the same way regardless of whether the scatter is random, the two-halo term (which arises from the clustering of haloes, and dominates the galaxy power spectrum on sufficiently large scales) can be affected if the weight is not solely dependent on halo mass.
In the case of random scatter, the ensemble average of the two-halo term can be characterised by the mean relation between halo mass and weight \citep[see e.g.][]{Schaan+21-multi}.
However, if the scatter is not random -- but instead correlated with halo bias -- then the two-halo term will be sensitive to the nature of this so-called \lq secondary bias\rq. Although secondary bias (or secondary halo bias) is commonly used to refer to the dependence of halo bias on secondary halo properties, we use the term secondary bias more broadly to also include the correlation of galaxy properties with halo bias at fixed halo mass.

One well-studied example of secondary bias is assembly bias,\footnote{Although many secondary properties have been found to be correlated with the assembly history of the dark matter halo, this does not necessarily need to be the case. Therefore, the term secondary (halo) bias is preferred for generality (see e.g. \citealt{Mao_2018}).}
where the clustering of haloes of a given mass depends on their formation history \citep{Gao_2007}. 
While halo clustering is primarily a function of mass -- with more massive haloes more strongly biased than less massive ones \cite[e.g.][]{Kaiser_1984, Cole89} -- it has been found for low mass haloes that, at fixed mass, those that formed earlier tend to be more strongly biased than those that formed later \citep{Gao_2004}. In the \eagle{} simulations \citep{Schaye15, Crain15}, those haloes that form earlier also host more massive galaxies \citep{Matthee17}, in agreement with the findings by \citet{Xu_2020} for the \illus\ simulation. Therefore assembly bias together with a dependence of galaxy properties on the assembly history will affect how galaxies of a given mass are clustered. 
In addition to assembly bias, other examples of secondary bias include the systematic dependence of galaxy properties on factors such as spin, shape, amount of substructure, and environment \citep[e.g.][]{Mao_2018,Montero_2024,Balaguera_2024}.
Studies have shown that {\sc hod} or {\sc sham} models that neglect secondary bias can introduce offsets in the clustering signal, leading to systematic errors in galaxy-halo connection and precision cosmology studies \citep[e.g.][]{Zentner_2014,Chaves-Montero_2016, Yuan_2022}.

\citet{Xu_2020} investigate central galaxies in the \illus\ simulations and find that, while {\em stellar mass} correlates strongly with several secondary halo properties, the galaxy's \sfr{} shows no significant correlation with these secondary properties.
Likewise, attempts to use machine learning to model the \sfr\ and its scatter by including properties in addition to the halo mass have not been particularly successful \citep[e.g.][]{Jespersen_2022,Chittenden_2023, Hernandez_2023}. Currently, models that use the full merger tree of the host halo do not predict the galaxy's \sfr{} better than models that just use the final halo mass.

In this paper, we demonstrate how secondary bias in the scatter of the luminosity–halo mass relation impacts the galaxy power spectrum when galaxies are weighted by their luminosity, as is appropriate for \lim. For simplicity, we consider all galaxies at a single redshift such that the luminosity-weighted power spectrum is equivalent to the flux-weighted one.
We first demonstrate this mathematically for luminosity-weighted power spectra in general, before examining the behaviour of secondary bias in the \illus{} simulations \citep{Illustris_TNG_Nelson_2018,IllustrisTNG_Pillepich_2018,IllustrisTNG_Springel_2018,IllustrisTNG_Marinacci_2018,IllustrisTNG_Naiman_2018} when considering the \sfr{}-weighted power spectrum.
We also use the \eagle{} simulations \citep{Schaye15, Crain15} for some additional tests.
We investigate the power spectrum in 3D real space to enable us to discuss the effects on the two- and one-halo terms separately.
A realistic \lim{} mock should also account for other effects that we do not consider here, such as the impact of interlopers, or the relation of the \sfr{} to observed flux for different emission lines: we limit ourselves to secondary bias for clarity and simplicity. Although our analysis is general, we focus mostly on the \sfr{} in galaxies at redshift $z \sim 1.5$, which is particularly relevant for \lim{} in the H$\alpha$ line. H$\alpha$ can also be emitted by active galactic nuclei (\agn), but we will neglect this complication in our analysis.

The paper is organised as follows. In \cref{sec:power_spectrum}, we introduce the basics of the weighted power spectrum, briefly reviewing the analysis presented in \cite{Jun_2024}. In \cref{sec:sfr-mass}, we examine the \sfr{}-halo mass relation in \illus{}. In \cref{sec:assignment_effect}, we show, both mathematically and through toy models, how the scatter in the luminosity-halo mass relation affects the power spectrum. In \cref{sec:shuffling_method}, we use the \illus{} simulations to illustrate the effect of secondary bias on the \sfr{}-weighted power spectrum, and discuss the incorporation of secondary properties to reduce the offset introduced. In \cref{sec:discussion}, we discuss the implications of the results, the limitations of our method and suggest possible avenues for future work. Finally, we summarise our findings in \cref{sec:summary}.
Much of our analysis is based on the \illus{} simulations. We provide a brief overview of these cosmological hydrodynamical simulations in \cref{app:tng}, together with a description of how we assign line luminosities to the simulated galaxies.

In the following, unless otherwise stated, we will express distances such as $\bm{r}$ in comoving $\unit{h^{-1}Mpc}$, wave-vectors $\bm{k}$ in the inverse of that, and masses in units of \massunit. The Hubble constant at redshift $z=0$ is $H_0=100h~\unit{km~s^{-1}~Mpc^{-1}}$, where $h = 0.6774$ for \illus{}.

\ifSubfilesClassLoaded{%
  \bibliography{bibliography}%
}{}

\end{document}

\section{Power spectrum}

We begin this section with a brief overview of how the 3D power spectrum of galaxies is related to that of haloes, as discussed in more detail in \citet{Jun_2024} (hereafter \citepaperone; see also references therein). Consider a density field $\rho(\bm{r})$, where $\bm{r}$ is comoving position. We relate the Fourier transform of the overdensity, $\delta(\bm{r})=(\rho(\bm{r})-\bar{\rho}(\bm{r}))/\bar{\rho}(\bm{r})$ to its power spectrum, $P(k)$, in the usual way, as
\begin{equation}\label{delta_ps}
P(k) = V\langle |\delta(\bm{k})|^2 \rangle\,.
\end{equation}
Here, $V$ is the volume of the Universe over which the field is assumed to be periodic (or the computational volume in the case of a simulation), and $\langle\cdot\rangle$ denotes the ensemble average. 
We use the usual Fourier convention,
\begin{equation}
\delta(\bm{k}) = \frac{1}{V} \int \delta(\bm{r}) e^{-i \bm{k\cdot r}} \mathrm{d}^3 \bm{r}\,,
\end{equation}
following \cite{Peebles80}. 

In \citepaperone, we derived the following expression for the 3D galaxy power spectrum in real space,
\begin{align} \label{eq:galaxy_ps_all}
    P^\mathrm{gal}_{\mathrm{tot}}(k)
    &= \underbrace{P_{2h}(k)}_{\hbox{two-halo term}} 
    + \underbrace{U(k)^2(P_{\rm shot}^{\rm halo} - P^{\mathrm{gal}}_{\rm shot}) + P^{\mathrm{gal}}_{\rm shot}}_{\hbox{\rm one-halo term}}\,.
\end{align}
The two-halo term depends on the clustering of haloes, and the one-halo term
depends on the distribution of galaxies inside haloes. The galaxy shot noise term, $P_{\rm shot}^{\rm gal}$, depends on the number density and weights of galaxies, and the halo shot noise term, $P^{\mathrm{halo}}_{\rm shot}$, depends on the number density and weights of the haloes; both are independent of $k$. The function $U(k)$ quantifies the spatial distribution of galaxies inside individual haloes; $U(k) \to 1$ as $k \to 0$ and $U(k) \to 0$ as $k \to \infty$. 
As a consequence, the one-halo term tends to $P_{\rm shot}^{\rm halo}$ as $k \to 0$, and $P_{\rm shot}^{\rm gal}$ as $k \to \infty$. Loosely speaking, $U(k)$ represents a weighted average of (the Fourier transforms of) the halo \lq profiles\rq\ -- i.e. it quantifies how galaxies are distributed in terms of a central galaxy with associated satellites.\footnote{The profile refers to how galaxies are distributed inside a halo and may be different to how mass is distributed inside that halo.}
Several papers present in-depth reviews of this halo model of galaxy clustering; see, for example, \cite{Cooray02} and \cite{Asgari23}.

For the purposes of this study, it is sufficient to understand that $U(k)$ changes if the distribution of galaxies within a halo changes (an aspect we'll examine in more detail in \cref{sec:central_satellite}). For details of the derivation of Eq.~(\ref{eq:galaxy_ps_all}), citations to original work, and for tests of how well this equation describes the distribution of galaxies in the \illus\ simulation, see \citepaperone.

The two-halo term depends on how haloes are clustered.\footnote{On scales close to the size of haloes, the two-halo term will also be affected by the distribution of galaxies within haloes but we ignore that effect for simplicity (see \citepaperone).} Clustering of haloes can be related to that of the underlying mass distribution by introducing a bias factor, $b$, which relates the halo overdensity, $\delta$, to the matter overdensity, $\delta_\mathrm{m}$, as \citep[e.g.][]{Cole89, Tinker10}
\begin{align}
    \delta &= b \delta_\mathrm{m}\,.
\end{align}
The bias factor depends on halo mass and scale, but there may also be a dependence on secondary halo parameters -- the subject of this paper.

For a sample of $N$ haloes with different $b$, the average halo overdensity field is related to that of the matter overdensity field by the mean weighted bias,
\begin{align}\label{eq:bias}
    \bar{b} &= \frac{\sum_i^N{b_i W_i}}{\sum_i^N{W_i}}\,,
\end{align}
where $W_i$ is the weight each halo contributes to the power spectrum. In a standard galaxy survey, each galaxy is typically weighted equally; therefore, the weight of the halo is proportional to the number of galaxies it hosts. If only the central galaxy is observed, then the net bias is the arithmetic mean bias of the haloes, $\bar{b} = \sum_i^N{b_i}/N$.

The two-halo term in \cref{eq:galaxy_ps_all} can be expressed in terms of the bias $\bar{b}$ and the matter power spectrum $P_\mathrm{m}(k)$ as follows:
\begin{align}\label{eq:2halo}
    P_{2h}(k) &= \bar{b}(k)^2 P_\mathrm{m}(k).
\end{align}
In addition to the two-halo term, the power spectrum of a set of discrete objects (such as haloes) that are sampled from a continuous underlying distribution (the continuous matter density field), has a shot noise component. 
This extra term arises from the inherent randomness of the sampling process, which introduces statistical fluctuations in the distribution of the discrete objects. If the objects are independently sampled from a probability distribution with no additional constraints, then Poisson sampling is an appropriate approximation for describing this \lq stochasticity\rq\ (see also \citepaperone{} and \citealt{Baldauf_2013}). The Poisson 
shot noise term is given by 
\begin{align}\label{eq:shotnoise_discrete}
    P_{\mathrm{shot}} &= V \frac{\sum_i{W_i^2}}{(\sum_i{W_i})^2}\,.
\end{align}

Since galaxies reside within haloes, the galaxy power spectrum receives contributions from both the stochasticity of haloes and that of galaxies.
In addition, haloes are not sampled independently, as they exclude each other, introducing additional contributions to the stochasticity in addition to the Poisson shot noise (see \citepaperone). In this paper, the effects of halo exclusion are included in the two-halo term of the power spectrum, rather than treated separately.

In \lim, we obtain a map of voxels, where the \emph{specific} intensity (intensity per unit frequency) in each voxel is given by
\begin{align}\label{intensity discrete}
I &= \frac{\sum_i{F_i}}{\mathrm{d}\Omega_{\mathrm{vox}}\mathrm{d}\lambda_{\mathrm{obs}}} = \frac{\sum_i{L_i}}{{\mathrm{d}\Omega_{\mathrm{vox}}\mathrm{d}\lambda_{\mathrm{obs}}\,4\pi D^2_L}},
\end{align}
where d$\Omega_{\mathrm{vox}}$ is the solid angle extended by the voxel and d$\lambda_{\mathrm{obs}}$ is the wavelength extent of the voxel. The source flux is given by $F_i = L_i/\,4\pi D^2_L$, where $L_i$ is the luminosity and $D_L$ is the luminosity distance. The sum is over all the sources that contribute to the voxel. 

Similarly, the specific mean intensity in a survey volume can be computed by replacing d$\Omega_{\rm vox}$ with the solid angle of the survey, d$\Omega$, and summing the line fluxes of all galaxies in the survey volume:
\begin{align}\label{eq:mean_intensity_discrete}
\bar{I} &= \frac{\sum_i{F_i}}{{\mathrm{d}\Omega \mathrm{d}\lambda_{\mathrm{obs}}}} = \frac{\sum_i\,L_i}{\mathrm{d}\Omega \mathrm{d}\lambda_{\mathrm{obs}}\,4\pi D^2_L}\,.
\end{align}
This also equals the mean value of $I$ over all voxels.
In \lim, although we can only measure the specific intensity of the voxels, the weight $W_i$ in \cref{eq:bias} \& \cref{eq:shotnoise_discrete} is given by the flux of objects, and not the specific intensity of voxels (see \citepaperone{}). To focus on the effects of scatter and secondary bias, we consider galaxies at a single redshift in this study, in which case the flux is related to the luminosity by a constant factor. However, when comparing to observations where galaxies span a range of redshifts, it is necessary to account for the redshift dependence of the luminosity distance.

It is common practice to compute the power spectrum of an overdensity field, rather than of the field itself. In \lim{}, that would correspond to computing the power spectrum of $I/\bar{I}-1$. However, measuring $\bar{I}$ can be difficult \citep{Schaan+21-galaxy}, and the power spectrum quoted in \lim{} surveys is often
that of $I$ itself. That power spectrum is given by $P_{I} (k) = V\bar{I}^2\langle |\delta(\bm{k})|^2 \rangle$. The corresponding two-halo term is then
\begin{align}\label{eq:2halo_unnormalised}
    P_{2h,I} &= \bar{I}^2 \bar{b}(k)^2 P_\mathrm{m}(k),
\end{align}
and the shot noise term $P_{\mathrm{shot},I} = \bar{I}^2 V \sum_i{L_i^2}/(\sum_i{L_i})^2$.
We will refer to the power spectrum $P$ of $I/\bar{I}-1$ simply as \lq the power spectrum\rq, whereas that of $I$ itself as the \lq unnormalised power spectrum\rq, $P_I$. This distinction is helpful for discussing changes in the amplitude and shape of the power spectrum separately.

\ifSubfilesClassLoaded{%
  \bibliography{bibliography}%
}{}

\end{document}\label{sec:power_spectrum}

\section{The SFR-halo mass relation in the IllustrisTNG simulation }\label{sec:sfr-mass}

\begin{figure*}
\centering
\begin{subfigure}[t]{.5\textwidth}
    \centering
    \includegraphics[width=\linewidth]{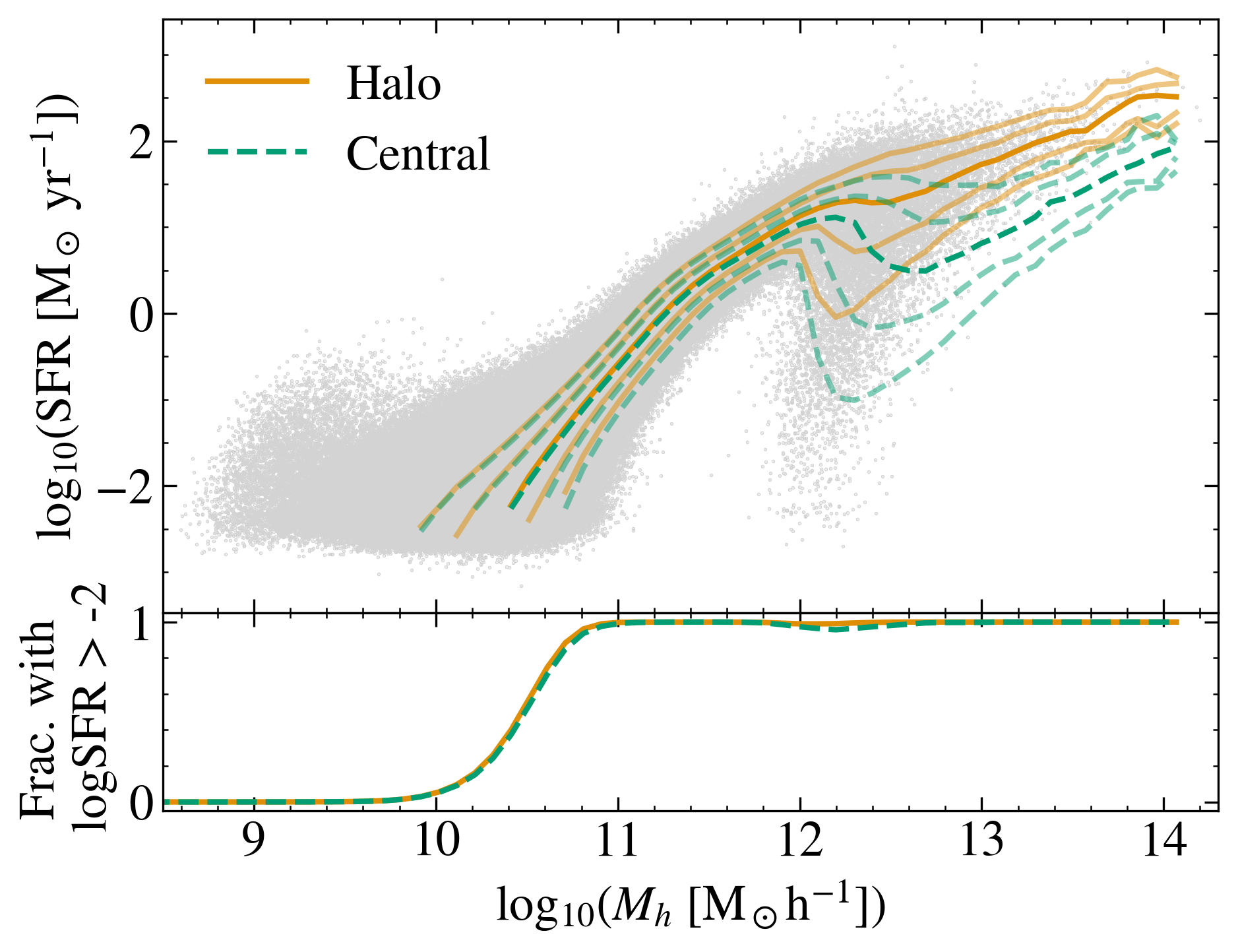}
    \label{fig:group_sfr_mass}
\end{subfigure}%
\begin{subfigure}[t]{.5\textwidth}
    \centering
    \includegraphics[width=\linewidth]{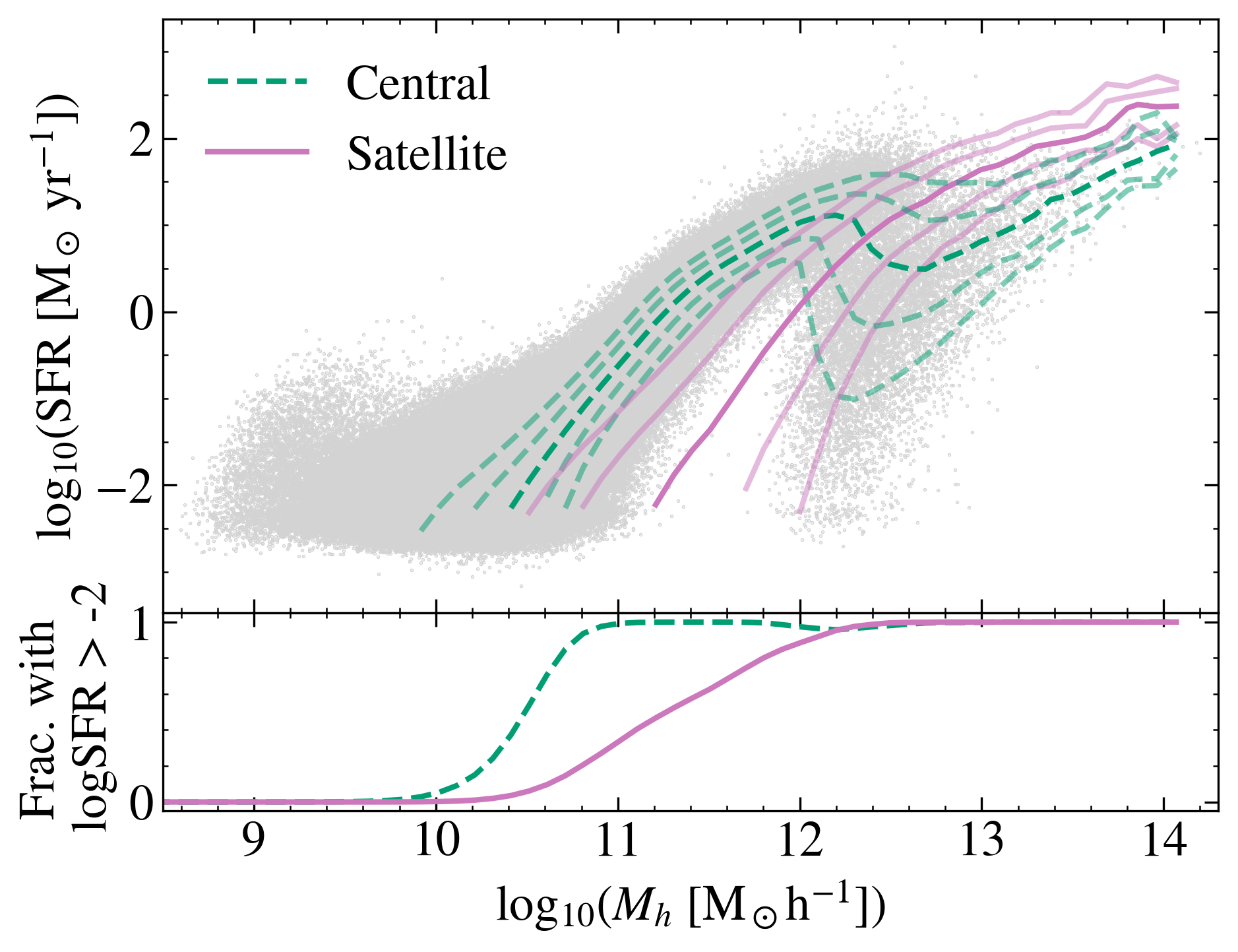}
    \label{fig:cent_sfr_mass}
\end{subfigure}
\caption{\sfr{}-virial mass relation for haloes in TNG300-1 at $z = 1.5$ for the halo \sfr{} (\emph{orange}), central \sfr{} (\emph{green}) and satellite \sfr{} (\emph{violet}). The \sfr{}s of all satellite subhaloes of a given friends-of-friends halo are summed to give the satellite \sfr{} of the halo. 
The $\{5,25,50,75,95\}^{\mathrm{th}}$ percentiles are plotted from bottom to top. The lower panels show the fraction of haloes with \sfr{} $> 10^{-2}$ \sfrunit. \\
\textbf{\textit{Left panel}}: The halo \sfr{} (\emph{orange}) largely follows the central \sfr{} (\emph{green}) relation up to $\log M_h \sim 12$, above which the group \sfr{} becomes noticeably larger than the central \sfr{}. The central \sfr{} relation decreases due to \agn{} feedback quenching the \sfr{} of central galaxies. Each \emph{grey dot} represents the total \sfr{} of a halo. \\
\textbf{\textit{Right panel}}: The total satellite \sfr{} (\emph{violet}) is lower than the central \sfr{} (\emph{green}) relation for $\log M_h \lesssim$ 12, and increases to be higher than the central \sfr{} relation for $\log M_h \gtrsim 12$. Each grey dot represents the central \sfr{} of a halo. \\
\textbf{\textit{Summary}}: While there is a general relation between \sfr{} and halo mass, there is significant scatter around this relation.
}
\label{fig:sfr-mass}
\end{figure*}

\begin{figure}
    \centering
    \includegraphics[width=\linewidth]{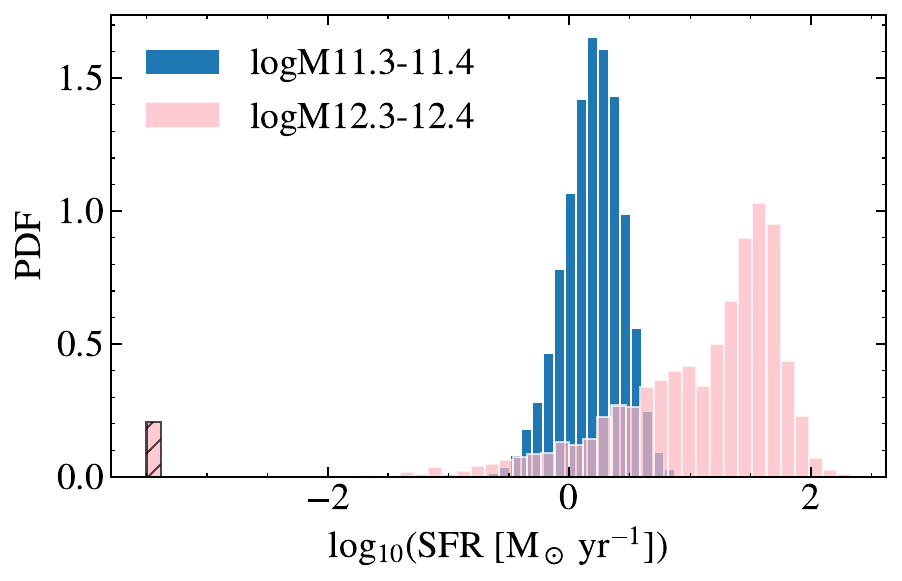}
    \caption{Probability distribution functions (PDFs) of \sfr{}s for haloes of a given mass, defined as PDF = ($n_{\mathrm{halo}}$ in \sfr{} bin)/($n_{\mathrm{halo}}$ in mass bin $\times$ \sfr{} bin width). The halo masses considered are $\log M_h \in (11.3,11.4)$ (blue) and $\log M_h \in (12.3,12.4)$ (pink). The hatched area corresponds to haloes with \sfr{} = 0 in the simulation. The distribution of \sfr{}s is approximately Gaussian for lower halo masses but is non-Gaussian for higher halo masses due to \agn{} quenching.
 }
    \label{fig:sfr_dist}
\end{figure}

\begin{figure}
    \centering
    \includegraphics[width=\linewidth]{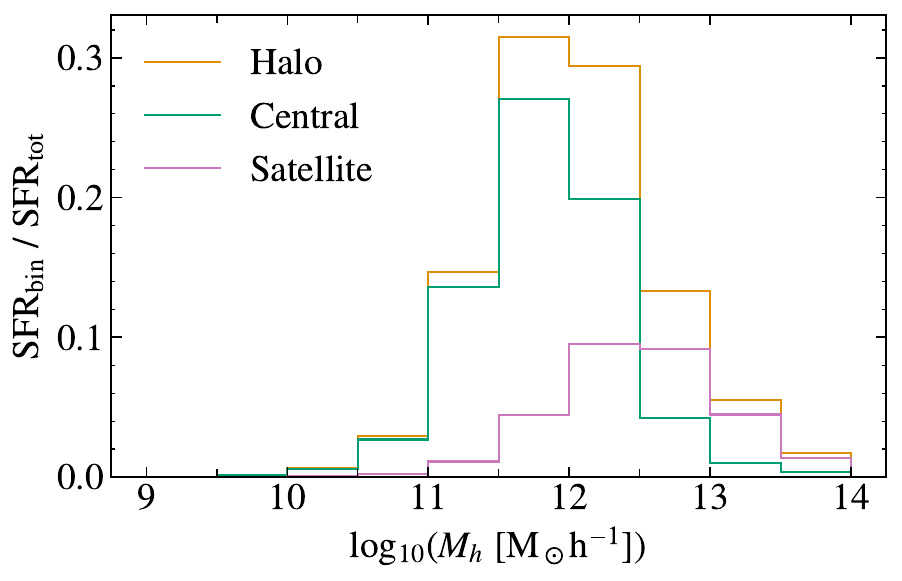}
    \caption{The contribution from different halo mass ranges to the total \sfr{} in \tng{}.
    The fraction of the total \sfr{} contributed by central galaxies (\emph{green}), satellite galaxies (\emph{pink}), and their combined contribution (\emph{orange}) are shown. The contribution shown is from halo mass bins of width $\Delta \log_{10}(M_h\ [\massunit]) = 0.5$ dex; these are the bins we use to investigate the power spectrum in \cref{sec:shuffling_method}. 
 }\label{fig:contribution_mass_bins}
\end{figure}

In the previous section, we showed that the galaxy power spectrum measured in \lim{} depends on line luminosities. 
It is common to assume that the luminosities of several emission lines, including H$\alpha$, [\ion{O}{III}] and [\ion{O}{II}], scale linearly with \sfr{} \citep[e.g.][]{Kennicutt98,Gong_2017}:
\begin{align}
    L (\unit{erg\ s^{-1}}) &= K\ \mathrm{SFR} (\unit{M_\odot\ yr^{-1}})\,,
\end{align}
where $K$ is a constant.
Even if the relation is not strictly proportional, \sfr{} remains a key component in modelling line luminosities for many lines \citep[e.g.][]{Fonseca_2017,Sato-Polito_2023}. Therefore, we focus on investigating the \sfr{}-halo mass relation as a basis for the luminosity-halo mass relation.
See \cref{app:sfr-luminosity} for a more detailed discussion of the relation between \sfr{} and luminosity, with particular focus on the \ha{} line.
In this section, we examine the \sfr{}-halo mass relation in the TNG300-1 (hereafter \tng) hydrodynamical simulation of the \illus{} project \citep{Illustris_TNG_Nelson_2018,IllustrisTNG_Pillepich_2018,IllustrisTNG_Springel_2018,IllustrisTNG_Marinacci_2018,IllustrisTNG_Naiman_2018}. We will investigate the impact of scatter in this relation on the power spectrum in \cref{sec:shuffling_method}.
\Cref{app:tng} provides a brief overview of the simulation and defines the quantities we use from the simulation. 

We plot the \sfr{}-halo mass relations for \tng{} at $z = 1.5$ in the upper panels of \cref{fig:sfr-mass}. 
In this paper, we adopt the virial mass for the dark matter haloes from \tng{}. Since our results are not significantly affected by the specific mass definition, we use $M_h$ to refer to the halo mass in general.
Grey dots show the \sfr{} summed over all galaxies in a given halo (hereafter \lq halo\rq\ \sfr{}; left panel) and that of the central galaxies alone (right panel). The lines show percentiles of \sfr{} in bins of halo mass for the halo \sfr{} (orange), the \sfr{} in central galaxies (green), and the total \sfr{} of all satellites in a halo (violet). The lower panel shows the fraction of haloes for which the halo \sfr{} is $> 10^{-2}$ \sfrunit: close to and below this value, the \sfr{} is unlikely to be numerically well-resolved in the simulation. 

\Cref{fig:sfr-mass} shows that the \sfr{}-halo mass relation is approximately a power law up to a halo mass of $\log M_h\sim 12$ but with significant scatter. The fraction of haloes with \sfr{} is $> 10^{-2}$ \sfrunit{} decreases towards lower halo masses, becoming approximately zero below $\log M_h \sim 10$. There is a characteristic feature in the \sfr{}-halo mass relation at $\log M_h\sim 12$, where the scatter in the relation increases significantly and where the median \sfr{} in the central galaxy drops significantly. This feature is due to the onset of efficient \agn{} feedback
in the simulations \citep{Weinberger_2017}. Such large scatter in the \sfr{} of central galaxies in clusters of galaxies is also seen in observations \cite[e.g.][]{Scholtz18}.

The \agn{} feature at $\log M_h \sim 12$ is less noticeable for the total \sfr{} of satellite galaxies (violet line, right panel), but the slope in the \sfr{}-halo mass relation for satellites also becomes shallower at around this halo mass. The quenching mechanism for satellites is the combined effect of tidal stripping, ram-pressure stripping and galaxy harassment \citep[e.g.][]{Bahe17, Dawes_Review9}. The number of satellites is large for these more massive haloes and, although the severity of quenching may vary between satellites, summing the \sfr{}s of a larger number of satellite galaxies reduces the variation from halo to halo. If individual satellites do not need to be mapped, then taking the total satellite \sfr{} can reduce the variance in the \sfr{}-halo mass relation.\footnote{
The distribution of satellite \sfr{} typically affects the power spectrum on scales $k \gtrsim 1\ \kunit{}$ (see \citepaperone).}
\Cref{fig:sfr-mass} also shows that the median total satellite \sfr{} exceeds that of the central galaxy for haloes with $\log M_h\gtrsim 12.5$. We investigate the role of satellites in contributing to secondary bias in the power spectrum in \cref{sec:shuffling_method}. Lower mass haloes, $\log M_h\lesssim 12$, have fewer satellites and some may host no satellites at all. This increases the scatter in the total satellite \sfr{} in this mass range.

The scatter in the \sfr{}-halo mass relation is often assumed to be lognormal \cite[e.g.][]{Conroy_2009}.
We plot the probability distributions of \sfr{}s for haloes of mass $\log M_h \sim 11$ and $\log M_h \sim 12$ in \cref{fig:sfr_dist}. While the distribution of $\log$ \sfr{} at $\log M_h \sim 11$ (blue) is indeed approximately Gaussian, the distribution at $\log M_h \sim 12$ (pink) is far from Gaussian, with a long tail towards \sfr{}s much below the mean. This tail is a consequence of the onset of \agn{} feedback. In \cref{sec:lum_assignment_schemes}, 
we discuss whether the shape of the distribution affects the power spectrum.

The shape and level of scatter in the \sfr{}-halo mass relation may be dependent on the galaxy formation model. We examine this relation in the {\sc eagle} simulation in \cref{app:sfr-halo_mass_eagle_vs_tng} and find that, despite differences in galaxy formation and \agn{} feedback models between \tng{} and \eagle{}, the trends are remarkably similar.

Figure 5 of \citepaperone{} showed that the contribution from central galaxies to the total \sfr{} reaches a peak at $\log M_h \sim 12$ and that satellite galaxies start contributing more than central galaxies for $\log M_h \gtrsim 12.5$. \Cref{fig:contribution_mass_bins} presents a similar figure but here we consider coarser mass bins of width $\Delta \log_{10}( M_h\ [\massunit]) = 0.5$ dex. These are the bins used in our power spectrum analysis, and we will refer back to this figure in \cref{sec:shuffling_method}.

\ifSubfilesClassLoaded{%
  \bibliography{bibliography}%
}{}

\end{document}

\section{Impact of luminosity assignment scheme on the power spectrum}\label{sec:assignment_effect}

In the previous section, we examined the \sfr{}-halo mass relation in \tng{} to show its shape and scatter, which will translate into the luminosity-halo mass (\lm{}) relation for some emission lines. In this section, we consider the \lm{} relation more generally, without specific focus on the \sfr{}. The following analysis is therefore applicable to any emission line.

To investigate the impact of secondary bias on the power spectrum, it is essential to understand how scatter in the \lm{} relation affects the power spectrum. 
We express the components of the power spectrum in terms of luminosities as a function of halo mass in \cref{sec:ps_components_L(M)} and use those expressions to help us understand the impact of scatter on the power spectrum when varying the scatter model in \cref{sec:lum_assignment_schemes}.
\citet{Schaan+21-multi} showed that the two-halo term depends only on the mean luminosity.
We will demonstrate what this implies when scatter is introduced into the \lm{} relation;
we will show that, except in the presence of secondary bias, the impact on the ensemble-averaged two-halo term can be fully accounted for by changes in the mean relation, even when the scatter is mass-dependent.

In the following, we neglect the distribution of galaxies within haloes, and simply consider the \lq total\rq{} luminosities of haloes (\lq halo\rq{} luminosity), which is obtained by summing the luminosity of the halo's central galaxy and all its satellites. 
The unnormalised power spectrum of haloes weighted by luminosity, is
\begin{align}
    P^{\mathrm{halo}}_{\mathrm{tot},I}(k) = P_{2h,I}(k) + P^{\mathrm{halo}}_{\mathrm{shot},I}\,,
\end{align}
where $P_{2h,I}(k) = \bar{I}^2 \bar{b}(k)^2 P_\mathrm{m}(k)$ (\cref{eq:2halo_unnormalised}).
This is a good description of the galaxy power spectrum on scales beyond the size of haloes, such that $U(k)\approx 1$ in Eq.~(\ref{eq:galaxy_ps_all}), as shown in \citepaperone. We will come back to the separate contributions of centrals and satellites in \cref{sec:shuffling_method}.

\subsection{Dependence of power spectrum on luminosity-halo mass relation}\label{sec:ps_components_L(M)}
As shown in \cref{eq:2halo_unnormalised}, the unnormalised two-halo term can be written as the product of the squared mean intensity, the squared effective bias, and the matter power spectrum. In this section, we express the mean intensity and the bias in terms of the luminosity of haloes, so that we can understand how the two-halo term is affected by the \lm{} relation. The bias $\bar{b}$ can depend on $k$, but we omit this dependence here for brevity.

We denote the total luminosity of haloes in a narrow bin of halo mass centred at $M_j$ by
\begin{align}
    L_\mathrm{T}(M_j) &= \sum_{q=1}^{N(M_j)} L_{j,q}\,,
\end{align} 
where $L_{j, q}$ is the halo luminosity of the $q$-th halo in the mass bin $M_j$, and the sum runs over all $N(M_j)$ haloes within the bin. The average luminosity of a halo with mass $M_j$ is then
\begin{align}
    \bar{L}(M_j) &= \frac{L_\mathrm{T}(M_j)}{N(M_j)}\,.
\end{align}
Summing over all mass bins, the total luminosity contributed by all haloes is
\begin{align}
    L_{\mathrm{TOT}} &= \sum_j L_\mathrm{T}(M_j)\,.
\end{align}
Using this notation, the mean intensity, defined in \cref{eq:mean_intensity_discrete}, can be written as
\begin{align}\label{eq:mean_intensity_L}
    \bar{I} &= A \sum_j \sum_q L_{j,q} = A \sum_j L_\mathrm{T}(M_j) = A\sum_j \bar{L}(M_j) N(M_j)\,,
\end{align}
where $A \equiv  1/ (\mathrm{d}\Omega\mathrm{d}\lambda_{\mathrm{obs}}\,4\pi D_L^2)$. 

The luminosity-weighted mean bias for haloes with halo mass $M_j$ can be written as
\begin{align}\label{eq:b-bar_M}
    \bar{b}({M_j}) &= \frac{\sum_q{b_{j,q} L_{j,q}}}{\sum_q{L_{j,q}}} 
    = \frac{\sum_q{b_{j,q} L_{j,q}}}{L_\mathrm{T}({M_j})} 
    =\sum_q b_{j,q} w_{j,q}\,,
\end{align}
where $b_{j,q}$ is the bias of the q-th halo and $w_{j,q} = L_{j,q}/ L_\mathrm{T}({M_j})$. 
The luminosity-weighted mean bias of the entire halo population is then
\begin{align}\label{eq:b-bar_whole}
    \bar{b} = \sum_j{\overline{b}(M_j) f_j}\,,
\end{align}
where $f_j = L_\mathrm{T}(M_j)/ L_{\mathrm{TOT}}$ is the fractional contribution of the halo mass bin $j$ to the total luminosity. 

Finally, the shot noise component of the power spectrum, \cref{eq:shotnoise_discrete}, can be rewritten as
\begin{align} \label{eq:shot_L(M)}
    P_{\mathrm{shot}, I}
    &= \bar{I}^2 V\frac{\sum_j\sum_q L_{j,q}^2}{L_{\mathrm{TOT}}^2} \nonumber \\
    &= \bar{I}^2\sum_j P_{\rm shot}(M_j) f_j^2,
\end{align}
where 
\begin{align} \label{eq:shot_normalized_L(M)}
    P_{\rm shot}({M_j}) 
     &\equiv V \frac{\sum_q L_{j,q}^2}{L_\mathrm{T}(M_j)^2}\,
\end{align}
is the contribution to the shot noise from haloes in halo mass bin $j$. In the case where all haloes of mass $M_j$ have the same luminosity, the shot noise is $P_{\rm shot}({M_j})=V/N(M_j)$, the inverse of the mean number density of such haloes. This value represents the minimum possible value of the shot noise for these haloes: the larger the scatter in luminosity, the larger the shot noise.

\subsection{Effect of scatter in the luminosity-halo mass relation on the power spectrum}\label{sec:lum_assignment_schemes}
\begin{figure*}
        \centering
        \begin{subfigure}[b]{0.49\textwidth}
            \centering
            \includegraphics[width=\textwidth]{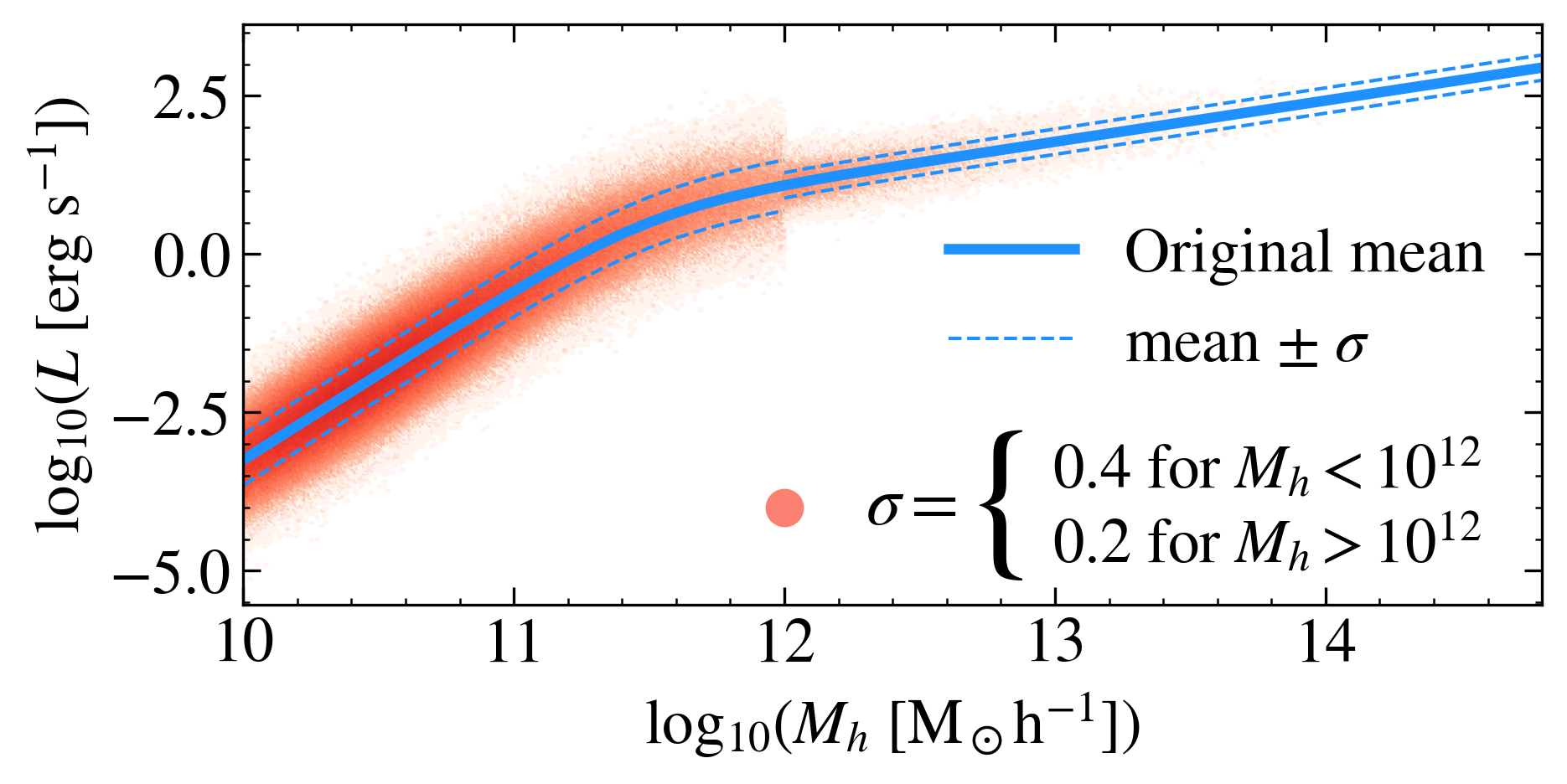}
        \end{subfigure}
        \hfill
        \begin{subfigure}[b]{0.49\textwidth}  
            \centering 
            \includegraphics[width=\textwidth]{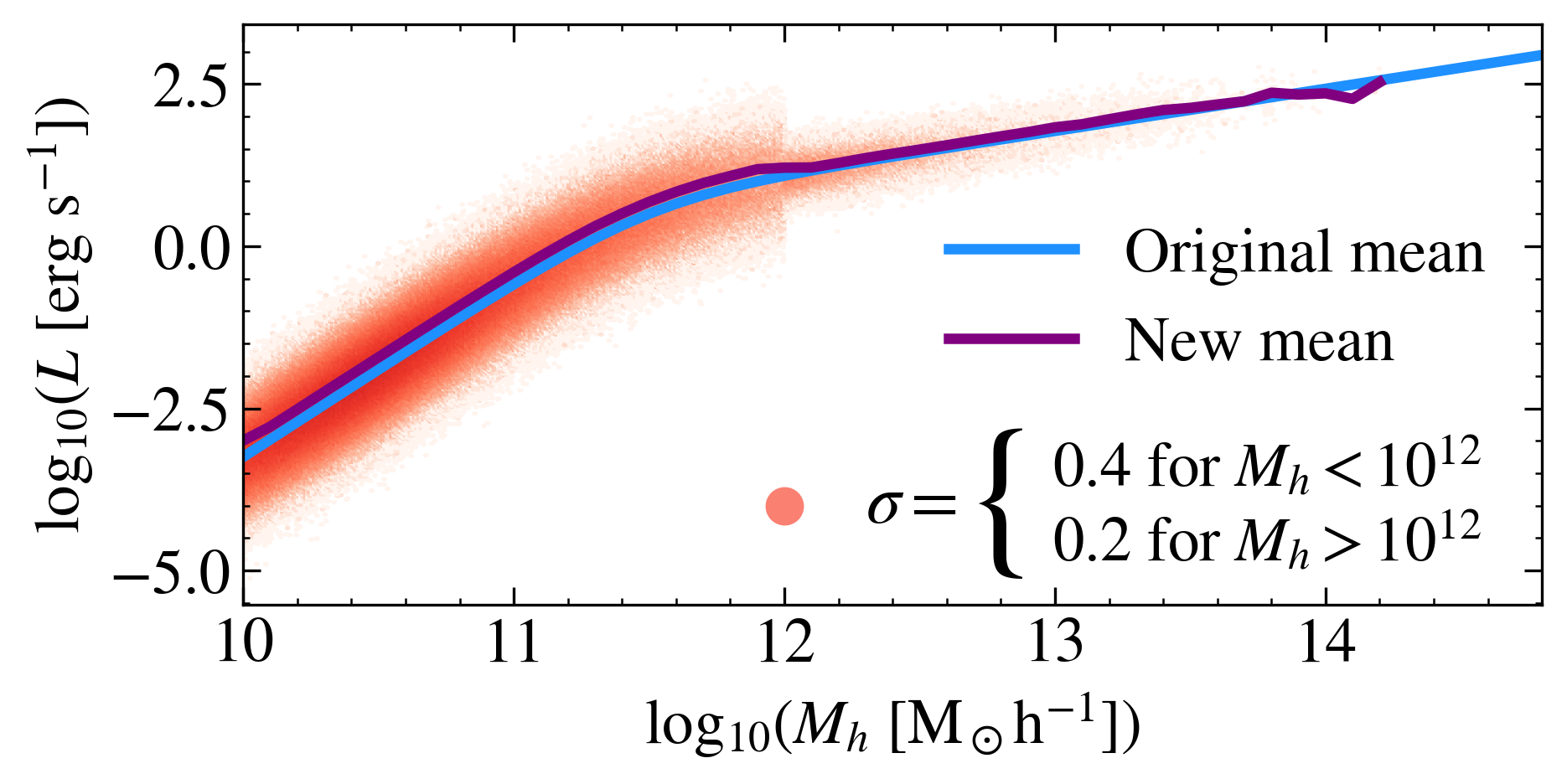}
        \end{subfigure}
        \vskip\baselineskip
        \begin{subfigure}[b]{0.49\textwidth}   
            \centering 
            \includegraphics[width=\textwidth]{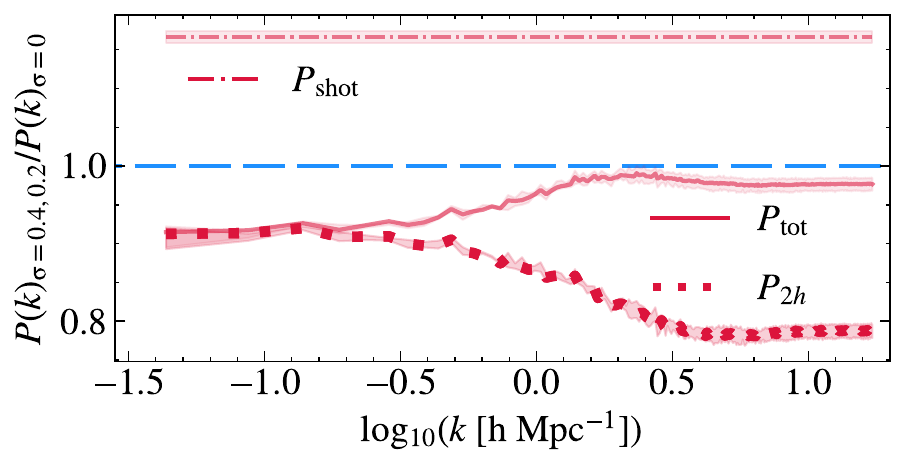}
        \end{subfigure}
        \hfill
        \begin{subfigure}[b]{0.49\textwidth}   
            \centering 
            \includegraphics[width=\textwidth]{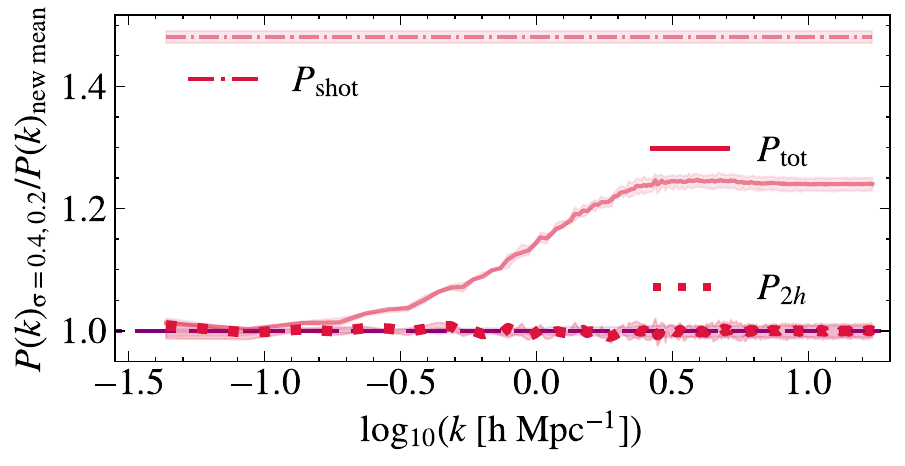}
        \end{subfigure}
        \caption
        {\textbf{\textit{Upper left panel}}: The \emph{blue solid line} shows the \lm{} relation given by \cref{eq:luminosity_mhalo_param}. 
        The \emph{red dots} are obtained by applying a lognormal scatter to this relation with $\sigma = 0.4$ for $\log M_h < 12$ and $\sigma = 0.2$ for $\log M_h > 12$. 
        The \emph{dashed blue lines} indicate the standard deviation around the mean.
        \textbf{\textit{Upper right panel}}: Same as upper left panel, with the addition of a \emph{purple solid line} which shows the new linear mean \lm{} relation after applying the scatter (see text). 
        \textbf{\textit{Lower left panel}}: The ratio of the power spectra for the case of mass-dependent lognormal scatter 
        (\emph{red dots} in upper panel) against that for a sample with a one-to-one \lm{} relation given by \cref{eq:luminosity_mhalo_param} (\emph{blue line} in upper panel) computed using the \tng{} halo catalogue. The \emph{red lines} represent the total power spectrum (\emph{solid}), two-halo term (\emph{dotted}), and shot noise (\emph{dash-dotted}) averaged over five realisations of scatter. The \emph{shaded bands} represent the 25th-75th percentile level.
        \textbf{\textit{Lower right panel}}: 
        Rather than taking the ratio with respect to the power spectrum for the case with the original logarithmic mean \lm{} relation (as in the lower left panel), the ratio is computed relative to that for a sample with the new (linear) mean \lm{} relation obtained after applying the mass-dependent scatter (purple line in the upper right panel).
        \textbf{\textit{Summary}}: The mean two-halo term depends only on the linear mean \lm{} relation relation, regardless of the scatter (provided the scatter is random).
        } 
        \label{fig:mass_dependent_scatter}
\end{figure*}

In this subsection, we will use the equations presented in \cref{sec:ps_components_L(M)} to help us understand the impact on the power spectrum of varying the scatter in the \lm{} relation. Note that the bias $b_{j,q}$ is set by the halo, and is unaffected by the luminosity. Therefore changing the \lm{} relation does not affect $b_{j,q}$. 

Before discussing the effects of scatter, let us briefly consider how modifications to a one-to-one \lm{} relation affects the power spectrum. A one-to-one \lm{} relation means that $L_{j,q}$ in \cref{eq:b-bar_M} is the same for all haloes of the same mass. If the original relation is simply multiplied by a constant factor $C$ then the new one-to-one relation is given by $L'(M) = C \cdot L(M)$: the total mean bias $\bar{b}$ (\cref{eq:b-bar_whole}) as well as the normalised shot noise $P_{\rm shot}$ are unchanged, and only the specific mean intensity $\bar{I}$ will change.\footnote{The same result follows when haloes do not follow a one-to-one \lm{} relation. If all haloes are multiplied by the same factor, then their relative weighting is unchanged, therefore the mean bias \rev{and shot noise are} unchanged.} However, if the shape of $L(M)$ changes, the relative weighting $f_j$ changes in \cref{eq:b-bar_whole} and \cref{eq:shot_L(M)}, leading to changes in both the two-halo term and shot noise. 

Now let us consider how adding scatter to a given one-to-one relation affects the power spectrum.

\vspace{1em}
\noindent {\bf Adding random scatter (linear).}
\hspace{1em}
    Consider adding random scatter to a given mean relation, so that $L'_{j,q} = \bar{L}(M_j) + \mathrm{scatter}$, with $\braket{L'_{j,q}}=\bar{L}(M_j)$.
    Although each random realisation of scatter will result in a slightly different power spectrum, the \emph{ensemble-averaged} two-halo term (the average over many realisations) depends only on the mean \lm{} relation, $\bar{L}(M)$, and not the variance \emph{if the scatter follows a random probability distribution}. 
    For completeness, we demonstrate this mathematically in \cref{app:scatter_maths} (see also \citealt{Schaan+21-multi}). 
    In contrast, the shot noise term is not linearly dependent on luminosity, and adding random scatter \emph{will} increase the shot noise, even if $\bar{L}(M)$ is unchanged.
    
\vspace{1em}
\noindent {\bf Adding random scatter (logarithmic).}
\hspace{1em}
Instead of linear scatter, it is more common to assume Gaussian scatter in $\log (L)$, i.e. lognormal scatter in $L$. While $\overline{\log L}$ is unchanged before and after adding this scatter, since $\overline{\log L} \neq \log\overline{L}$, the added scatter will change the \emph{linear} mean \lm{} relation --
and hence both the normalised and unnormalised power spectra. 

Consider a set of haloes of given mass, and assign luminosities to them according to a lognormal distribution with variance $\sigma^2$. 
The probability density function of the lognormal distribution is given by
\begin{align}
    P(L) &= \frac{1}{\ln(10)\,L \sigma\sqrt{2\pi}}\exp\left(-\frac{(\log_{10} (L/L_0))^2}{2\sigma^2}\right)\,,
\end{align}
where $L_0$ is the mean luminosity in the absence of scatter. 
The mean luminosity in the presence of scatter is proportional to $L_0$ \citep[see also][]{Sun_2019}, 
\begin{align}\label{eq:L_mean}
    \bar{L} &=  10^{\frac{1}{2} \sigma^2 \ln(10)} L_0\,.
\end{align}
When a lognormal scatter is applied, the mean luminosity increases, therefore the mean intensity will increase as well. The shot noise also increases as the variance of luminosities increases. 

When $\sigma$ is constant with respect to mass, the shape of $\bar{L}(M)$ does not change, and the mean relation is simply multiplied by a constant. Therefore the ensemble-averaged two-halo term is unchanged.
On the other hand, if $\sigma$ depends on mass, the contribution to the bias from each mass changes, and therefore the two-halo term changes \citep[see also][]{Murmu_2023}. 
We stress, however, that the change in the two-halo term arises from the change in the linear mean \lm{} relation when lognormal scatter is applied. The ensemble-averaged two-halo term in the presence of lognormal scatter can still be fully characterised by the (new) linear mean \lm{} relation.

We illustrate the effect of mass-dependent scatter using a toy model where a lognormal scatter with variance $\sigma = 0.4$ is applied to haloes (in \tng) with mass $\log M_h < 12$, and $\sigma = 0.2$ to those with $\log M_h > 12$. 
In the upper left panel of \cref{fig:mass_dependent_scatter} we plot the assumed $L(M)$ relation without scatter (blue line) and with scatter (red points). 
The equation for the blue line is the following double power-law,\footnote{This relation fits the H$\alpha$ luminosity-to-mass relation measured in the \tng{} simulation at $z=1.5$.}
\begin{align}\label{eq:luminosity_mhalo_param}
    L(M_h) &= 2 L_{1}\,\frac{M_h}{M_1} \left[ \left(\frac{M_h}{M_1}\right)^{-a} + \left(\frac{M_h}{M_1}\right)^b \right]^{-1}\,,
\end{align}
where $a = 1.7,\ b = 0.35,\ M_1 = 10^{11.5} \massunit$, and $L_{1}=6.32 \times 10^{41}\unit{erg~s^{-1}}$.
The upper right panel of \cref{fig:mass_dependent_scatter} shows that the lognormal scatter causes the linear mean \lm{} relation to change (purple line).

In the lower panels of \cref{fig:mass_dependent_scatter}, the ratios of the power spectra with and without scatter are shown for the two-halo term (dotted), shot noise (dash-dotted), and sum of them (solid). The original (left) and newly obtained (right) mean relations are used as the power spectra without scatter, respectively. 
While it appears at first sight that the two-halo term of the power spectrum changes (lower left panel), if the power spectrum with scatter is compared against that for the new mean relation, then the two-halo terms become identical. This confirms our earlier claim that the two-halo term changes due to the lognormal scatter changing the mean luminosity of galaxies hosted by a halo of given mass. 

For all cases, it holds that the ensemble-averaged two-halo term depends only on the linear mean \lm{} relation, {\em provided the scatter is random}.
We discussed this for the case of lognormal scatter, but in fact the same argument can be used regardless of the nature of the scatter, as we illustrate in \cref{app:tng_gaussian}. 
To summarise, random scatter affects the shot noise term but the ensemble-averaged two-halo term is unchanged.

\vspace{1em}
\noindent {\bf Luminosity correlated with bias}
\hspace{1em}
If the scatter is not random, but there is instead a correlation between the luminosity $L_{j,q}$ and bias $b_{j,q}$ of the halo, then the two-halo term is no longer solely determined by the mean \lm{} relation. For example, assume that haloes with higher $b_{j,q}$ also have higher $L_{j,q}$. In this case, the weighted mean bias $\bar{b}(M_j)$ (\cref{eq:b-bar_M}) is higher than the case where luminosity is uncorrelated with bias, and therefore the amplitude of the two-halo term increases (see \cref{eq:2halo}).
Such dependence of the bias on other halo properties
in addition to halo mass is known as secondary bias (the primary bias is simply the halo mass itself). 
Although galaxy assembly bias is sometimes used to refer to the correlation of galaxy properties with secondary halo properties or with halo bias, we refrain from using this term, as assembly bias is not the only form of secondary bias (see discussion in \citealt{Mao_2018}). 
Instead we use secondary bias as a more general term to include the correlation between luminosity and bias for haloes of a given mass.

To illustrate the impact of such correlated scatter, we consider a toy model in which half of the haloes in the mass bin $\log M_h \in (11, 11.1)$ are assigned a weight \(W = 1\), while the other half are assigned a weight \(W = 5\). 
As an example, we consider a proxy for halo concentration \citep[inspired by][]{Bose_2019},
\begin{align}\label{eq:conc_proxy}
    \tilde{c} &= \frac{V_\mathrm{max}/V_h}{R_\mathrm{max}/R_h} 
    = \frac{V_\mathrm{max}}{10\,H\,R_\mathrm{max}}\, ,
\end{align}
as a secondary halo property, where
$V_h=(GM_h/R_h)^{1/2}$ is the circular velocity at the virial radius $R_h$
of the halo with mass $M_h$, while $V_{\rm max}$ is the maximum circular velocity of the subhalo and $R_\mathrm{max}$ is the radius at which $V_{\rm max}$ is achieved (these are variables \texttt{SubhaloVmax} and \texttt{SubhaloVmaxRad} in the \tng{} database); $H$ is the Hubble constant. Note that we use the values for the central subhalo to compute $\tilde{c}$ for the halo.

Concentration has been found to positively correlate with halo bias for low mass haloes \citep[e.g.][]{Wechsler_2006}. We discuss in detail the nature of the relation between concentration, bias and \sfr{}, including their mass dependence, in \cref{sec:shuffling_method}. For the purposes of this toy model, we simply assume that a positive correlation between concentration \(\tilde{c}\) and bias exists for the haloes we are considering.

We examine three methods of assigning these weights:
\begin{enumerate}
    \item $W = 1\ (\tilde{c} < \tilde{c}_{50}), \, W = 5\ (\tilde{c} > \tilde{c}_{50})$: Higher weights (\(W = 5\)) are assigned to haloes with \(\tilde{c}\) values above the median \(\tilde{c}_{50}\), while lower weights (\(W = 1\)) are assigned to haloes with \(\tilde{c}\) values below the median.
    \item $W=1\ (\tilde{c}>\tilde{c}_{50}), W=5\ (\tilde{c}<\tilde{c}_{50})$: We reverse the assignment, giving higher weights (\(W = 5\)) to haloes with lower \(\tilde{c}\) and lower weights (\(W = 1\)) to haloes with higher \(\tilde{c}\).
    \item Random assignment: We assign weights randomly, independently of \(\tilde{c}\) -- given our previous arguments, this should have no affect on the two-halo term as long as the appropriate mean \lm{} relation is used.
\end{enumerate}

\begin{figure}
    \centering
    \includegraphics[width=\linewidth]{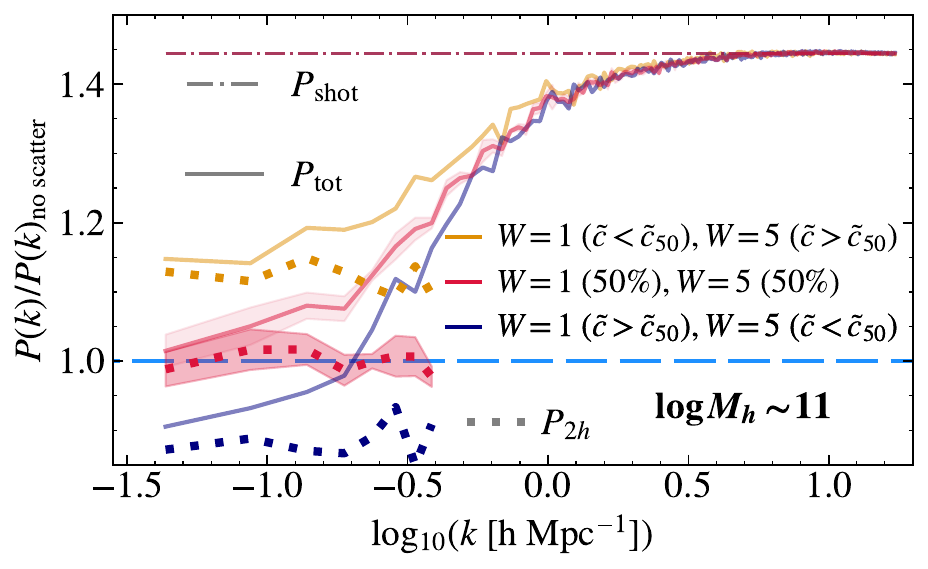}
    \caption{Ratio of the power spectra for various assignment schemes relative to the case where all haloes have equal weighting. The linestyles are the same as in \cref{fig:mass_dependent_scatter}. The \emph{red} colour corresponds to the case where weights are assigned randomly.  The \emph{gold} colour corresponds to the case where haloes with secondary property \vmax{} (correlated with halo bias) are assigned $W=1$ if \vmax{} is below the median and $W=5$ if above. The \emph{navy} colour represents the reverse weighting scheme, with $W=5$ for haloes below the median and $W=1$ for those above. Correlation of weight with bias affects the two-halo term. 
 }\label{fig:2nd_bias}
\end{figure}

\Cref{fig:2nd_bias} shows the ratio of the power spectra for these cases compared with the power spectrum of a scenario where all haloes have equal weighting. The ratios for the total power spectrum (solid), two-halo term (dotted), and shot noise (dash-dotted) are shown, respectively. As in the bottom panels of \cref{fig:mass_dependent_scatter}, the shot noise increases in all cases as the variance of the weights is higher compared to the reference case where all weights are equal.

For the case of random weight assignment (red line), the average two-halo term remains approximately unchanged, consistent with the bottom-right panel of \cref{fig:mass_dependent_scatter}.  
In contrast, when the weights are assigned based on \(\tilde{c}\), the two-halo term changes because the mean bias is now different. Specifically:
\begin{itemize}
    \item When haloes with higher \(\tilde{c}\) (and thus higher bias) are given higher weights (gold line), the mean bias of the sample increases, resulting in an increase in the amplitude of the two-halo term.
    \item Conversely, when haloes with lower \(\tilde{c}\) (and thus lower bias) are given higher weights (navy line), the mean bias decreases, leading to a decrease in the amplitude of the two-halo term.
\end{itemize}
These results demonstrate the impact of correlated weight assignments on the mean bias and hence on the amplitude of the two-halo term.

\ifSubfilesClassLoaded{%
  \bibliography{bibliography}%
}{}

\end{document}

\section{Impact of secondary bias on the power spectrum}\label{sec:shuffling_method}

In the previous section, we demonstrated that the ensemble-averaged two-halo term can be characterised by the mean luminosity-halo mass relation, even in the presence of scatter, provided that the scatter is random. 
In this section, we investigate whether the scatter in the \sfr{}-halo mass relation is random in the \tng{} simulation, and, if not, whether that does affect the galaxy power spectrum. Since the relation between luminosity and \sfr{} depends on the emission line, we focus on investigating the effect of scatter in the \sfr{}–halo mass relation on the power spectrum, rather than computing luminosities for different emission lines. For lines where luminosity is directly proportional to \sfr{}, this is equivalent to studying scatter in the luminosity–halo mass relation, as the normalised power spectrum is unaffected by constant factors (see \cref{sec:lum_assignment_schemes}). For emission lines whose luminosities depend strongly on \sfr{} but are not directly proportional, the results may still be qualitatively similar, though quantitative differences may arise.

We construct the scenario of random scatter by {\em shuffling} galaxies between haloes in narrow bins of halo mass in \tng{} (i.e., randomly reassigning galaxies among haloes of similar mass). 
The shuffling method preserves the mean \sfr{}-halo mass relation and its scatter, but any secondary bias that may be present in the simulation will be erased.

Shuffling has been adopted in previous studies \citep[e.g.][]{Croton_2007} to demonstrate that the clustering of galaxies is not solely dependent on host halo mass. More recently, \citet{Hadzhiyska_2021} 
used shuffling to investigate \tng{} galaxies selected to be similar to the luminous star-forming emission-line galaxies (\elg{}s) from the \desi{} survey \citep{Desi_2016}. 
While our investigation is similar to theirs, it differs in several aspects. The criteria in \desi{} correspond more closely to selecting galaxies above a threshold in specific star formation rate (\ssfr), rather than in \sfr{}, as we do here. Another key difference is the galaxy weight; in the \desi{} survey, each selected galaxy has a weight of one, whereas in our approach, which is designed to mimic \lim{}, galaxies contribute with a weight that is proportional to their \sfr{} (or line luminosity).

Our shuffling procedure works as follows. We first divide \tng{} haloes in bins of halo mass (width $\Delta \log_{10}(M_h\ [$\massunit$]) = 0.1$ dex).\footnote{The bin width is a compromise between avoiding mass dependence inside a given bin and having too few haloes per bin to draw meaningful conclusions.}
We then select all central galaxies of the haloes in a given bin, and randomly assign them to another halo within the same bin. In case the central galaxy has satellites, we can choose to move the satellites with their central, keeping relative positions the same -- or we can shuffle central and satellite galaxies separately. In both cases, we preserve the relative positions between satellites and their (new or original) centrals. We examine both choices in this section. In the first case, the one-halo term in the power spectrum is conserved by construction. 
In the second case, the combinations of centrals and their associated satellites are altered, and the one-halo term may change. For each test in this section, we shuffle the galaxy catalogues 100 times with different random seeds, generating 100 distinct shuffled catalogues. This allows us to estimate the variance of the shuffled catalogue.

\begin{figure}
    \centering
    \includegraphics[width=\linewidth]{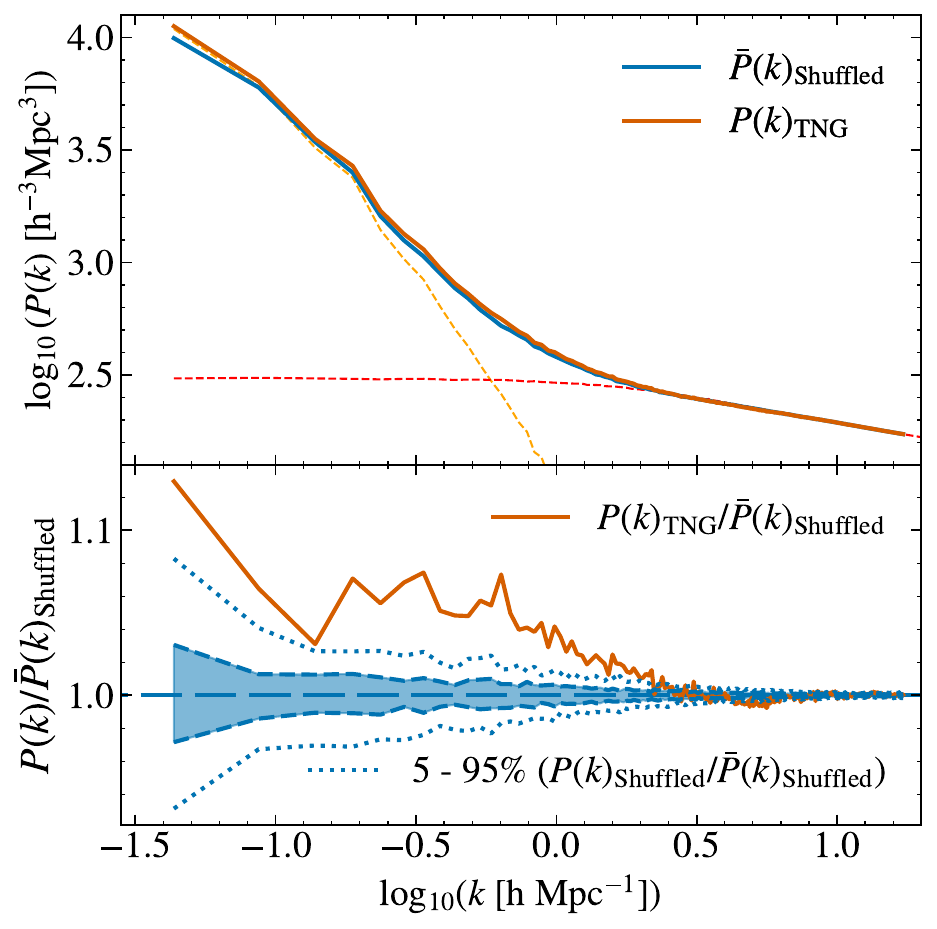}
    \caption{The upper panel shows the power spectrum of \tng{} galaxies in their fiducial positions (\tng{}; \emph{orange}) and the power spectrum of \tng{} galaxies with their positions shuffled amongst haloes of similar mass (Shuffled; \emph{blue}). In the shuffled case, \sfr{}s are distributed randomly among haloes within mass bins of d$\log M_h = 0.1$ dex, averaged over 100 random seeds. The \emph{dashed yellow} and \emph{red lines} show the two-halo and one-halo terms respectively for the fiducial \tng{} galaxy power spectrum.
    The \emph{solid orange line} in the bottom panel shows the ratio of the \tng{} power spectrum to the mean shuffled power spectrum. The \emph{blue colour band} bounded by the \emph{dashed lines} shows the 25th and 75th percentiles and the blue dotted lines show the 5th and 95th percentiles for the power spectra obtained from shuffling randomly. The \tng{} power spectrum is systematically higher suggesting that the \sfr{}s for a given halo mass are not distributed randomly, with higher \sfr{} haloes tending to be more clustered on average.
    }  \label{fig:shuffled_ps_main}
\end{figure}

For the first test, we moved satellites together with their centrals. 
The original \tng{} power spectrum (solid orange line) and the average power spectrum of the 100 shuffled catalogues (solid blue line) are shown in the upper panel of \cref{fig:shuffled_ps_main}, with their ratio plotted in the lower panel.
The dashed orange and dashed red lines represent the two-halo and one-halo terms, respectively, for the unshuffled case.
In the lower panel, the shaded region marks the 25th-75th percentiles, and the dotted lines indicate the 5th-95th percentiles for the 100 shuffled realisations.

On large scales, the original (unshuffled) power spectrum is systematically higher by about 5~per cent compared to the power spectrum of the shuffled catalogues. This demonstrates that the scatter in the \sfr{}, and hence in the weights of the galaxies, {\em is} correlated with the bias of the haloes.
The higher power spectrum indicates that, on average, galaxies with higher \sfr{} at a given halo mass are more strongly clustered than those with lower \sfr{}. Mock surveys that do not account for this secondary bias will yield biases in derived constraints. 

Since satellites move together with their central, the one-halo term of the original simulation is identical to that of all shuffled realisations. 
Consequently, the offset between the original power spectrum and that of the shuffled realisations decreases on scales $\log k \gtrsim -0.2$, 
where the contributions of the one-halo term to the total power spectrum becomes increasingly more dominant over the two-halo term (see the orange and red dashed lines in the upper panel). 

The variance in the amplitude of the power spectra between different random realisations of the shuffled cases is quite large on scales $\log k\lesssim -1$. On the largest scale probed by the simulation, the 25th-75th percentile (blue shading) is at the $\sim 3$ per cent level, while the 5th-95th percentile (blue dotted line) is at the $\sim 8$ per cent level (\cref{fig:shuffled_ps_main}). Even in the absence of secondary bias, such a high level of variance limits the ability to constrain cosmological parameters accurately. A larger computational volume is necessary to reduce this variance.

We concluded from \cref{fig:shuffled_ps_main} that secondary bias between the \sfr{} of galaxies and the spatial bias of their host halo increases the clustering signal compared to a shuffled distribution. \citet{Croton_2007} studied the clustering of galaxies in subhaloes above a given mass threshold, finding that they too show evidence of secondary bias. They found that the level of bias also depended on galaxy colour, with redder galaxies more clustered, and blue galaxies less clustered, compared to the shuffled case. \citet{Hadzhiyska_2021} similarly find a higher bias when considering mass-selected galaxies compared to a shuffled distribution, with the effect being smaller for colour- and \ssfr{}-selected galaxies.

\subsection{Secondary bias: which halo masses affect the power spectrum more?} \label{sec:halo_mass_range}
\begin{figure*}
    \centering
    \includegraphics[width=\linewidth]{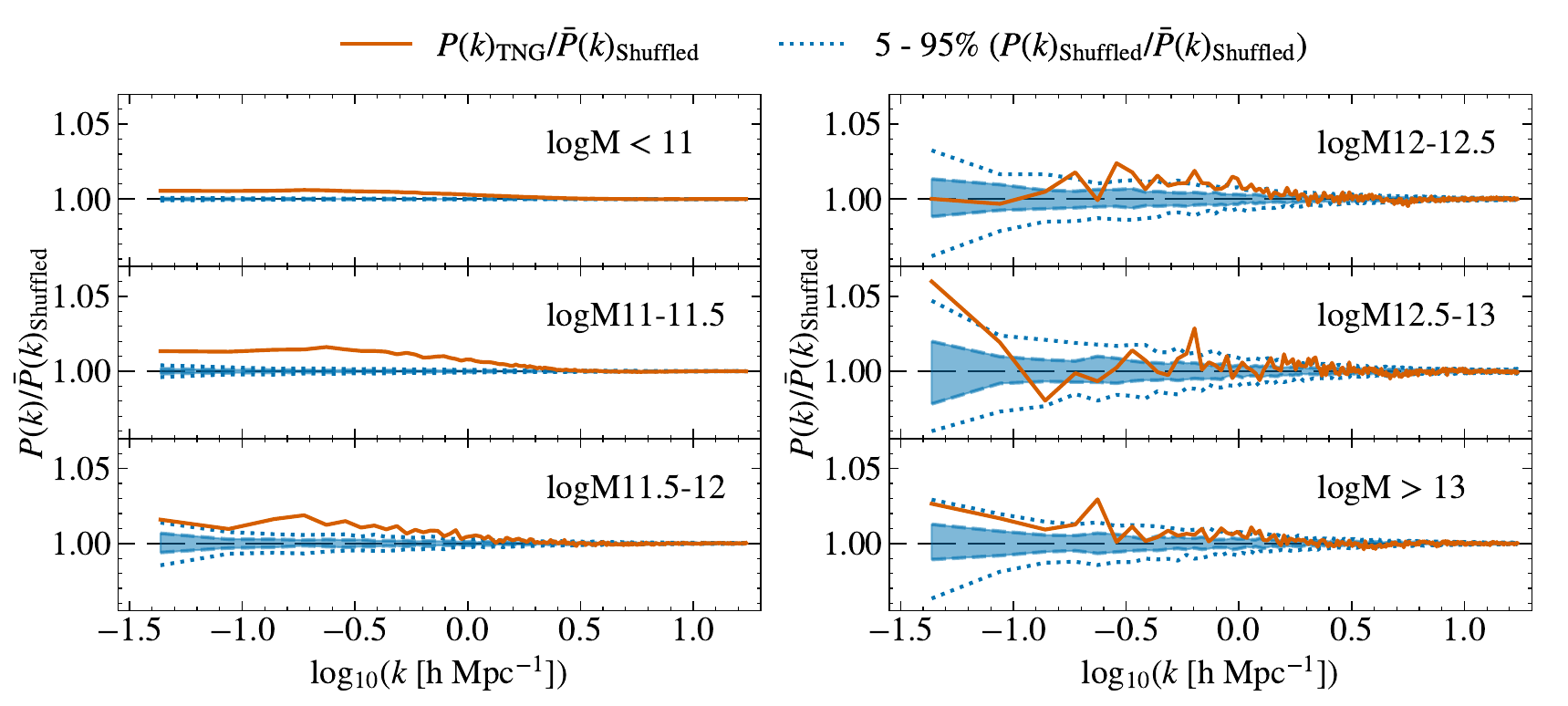}
    \caption{The ratio of the \tng{} power spectrum relative to the shuffled power spectrum for different halo mass ranges. The linestyles and colours are the same as in \cref{fig:shuffled_ps_main}. The galaxies are still shuffled in bins of d$\log M_h = 0.1$ dex, but, for each panel, only the \sfr{}s of haloes within the labelled mass range are shuffled, while all other haloes are fixed. 
    The \tng{} power spectrum has a positive bias relative to the shuffled case for all mass ranges. The intermediate mass ranges contribute most to the bias partly because they contribute most to the total \sfr{}.
    }  \label{fig:shuffled_ps_logMs}
\end{figure*}

To better understand what causes the offset in the power spectrum seen in \cref{fig:shuffled_ps_main}, it is useful to investigate whether different halo mass ranges are affected differently. This investigation is motivated by the fact that different physical processes may dominate in haloes of different masses. 

We investigate the effect of the shuffling on different mass ranges in \cref{fig:shuffled_ps_logMs}, which has a similar format to \cref{fig:shuffled_ps_main}. Here, we shuffle galaxies between haloes within a single bin of halo mass (bin width d$\log M_h=0.1$ dex, as before), keeping the \sfr{} of all other galaxies fixed; different panels in \cref{fig:shuffled_ps_logMs} correspond to different halo mass bins in which galaxies are shuffled. As before, we keep centrals and satellites together when shuffling.

The higher power spectrum of the \tng{} simulation compared to the case when the simulation's galaxies are shuffled is clear up to $\log M_h \sim 12$. For higher halo masses, it could be argued that the original (unshuffled) \tng{} power spectrum is still close to the 95th percentile of the shuffled realisations, and thus could result from random sampling. However, the fact that the offset consistently goes in the same direction across halo masses suggests that the offset may, nevertheless, be systematic. An extreme case is found for haloes in the mass range $\log M_h \in (12.5,13)$, which contributes to nearly half of the high bias observed in the total power spectrum on larger scales,\footnote{The galaxies in those haloes tend to have larger \sfr{}s and hence their impact on the power spectrum is larger.} as seen in \cref{fig:shuffled_ps_main}. A larger computational volume is needed to verify that this high bias is indeed systematic. 

\subsection{Secondary bias: impact of central and satellite galaxies on the power spectrum} \label{sec:central_satellite}

\begin{figure*}
\centering
\begin{subfigure}[t]{.5\textwidth}
    \centering
    \includegraphics[width=\linewidth]{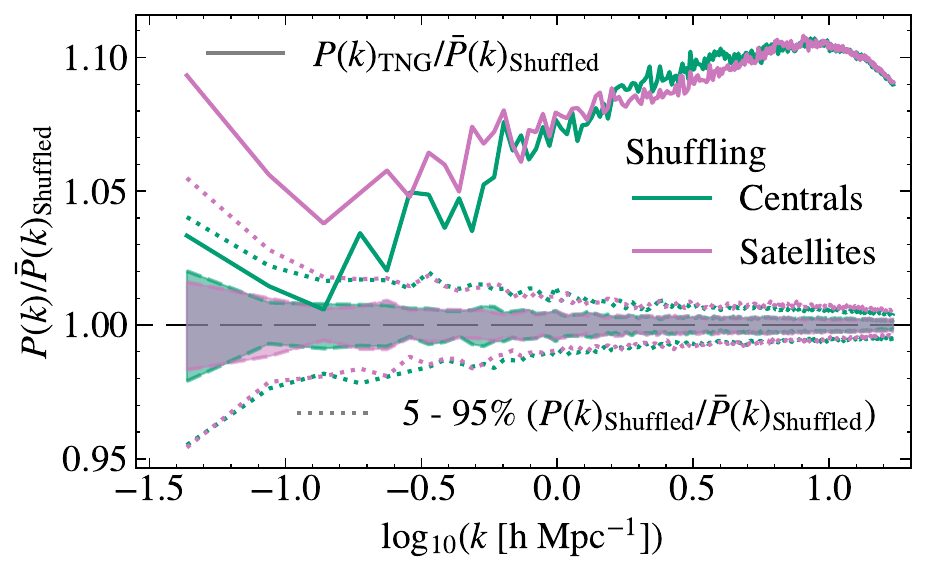}
    \caption{Galaxy power spectrum}
    \label{fig:cent_vs_sat_shuffled_all}
\end{subfigure}%
\begin{subfigure}[t]{.5\textwidth}
    \centering
    \includegraphics[width=\linewidth]{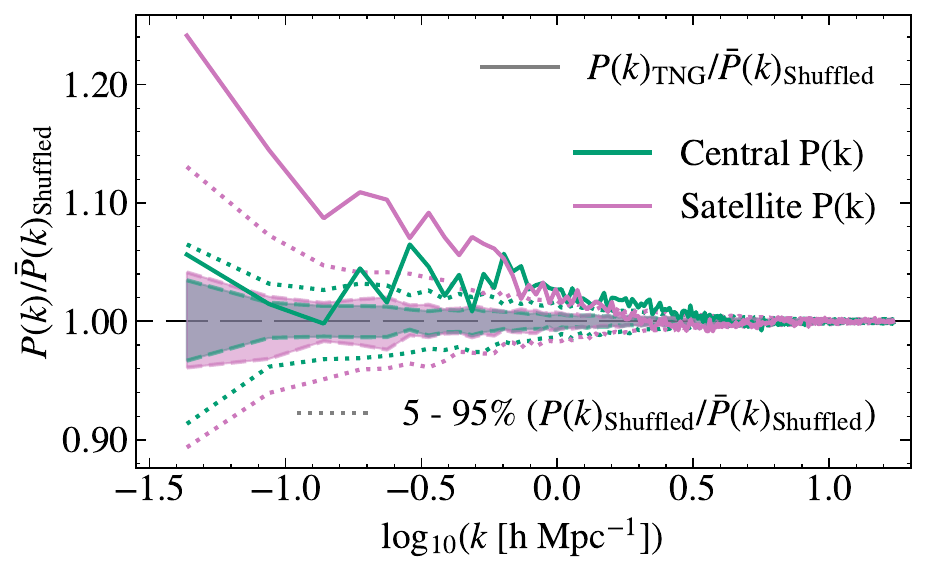}
    \caption{Central/Satellite power spectrum}
    \label{fig:cent_vs_sat_shuffled_only}
\end{subfigure}
\caption{The ratio of the \tng\ power spectrum relative to the shuffled power spectrum when shuffling central (\emph{green}) or satellite (\emph{violet}) galaxies only. The linestyles are as in \cref{fig:shuffled_ps_main}. \\
\textbf{\textit{Left panel}}: The galaxy power spectrum where satellite galaxies are fixed while shuffling central galaxies (\emph{green}), and central galaxies are fixed while shuffling satellite galaxies (\emph{violet}). The small-scale bias of \tng\ is higher than the shuffled case, suggesting that the \sfr\ of central and satellite galaxies belonging to the same halo are correlated. \\
\textbf{\textit{Right panel}}: The case where the power spectrum of central galaxies only is computed when shuffling central galaxies (\emph{green}), and the power spectrum of satellite galaxies only is computed when shuffling satellite galaxies (\emph{violet}). \\
\textbf{\textit{Summary}}: A significant amount of the secondary bias seen on large scales is from satellite galaxies.}
\label{fig:cent_vs_sat_shuffled}
\end{figure*}

We examine the effect of shuffling central and satellite galaxies separately, instead of moving all the galaxies within the same halo together. The result is shown in the left panel of \cref{fig:cent_vs_sat_shuffled}. When the satellite galaxies are shuffled, the central galaxies are fixed, and vice versa.
On the largest scales ($\log k\lesssim -0.5$), the \tng{} power spectrum is
almost 10~per cent larger than the case when satellites are shuffled (violet line), but this reduces to 4~per cent if only centrals are shuffled (green line). This suggest that much of the effect of secondary bias on the power spectrum is due to satellite galaxies. 

\begin{figure}
    \centering
    \includegraphics[width=.5\textwidth]{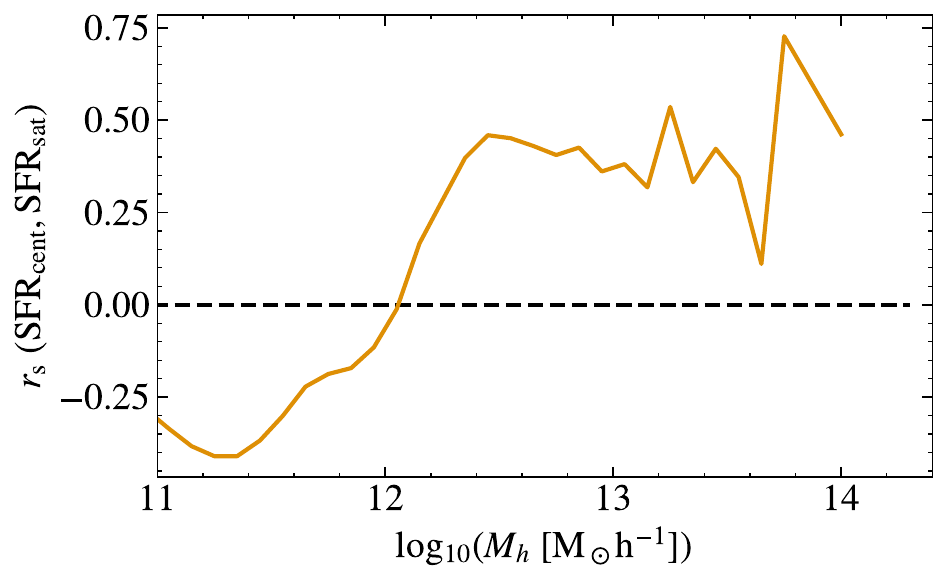}
    \caption{Spearman correlation coefficient between the \sfr{} of central and satellite galaxies hosted by the same halo. \sfr{}$_{\mathrm{sat}}$ is given by the sum of the \sfr{}s of the satellite galaxies within the same halo. There is positive correlation between central and satellite \sfr{} for high mass haloes.}
    \label{fig:corr_sfr_cent_sat}
\end{figure}

To confirm that the increase in the amplitude of the power spectrum on large scales due to secondary bias is not driven by changes in the  halo \sfr{} (the sum of the \sfr{}s of central plus satellites), we consider a catalogue consisting of central galaxies only and shuffle their positions, and similarly a catalogue with satellite galaxies only which are then shuffled; the results are shown in the right panel of \cref{fig:cent_vs_sat_shuffled}. When neglecting central galaxies, the large-scale bias becomes even more pronounced: although the variance from randomly shuffling satellite galaxies is also larger than that of central galaxies, the bias in the satellite galaxy power spectrum beyond the 95th percentile level is much clearer, reinforcing the suggestion that a significant contribution to the large-scale bias is due to satellite galaxies.

The impact of secondary bias on satellite galaxies was also investigated by \citet{Croton_2007}. They detect a non-negligible but relatively weak effect, which is stronger for red satellites than for blue satellites. We recall that all galaxies have equal weight in their investigation, whereas we weigh galaxies by their \sfr{}.

The left panel of \cref{fig:cent_vs_sat_shuffled} also shows an offset on small scales ($\log k \gtrsim 0.0$), implying that shuffling centrals separately from their satellites changes the one-halo term. The fact that this affects the power spectrum indicates that there is correlation between the \sfr{}s of central galaxies and that of their satellites -- a correlation sometimes referred to as (one-halo) galactic conformity \citep{Weinmann06,Hearin_2015,Ayromlou23}. The presence of this correlation in \tng{} is confirmed in \cref{fig:corr_sfr_cent_sat}, which shows the Spearman correlation coefficient between the \sfr{}s of centrals and their satellites as a function of halo mass. The correlation is positive for haloes with $\log M_h\gtrsim 12$, and weakly negative for lower mass haloes.
 
Changing the relation between centrals and their satellites changes the halo profiles, $U(k)$, as well as the halo shot noise term, $P^{\mathrm{halo}}_{\mathrm{shot}}$ (see  \cref{eq:galaxy_ps_all}). The halo shot noise term changes, because the total \sfr{} of individual haloes changes as different central galaxies are paired with different sets of satellites in the shuffled case: the halo shot noise is 304 \pkunit\ for \tng{} compared to 280 \pkunit\ after shuffling. 
The higher halo shot noise in \tng{} arises from the correlation between the \sfr{} of centrals and satellites, which leads to increased variance in the total halo \sfr{}.
The {\em galaxy} shot noise, however, is unchanged, as only the positions of galaxies change, while the individual \sfr{}s within the volume remain the same. Although the power spectrum is dominated by the galaxy shot noise on small enough scales, the smallest scale shown in \cref{fig:cent_vs_sat_shuffled} has not yet reached the scale where that shot noise dominates (see \citepaperone): the differences seen on the smallest scale are a consequence of differences in the halo profile, $U(k)$, and $P^{\mathrm{halo}}_{\mathrm{shot}}$.

\begin{figure*}
    \centering
    \includegraphics[width=\linewidth]{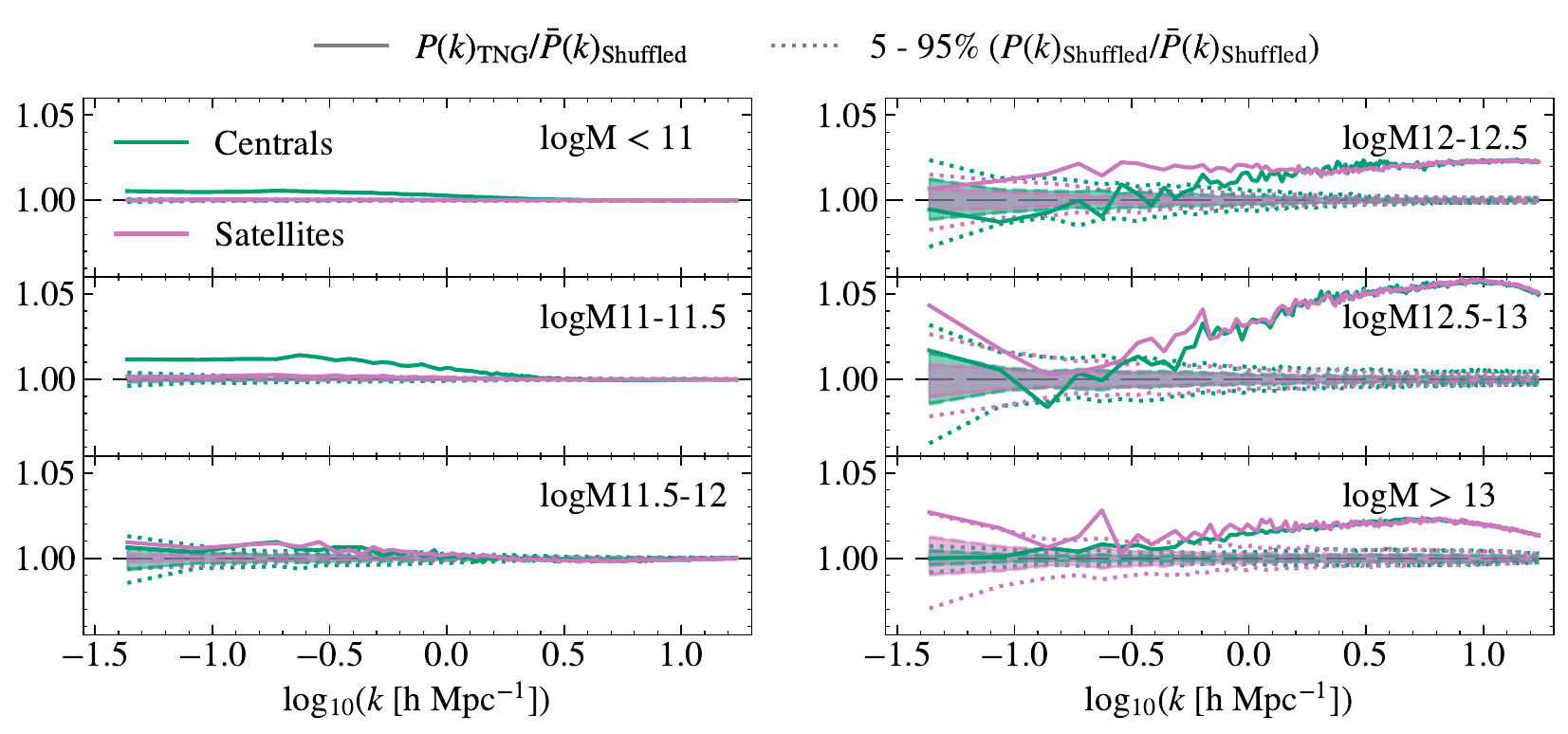}
    \caption{The ratio of the \tng{} power spectrum relative to the shuffled power spectrum when shuffling either central or satellite galaxies only. The colours and linestyles as the same as in \cref{fig:cent_vs_sat_shuffled_all} but only centrals/satellites within the indicated mass range are shuffled. The large-scale bias of \tng{} relative to the case when shuffling central galaxies is only significant for $\log M_h < 12$. The effect of satellite galaxies on the bias is larger for higher masses. 
    }  \label{fig:cent_vs_sat_shuffled_logM}
\end{figure*}

We further investigate the extent to which galactic conformity affects the power spectrum as a function of halo mass in \cref{fig:cent_vs_sat_shuffled_logM}. We shuffle galaxies and centrals separately, as before, but restrict shuffling to galaxies in one bin of halo mass at a time. The offset on small scales ($\log k \gtrsim 0$) is large for massive haloes ($\log M_h > 12$), and negligible in less massive haloes. This is not surprising, since massive haloes host many satellites, whereas lower mass haloes host few or none. 
Due to the positive correlation between central and satellite \sfr{}s for $\log M_h > 12$ (\cref{fig:corr_sfr_cent_sat}), the \tng{} power spectrum has a higher amplitude on small scales compared to the shuffled case.

On large scales ($\log k \lesssim 0$) and for low mass haloes ($\log M_h < 11.5$), central galaxies are the main contributor to the offset of the \tng{} power spectrum relative to the shuffled case. This is expected since most of the contribution to the halo \sfr{} is from central galaxies in this mass range (see \cref{fig:contribution_mass_bins}). For the mass range $\log M_h \in (11.5,12)$, central galaxies still contribute significantly more to the total \sfr{} than satellites. Even so, shuffling satellites in this mass range has a comparable or even larger effect on the power spectrum than shuffling centrals. This suggests that satellite galaxies may generally be more strongly affected by secondary bias, even in lower mass haloes (provided they are present).

The halo \sfr{} in the mass range $\log M_h \in (12,12.5)$ is still dominated by that of their central galaxies, yet the impact of shuffling centrals on the power spectrum is small (unlike the case of lower mass haloes).
The \sfr{} of those central galaxies is strongly affected by \agn{} feedback in \tng{} (see \cref{fig:sfr-mass}), and this quenching may be responsible for suppressing the impact of secondary bias on centrals in this halo mass range. In fact, the effect of secondary bias on centrals is to {\em decrease} the amplitude of the power spectrum: more clustered haloes in this mass range tend to have centrals with {\em lower} \sfr{}, on average. On the other hand, the offset caused by satellites is significant in this mass range, with more biased haloes having a higher \sfr{} in their satellites. The net result of more biased haloes having centrals with lower \sfr{} yet satellites with higher \sfr{}, is a slight suppression of the large-scale power spectrum for those haloes, as shown in the top right panel of \cref{fig:shuffled_ps_logMs}.

For haloes in the mass range $\log M_h \in (12.5,13)$, the \tng{} power spectrum is more than 4~per cent higher than that of the shuffled satellite power spectrum at the largest scales, and this bias contributes significantly to the total bias at that scale. For $\log M_h > 12.5$, satellites dominate the halo \sfr{}, and the net effect of shuffling central galaxies on the power spectrum is small.

Independently of halo mass, the total satellite \sfr{} in \tng{} is higher in more biased haloes. The effect of this secondary bias on the power spectrum is significant for haloes with mass $\log M_h>12$, where the contribution from satellite galaxies to the halo \sfr{} becomes significant. 

\subsection{Origin of the scatter in the SFR-halo mass relation}\label{sec:corr_coef}
The fact that shuffling the \sfr{} of galaxies affects the galaxy power spectrum implies that the scatter in the \sfr{}-halo mass relation is not random, but shows evidence for secondary bias. 
Several secondary properties have been studied in the context of secondary bias, including halo concentration, formation time, spin, and environment. 
Here, we focus on concentration, which is one of the most commonly studied examples (e.g. \citealt{Contreras_2019}).
In \cref{sec:central_satellite}, we found that satellite galaxies contribute significantly to the secondary bias, motivating us to also consider the (total) satellite mass.
In this section, we investigate how \sfr{} correlates with these two secondary halo properties. In \cref{sec:reducing_bias}, we will examine whether these two parameters contribute to the secondary bias signal.
These parameters are defined and computed as follows:
\begin{itemize}
\item Concentration, $\tilde{c}$: a proxy for halo concentration, as defined in \cref{eq:conc_proxy}.
\item Total satellite subhalo mass, \msat{}: for a given {\sc fof} halo, we sum up the dark matter masses of all satellite subhaloes (variable \texttt{SubhaloMass} in the \tng{} database) identified in the halo by {\sc subfind}.
\end{itemize}

We investigate the correlation between $\tilde{c}$, \msat{}, and the \sfr{} for haloes of a given mass, by computing the Spearman correlation coefficient. We use halo mass bins of width d$\log M_h=0.1$ dex, with the exception of the final bin for which we use $\log M_h \in (13.8,14.2)$ (to account for the lack of high mass haloes due to limited extent of the computational volume of \tng{}). We discard haloes for which the \sfr\ is zero. 
    
\begin{figure}
    \centering
        \begin{subfigure}[t]{0.49\textwidth}  
            \centering 
            \label{fig:corr_coef_msat_vmax}
            \includegraphics[width=\textwidth]{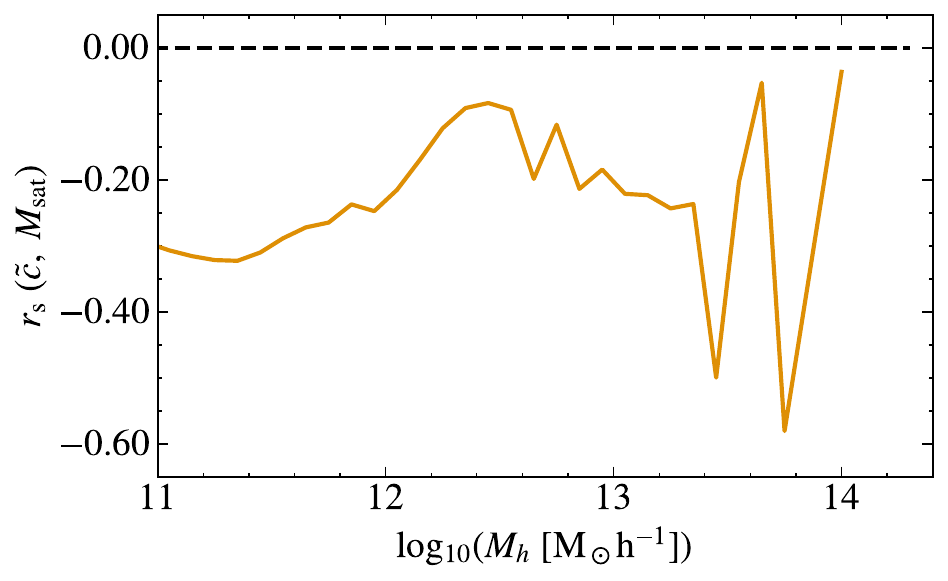}
        \end{subfigure}
        \caption
        {Spearman correlation coefficient, $r_s$, between concentration, $\tilde{c}$, and total satellite subhalo mass, $M_\mathrm{sat}$, 
        as a function of halo mass, $M_h$, in logarithmic bins with d$\log M_h=0.1$ dex. There is a weak negative correlation, with more concentrated haloes having a lower satellite mass, for all halo masses.
    }  \label{fig:corr_msat_vmax}
\end{figure}

We first show that \vmax{} and \msat{} are themselves weakly anti-correlated (\cref{fig:corr_msat_vmax}); at given halo mass, more concentrated haloes tend to have a lower satellite mass.
This result is consistent with the findings by \citet{Gao_2004} and \citet{Bosch_2005}, who show that concentration and formation time are negatively correlated with substructure mass, and by \citet{Bose_2019}, who similarly find that the number of satellites (with stellar mass above a threshold) is lower when the concentration of the host halo is higher. A plausible explanation is that more concentrated haloes form earlier on average, and although their subhaloes may be more concentrated as well, there is more time for the subhaloes to be tidally disrupted or merge with the main halo.

\begin{figure*}
        \centering
        \begin{subfigure}[t]{0.49\textwidth}   
            \centering 
            \label{fig:corr_coef_conc}
            \includegraphics[width=\textwidth]{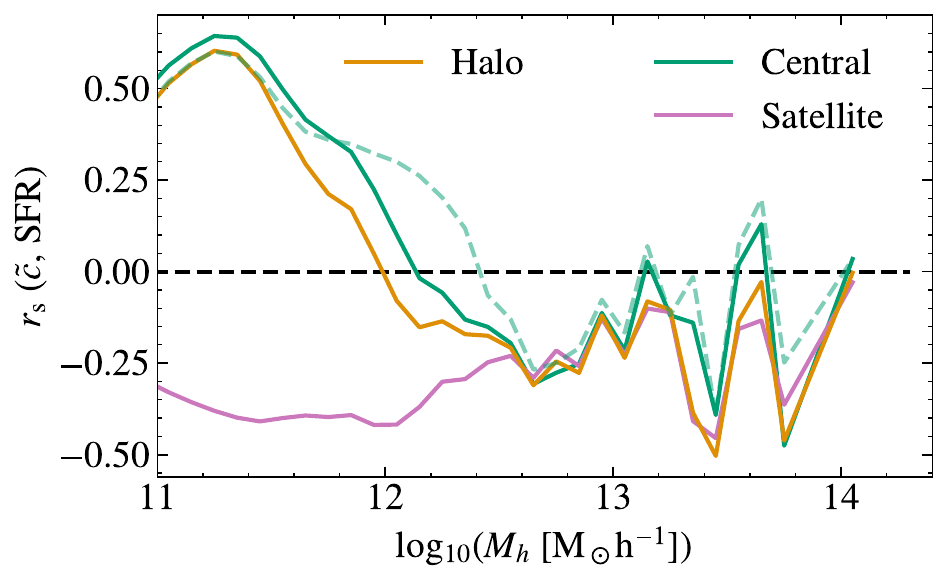}
        \end{subfigure}
        \hfill
        \begin{subfigure}[t]{0.49\textwidth}
            \centering
             \label{fig:corr_coef_msat}
            \includegraphics[width=\textwidth]{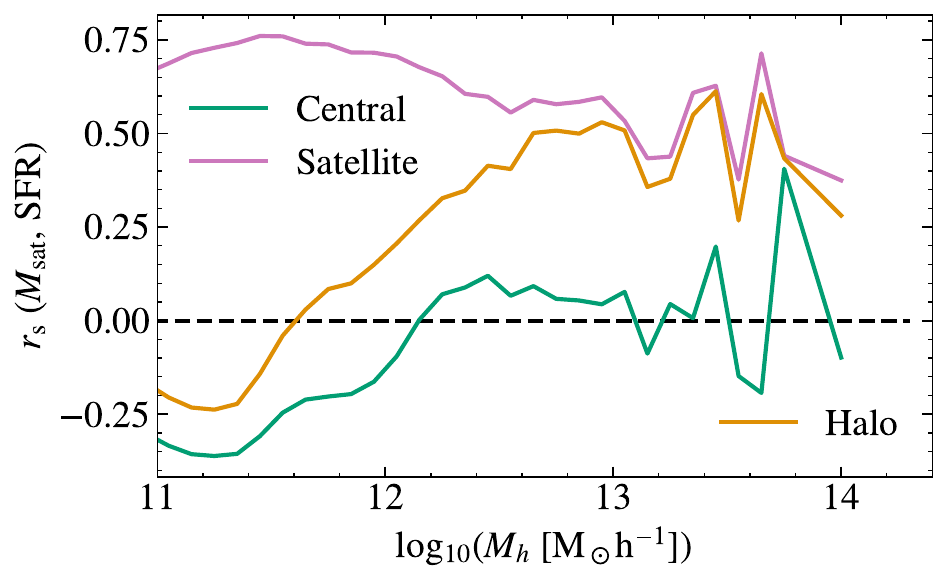}
        \end{subfigure}
        \caption
        {Spearman correlation coefficients for different halo mass bins with d$\log M_h=0.1$ dex. The colours represent halo (\emph{orange}), central (\emph{green}) and satellite (\emph{violet}) \sfr{}s. 
        \textbf{\textit{Left panel}}: Correlation coefficients between \vmax\ and \sfr{}. The \emph{dashed green} line represents the case where the correlation coefficient has been computed after excluding central galaxies with \sfr{}s below a threshold (computed by subtracting the difference between the 50th and 99th percentiles of the \sfr{} distribution from the median \sfr{}). The correlation between \vmax\ and central \sfr{} is strong for lower masses, but weak for higher masses.
        \textbf{\textit{Right panel}}: Correlation coefficients for total satellite subhalo mass and \sfr{}. There is strong positive correlation between satellite mass and satellite \sfr{} for all halo masses. 
        }  
    \label{fig:corr_coefs}
\end{figure*}

Next, we plot the Spearman correlation coefficient between \sfr{} and concentration, $\tilde{c}$, in the left panel of \cref{fig:corr_coefs}. For central galaxies (green line), the correlation is positive for haloes with mass $\log M_h \lesssim 12$. Higher concentration implies a deeper potential well, and hence a higher \sfr{} when stellar feedback is regulating the \sfr{} of the galaxy (see the discussion around eq. 15 of \citealt{Sharma20}). At higher halo masses, \agn\ feedback becomes important, and the Spearman correlation coefficient becomes negative: more concentrated haloes have central galaxies with {\em lower} \sfr{}. A plausible reason is that the black hole in massive haloes with high $\tilde{c}$ forms earlier and is more massive, therefore \agn\ feedback from this more massive black hole decreases the central galaxy's \sfr{} more. To examine whether very strongly quenched galaxies affect the correlation between $\tilde{c}$ and the \sfr{} of the central galaxy, we recompute the Spearman correlation coefficient but without these strongly quenched galaxies (dashed green line). This has some effect on the correlation, but the broad trends remain the same: the correlation between \sfr{} and $\tilde{c}$ arises for the majority of galaxies, rather than some outliers for which the correlation is particularly strong.

The Spearman correlation coefficient $\tilde{c}$ of the host halo and the combined \sfr{} of all satellite galaxies in the halo (violet line) is negative for all halo masses: more concentrated haloes have less star formation in their satellites. It would be worth investigating whether this is due to the anti-correlation between subhalo mass and $\tilde{c}$ that we illustrated in \cref{fig:corr_msat_vmax}.

The right panel of \cref{fig:corr_coefs} shows the Spearman correlation coefficient, $r_s$, between \sfr{} and total satellite mass, $M_\mathrm{sat}$.
The positive correlation between $M_\mathrm{sat}$ and total satellite \sfr{} (violet) is strong for all halo masses ($r_S \gtrsim 0.5$ for $\log M_h \lesssim 13$). The correlation becomes noisy for $\log M_h \gtrsim 13$ because the number of haloes in the \tng{} simulation decreases rapidly; however, it remains clearly positive. This positive correlation can be explained as follows: when a galaxy falls into another halo and becomes a satellite, star formation quenching mechanisms such as strangulation, ram-pressure stripping and harassment decrease the \sfr{}, by stripping gas or reducing gas cooling. Tidal forces also cause the dark matter mass to be stripped, causing the subhalo mass to decrease. This reduces the potential well of the satellite subhalo, also causing the \sfr{} to decrease. Therefore, the satellite's \sfr{} and its subhalo mass decreases in tandem.

The correlation of the central galaxy's \sfr{} with $M_\mathrm{sat}$ (green) is slightly negative for $\log M_h \lesssim 12$, but there is no clear correlation for $\log M_h \gtrsim 12$. 
The anti-correlation between the central galaxy's \sfr{} and $M_\mathrm{sat}$ for $\log M_h \lesssim 12$ may be linked to the fact that the central \sfr{} positively correlates with \vmax{} (\cref{fig:corr_coefs}), while \vmax{} anti-correlates with $M_\mathrm{sat}$ (\cref{fig:corr_msat_vmax}).

In both the left and right panels of \cref{fig:corr_coefs}, the correlations for the halo \sfr{} (orange) follow those of the central galaxy for $\log M_h \lesssim 12$, as the central galaxy's \sfr{} dominates the total halo \sfr{} in this regime (see \cref{fig:contribution_mass_bins}). For $\log M_h \gtrsim 12$, the correlations become increasingly influenced by the satellite \sfr{}, which contributes more significantly at higher halo masses.

We further examine these correlations and also make a comparison to \eagle{} in \cref{app:sfr-c-msat-corr}.
The strong correlation between halo \sfr{} and $\tilde{c}$ at $\log M_h \lesssim 12$, and between halo \sfr{} and $M_\mathrm{sat}$ for $\log M_h \gtrsim 12$, suggests that concentration and satellite mass are promising secondary halo properties which may improve the modelling of \sfr{}.

\subsection{Accounting for the offset in the power spectrum due to secondary bias} \label{sec:reducing_bias}

In the previous section, we studied correlations between the \sfr{} and two potential secondary bias indicators of haloes: concentration, $\tilde{c}$, and satellite mass, $M_\mathrm{sat}$. 
However, even if these parameters are correlated with \sfr{}, this does not necessarily imply they influence the \lim{} power spectrum. To affect the power spectrum, their correlations with the galaxy property (\sfr{}) and with the halo bias must intersect (see \citealt{Mao_2018}).
In this section, we investigate if it is possible to reduce the offset in the power spectrum by introducing secondary halo properties.

Many studies have investigated which halo properties, in addition to halo mass, affect halo bias. 
\citet{Wechsler_2006} showed that, at a given halo mass, $M_h$, 
halo bias increases with concentration below a characteristic value of $M_h$, whereas it decreases with concentration above it. They also
found bias to be correlated with formation time, particularly for lower halo masses, and, in addition, found that bias correlates with the number of
subhaloes. \citet{Gao_2007} demonstrated that there is a correlation between bias and \vmax{}$'$$=V_{\rm max}/V_{\rm 200}$, as a proxy for concentration. They additionally consider two different measures of the amount of substructure. The two measures correlate with bias somewhat differently, but in both cases haloes with more substructure are more biased. \citet{Mao_2018} similarly found a strong correlation between bias and the number of subhaloes, for haloes with a given value of $M_h$. Spin is also commonly found to have strong correlations with clustering bias \citep[e.g.][]{Bett_2007,Lacerna_2012,Tucci_2021}, but we do not investigate this property in this work. We refer the reader to, e.g., \citet{Wechsler_2018} for a review on secondary bias.

\subsubsection{Impact of secondary bias on the power spectrum}
\begin{figure}
    \centering
    \includegraphics[width=\linewidth]{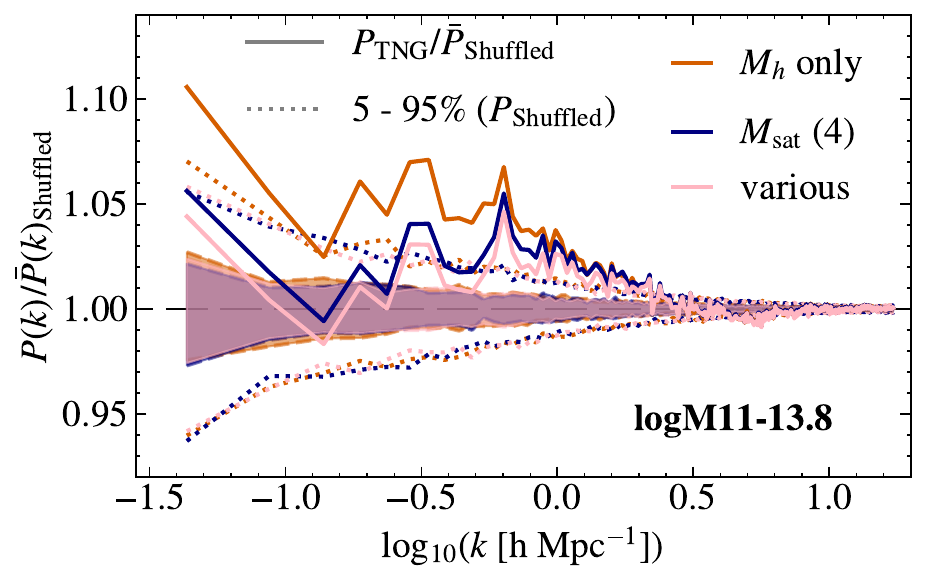}
    \caption{The ratio of the \tng{} power spectrum to the shuffled power spectrum when including secondary properties. The \emph{orange colour} represents the case where only halo mass is considered.
    The \emph{navy} colour represents the case where each mass bin is divided into 4 bins by satellite mass. For the \emph{pink} colour, logM11-11.5 haloes have been shuffled in 2 bins of \vmax, logM12-13.8 have been shuffled in 4 bins of satellite mass, and logM11.5-12 haloes have had centrals shuffled in 4 bins of \vmax, and satellites shuffled in 4 bins of satellite mass. The solid lines represent the mean ratio, while the shaded regions and dotted lines show the 25th-75th and the 5th-95th percentiles, respectively, for each shuffled case. The bias discrepancy between \tng{} and the shuffled case can be reduced, but not fully eliminated, by incorporating these secondary properties.
    }  \label{fig:property2_all}
\end{figure}

In \cref{fig:property2_all} we examine the effect of secondary bias on the power spectrum by shuffling galaxies
in bins of halo mass, for all mass bins in the range $\log M_h \in (11,13.8)$.\footnote{Secondary bias is small for haloes with $\log M_h < 11$, and the small number of haloes with $\log M_h > 13.8$ makes it difficult to distinguish between random variations and secondary bias.}

Using satellite mass, $M_\mathrm{sat}$, as the secondary parameter for all halo mass ranges (navy line) reduces the offset between the \tng\ and mean shuffled power spectrum to within the 95th percentile level on scales $\log k \lesssim -0.5$. In the next section, we show how \vmax{} captures secondary bias better for lower mass haloes, whereas $M_\mathrm{sat}$ is a better choice at higher mass. We attempt to combine the best of both worlds by computing the pink line labelled \lq various\rq. In this case, satellite mass is used as a secondary parameter 
for haloes with $\log M_h \in (12,13.8)$, and \vmax\ is used as a secondary parameter for $\log M_h \in (11, 11.5)$, in both cases shuffling centrals and satellites together. In the intermediate range, $\log M_h \in (11.5,12)$, central galaxies and satellites are shuffled independently, using \vmax\ for centrals and $M_{\rm sat}$ for satellites. 
The ratio between the \tng{} power spectrum and the mean of the \lq best case\rq\ scenario (pink line) falls
within the 25-75$^{\rm th}$ percentiles of the shuffled power spectra for most $k$'s in the range $\log k<-0.25$, and falls close to the 95$^{\rm th}$ percentiles for $\log k\in (-0.25, 0.25)$. At even higher $k$, the power spectra are similar since the one-halo term remains almost identical.
In the following section, we demonstrate the effects by considering different halo mass ranges separately, providing the motivation behind the \lq best case\rq{} scenario shown in \cref{fig:property2_all}.

\subsubsection{Impact of secondary bias on the power spectrum in bins of halo mass}
\begin{figure*}
        \centering
        \begin{subfigure}[b]{0.49\textwidth}
            \centering
            \includegraphics[width=\textwidth]{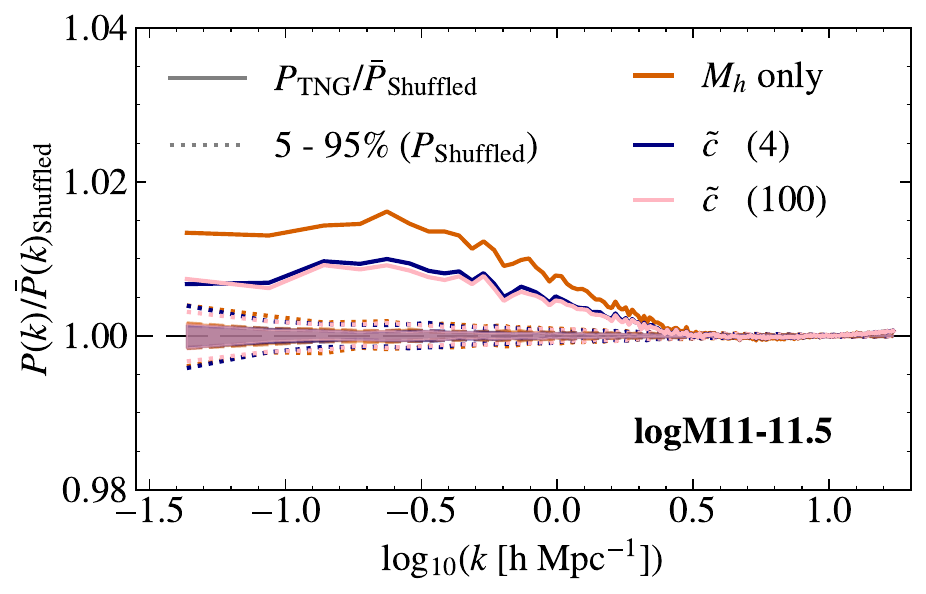}
        \end{subfigure}
        \hfill
        \begin{subfigure}[b]{0.49\textwidth}  
            \centering 
            \includegraphics[width=\textwidth]{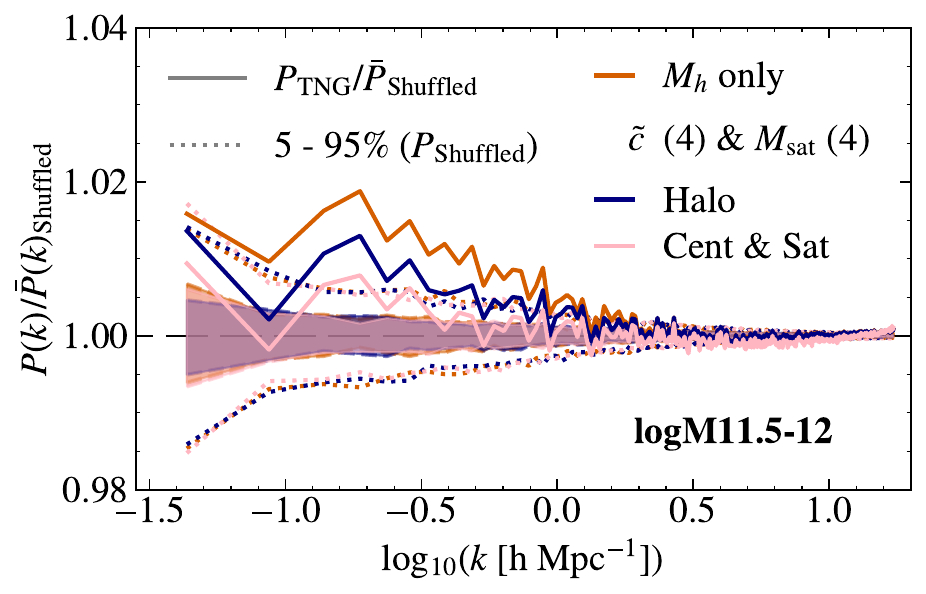}
        \end{subfigure}
        \vskip\baselineskip
        \begin{subfigure}[b]{0.49\textwidth}   
            \centering 
            \includegraphics[width=\textwidth]{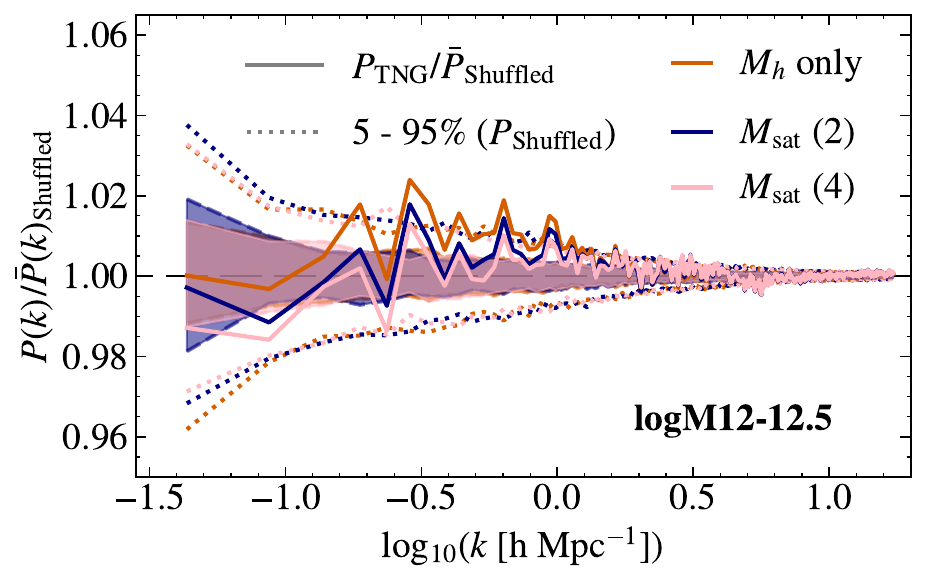}
        \end{subfigure}
        \hfill
        \begin{subfigure}[b]{0.49\textwidth}   
            \centering 
            \includegraphics[width=\textwidth]{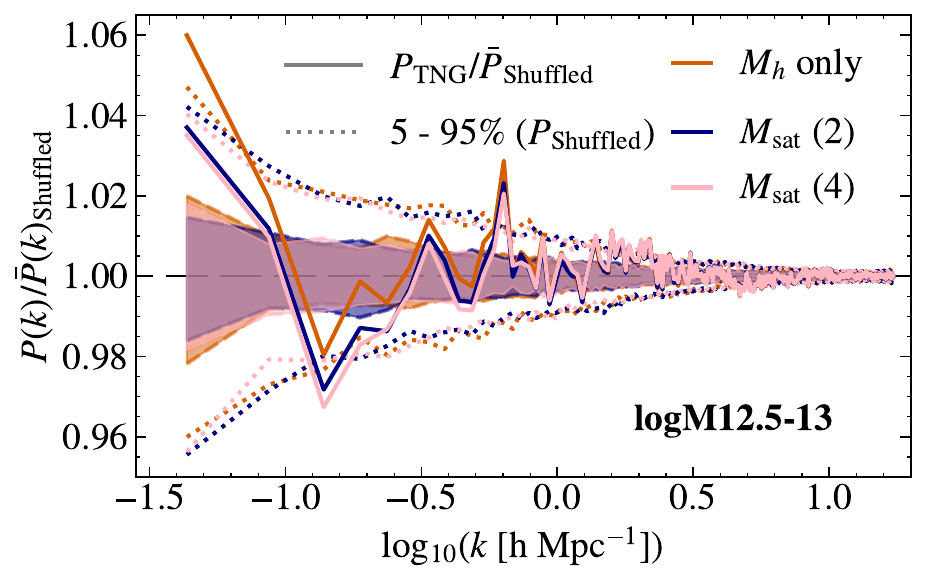}
        \end{subfigure}
        \caption
        {The ratio of the \tng{} power spectrum to the shuffled power spectrum when including secondary properties. Haloes with $\log M_h\in (11, 11.5)$ (upper left), $\log M_h\in(11.5,12)$ (upper right), $\log M_h\in(12,12.5)$ (lower left), and $\log M_h\in(12.5, 13)$ (lower right) are shuffled, with the remaining haloes fixed. The \emph{orange solid lines} are the same as in \cref{fig:shuffled_ps_logMs} for the respective mass ranges, representing the ratio when not considering secondary properties. The \emph{navy} and \emph{pink solid lines} represent the case where each halo mass bin within the labelled mass range is further divided into bins by a secondary property, the concentration proxy \vmax\ or the satellite mass $M_{\mathrm{sat}}$, before shuffling. The shaded regions show the 25th-75th percentiles while the dotted lines show the 5th-95th percentiles, for each shuffled case. The number in the parentheses represent the number of bins each halo mass bin is further divided into. 
        Different binning is applied for each halo mass range (see text).
        } 
        \label{fig:msat_conc_effect_logMs}
    \end{figure*}

In this section, we show the effect of including the secondary properties for different mass ranges. In \cref{fig:msat_conc_effect_logMs}, we examine the impact on the power spectrum of including either \vmax\ or the total satellite mass, \msat{}, as a secondary property by plotting the ratio of the \tng{} power spectrum to the power spectrum obtained after shuffling galaxies. As in \cref{fig:shuffled_ps_logMs}, we restrict shuffling to galaxies in given mass bins, with different panels corresponding to different bins, whilst keeping all other haloes fixed.

The orange solid lines in \cref{fig:msat_conc_effect_logMs}, labelled $M_h$, 
are the ratios of the \tng{} power spectra to the shuffled power spectra, not accounting for any secondary bias parameter: these lines are therefore identical to the orange solid lines in \cref{fig:shuffled_ps_logMs}. We then additionally restrict shuffling to haloes in bins of a secondary parameter, namely $\tilde{c}$ and/or $M_\mathrm{sat}$. 
The number of bins used for the secondary parameter is indicated in parentheses in the legend; the bins are chosen to contain equal numbers of haloes (i.e. we shuffle in bins defined by equal percentiles in the value of the parameter used for shuffling). As before, we generate 100 independent randomly shuffled realisations.
The solid lines represent the ratio to the mean, and the shaded region indicates the 25th–75th percentiles range, while the 5th–95th percentiles are shown as dotted lines, allowing us to assess the scatter among the random realisations.
We verified that the number of haloes in each bin is large enough to additionally restrict them in percentile bins.

\vspace{0.5em}\noindent
\underline{Halo mass bin $\log M_h\in (11, 11.5)$ (upper left panel)} Restricting shuffling in 4 percentile bins of $\tilde{c}$ (navy line) approximately halves the offset between the \tng{} and shuffled power spectra, thereby reducing the bias at the largest scales, $\log k \lesssim -1$, to around 1~per cent. Using 100 bins in $\tilde{c}$ (pink line) does not lead to any significant further improvement, suggesting that another secondary parameter contributes significantly to the remaining bias. Satellites do not contribute much to the \sfr{} in these low-mass haloes, therefore they are unlikely to play a significant role.

\vspace{0.5em}\noindent
\underline{Halo mass bin $\log M_h\in(11.5, 12)$ (upper right panel)}
For this mass range, we start by splitting each mass bin first into 4 percentile bins of \vmax{}, and then further subdivide each of these in another 4 percentile bins of \msat{}. Shuffling in these bins leads to a slight reduction in the offset between the \tng{} and the shuffled power spectra (represented by the navy line labelled ‘halo’). Next, we shuffle central and satellites independently: centrals are shuffled in 4 bins of \vmax{}, and satellites in 4 bins of $M_\mathrm{sat}$ (pink line labelled \lq Cent \& Sat\rq). This test is motivated by the fact that central and satellite galaxies exhibit opposite correlations with \vmax{} and \msat{}, as noted in \cref{fig:corr_msat_vmax}. Shuffling centrals and satellites separately reduces the power spectrum offset to less than 1~per cent on all scales. As we are not shuffling centrals and their satellites together, this type of shuffling affects the one-halo term. However, the panel shows that the change in the one-halo term affects the power spectrum by less than 0.2~per cent on scales $\log k \gtrsim 0$.

\vspace{0.5em}\noindent
\underline{Halo mass bin $\log M_h\in(12, 12.5)$ (lower left panel)} Shuffling in 2 percentile bins of $M_\mathrm{sat}$ (navy line) removes most of the offset between the \tng{} and shuffled power spectra. Using 4 bins in $M_\mathrm{sat}$ (pink line) does not significantly reduce the offset further. The amplitude of the \tng{} power spectrum is in fact slightly {\em below} that of the shuffled spectra on the largest scales, $\log k \lesssim -0.5$. 
The upper right panel of \cref{fig:cent_vs_sat_shuffled_logM} suggested that central galaxies with higher \sfr{} tend to be less biased for haloes with $\log M_h\in(12, 12.5)$. The lower amplitude of the \tng{} power spectrum compared to the shuffled case here may also be attributed to this anti-correlation of central \sfr{} with bias.

\vspace{0.5em}\noindent
\underline{Halo mass bin $\log M_h\in(12.5, 13)$ (lower right panel)}
The number of haloes in this bin is relatively low, yet these massive haloes contribute significantly to the power spectrum: this makes the ratio plot relatively noisy.
Shuffling in 2 percentile bins in satellite mass (navy line) reduces the bias on the largest scales (by $\sim 2$ per cent) yet the remaining offset is still $\sim 4$ per cent. This is within the 95th percentile of scatter between randomly shuffled realisations, meaning it is possible that a random assignment could also give this value of $P(k)$ at that $k$.

\vspace{0.5em}
At even larger halo masses, the number of haloes in the simulation is too low to convincingly distinguish between bias and random scatter.

\ifSubfilesClassLoaded{%
  \bibliography{bibliography}%
}{}
\end{document}

\section{Discussion}\label{sec:discussion}

\subsection{Impact of secondary bias on inferred astrophysical and cosmological parameters}

The amplitude of the large-scale power spectrum measured in a \lim\ survey is strongly affected by the specific mean intensity, $\bar{I}$, and the bias, $b$, in addition to the amplitude of the underlying matter power spectrum, $P_\mathrm{m}$. Neglecting {\em secondary bias} will bias the inferred values of $I$, $b$ and $P_\mathrm{m}$.
Any scale dependence of the secondary bias will further limit our ability to infer astrophysical and cosmological parameters from \lim. For instance, \citealt{Jimenez_2021} find scale-dependent secondary bias for \elg{}s using a semi-analytical model. However, the \tng{} simulation analysed here, with a linear extent of $205$\lenunit{}, is likely too small to distinguish scale dependence from sample variance. 

Cross-correlating a \lim\ survey with other surveys, where the galaxy bias ($b_g$) can be measured, can
put constraints on $I$ and $b$ \citep{Schaan+21-galaxy}. If these are well constrained, $P_\mathrm{m}$ can be extracted. However, the accuracy of this depends on the bias, $b_g$, estimated for the external survey catalogue, which itself may be affected by secondary bias. To 
break the degeneracy between secondary bias, the relation between the mean halo mass and the \sfr{} of its galaxies, and the parameters of the underlying cosmological model, additional statistics are required. 
A potential statistic is the voxel intensity distribution ({\sc vid}), which is the probability distribution function of the intensity measured within voxels \citep{Breysse_2017_pdf}. 

\subsection{Using secondary properties to reduce secondary bias}
Accounting for the halo concentration, \vmax{}, for haloes with mass $\log M_h \lesssim 12$, and subhalo mass, \msat{}, for higher mass haloes, significantly reduces but does not completely eliminate the impact of secondary bias (see \cref{fig:property2_all}).
As we explain in \cref{app:sat_ps}, much of the remaining offset in the power spectrum is likely due to secondary bias of central galaxies. To further reduce this bias
requires a better understanding of its origin.

Note, however, that whilst adding secondary properties can improve the modelling of the power spectrum, it may also introduce additional biases if the correlations that occur in simulations do not apply to the observed Universe.
It is therefore important to understand the impact on the power spectrum when including secondary properties in one's model.

\subsubsection{Other secondary properties}\label{sec:other_properties}

In this paper, we have only considered two properties, namely concentration and subhalo mass (\vmax{} and \msat{}).\footnote{Alternatives or proxies for \vmax{} and \msat{} may also be required if these quantities are not directly measurable from the dark matter simulation.}
In general, the dependence of {\em stellar mass} on secondary properties has been well studied, but quantifying and understanding the relationship with \sfr{} has proven more challenging, likely due to its stochastic variation \citep[e.g.][]{Caplar_2019,Matthee_2019}.

\citet{Xu_2021} find that, when selecting galaxies by stellar mass using a semi-analytic model, concentration accounts for some of the clustering bias (26\%), while the number of subhaloes per host halo accounts for a larger (30\%) amount of the bias. However, they also find that environmental properties can make an even larger contribution.
Many studies have also considered cosmic web locations or local density as secondary parameters to account for secondary bias in galaxies, which are often selected based on thresholds in properties such as stellar mass, specific star formation rate, or colour \citep[e.g.][]{Hadzhiyska_2021, Hasan_2023, Montero_2024}.

\citet{Gonzalez_2020} and \citet{Jimenez_2021} consider galaxies selected by cuts in \sfr{}, and in \ha, [\ion{O}{iii}] and [\ion{O}{ii}]. Their results suggest some correlation of the clustering of these galaxies with environment. While the power spectrum would be dominated by highly star-forming galaxies in both the luminosity-cut and weighting scenarios, it remains important to assess whether weighting by \sfr{} (or luminosity), as is relevant in \lim, reveals a distinct signal.

Spin also correlates with galaxy bias \citep{ Montero_Dorta_2020, Bose_2019}, and
the halo's merger history is another potential candidate for accounting for secondary bias. While the merger history goes some way into explaining the scatter in stellar mass at given halo mass, it has been less successful at predicting the \sfr{} \citep{Jespersen_2022}. Machine learning models which aim to reproduce galaxy properties also typically predict stellar mass more accurately than \sfr{} \citep[e.g.][]{Jo_2019, Hernandez_2023}.

\subsubsection{The dependence of secondary bias on halo mass}
An important note to take away is that we should not expect the same secondary bias parameter to work equally well for all halo masses. Although concentration is a commonly considered secondary property, we find that it is only effective for low-mass haloes. As satellite galaxies dominate the \sfr{} in higher-mass haloes, it is more appropriate to consider satellite-related properties for higher mass haloes and central-related properties for lower mass haloes. 

Previous studies have also shown that low-mass and high-mass haloes show different assembly bias signatures. For example, \citet{Gao_2005} show that the correlation of clustering with formation time is only significant for low-mass haloes.
\citet{Wechsler_2006} confirm that the correlation of clustering with formation time is stronger for low-mass haloes, and also find that the correlation with concentration is a function of halo mass -- they find a positive correlation for lower-mass haloes, with the trend reversing above a characteristic mass.
Consistent with this, we showed in \cref{sec:reducing_bias} that concentration accounts for some secondary bias for lower-mass haloes but has little effect at higher masses, due to quenching by \agn{} reducing the correlation of \sfr{} with concentration. \citet{Zehavi_2018} similarly find that early forming haloes are more likely to host central galaxies at lower halo mass. However, the reverse trend is found for satellites -- early forming haloes tend to have fewer satellites. \citet{Vakili2019} find that there is some correlation with concentration for central galaxies but not for satellite galaxies.

Therefore, when considering secondary properties it is important to consider different halo mass bins separately.

\subsubsection{Galactic conformity}
The term galactic conformity refers to correlations between the properties of nearby galaxies, for example in terms of colour or \sfr{}. In particular, the correlation between the properties of central and satellite galaxies within the same halo is often referred to as one-halo galactic conformity. 
If one assigns a single luminosity to a halo -- representing the combined emission from its central and satellite galaxies -- it may appear at first sight that we do not need to worry about galactic conformity.
However, even in this case, it may be valuable to consider what additional information can be obtained by considering centrals and satellites separately.
For example, central and satellite \sfr{}s show strong correlation in haloes with $\log M_h \gtrsim 12.5$ ($r_S \sim 0.4$, see \cref{fig:corr_sfr_cent_sat}), yet there is almost no correlation of central \sfr{} with satellite mass (\cref{fig:corr_coefs}). This suggests that there is another factor that drives conformity. We have not investigated this further in this work, but a better understanding of galactic conformity could help identify other causes of \sfr{} bias. 

In \cref{fig:cent_vs_sat_shuffled}, we demonstrated that the amplitude of the power spectrum on small scales, $\log k\gtrsim 0$, where the one-halo term dominates, is higher in \tng{} than in random realisations in which centrals and satellites are shuffled independently. A similar signal is found in the \desi{} One-Percent Survey \citep{desi_2024}. To account for this, \citet{Gao_2024} propose a subhalo abundance matching model in which the presence of a central \elg{} increases the probability that its satellites are also \elg{}s. Similarly, \citet{Reyes-Peraza_2024} consider an \hod{} model where the mean number of satellite galaxies is dependent on the properties of the central galaxy. 
See, e.g., \citet{Hartley_2015,Hadzhiyska_2023,Yuan_2025} for additional studies on galactic conformity.

\subsection{Caveats}

\subsubsection{Sample variance due to limited volume of the simulation}
Another source of uncertainty in our analysis is sample variance. The large scatter in both luminosity and bias at fixed halo mass means that different realisations of scatter can lead to significantly different power spectra -- even in the absence of secondary bias (see blue dashed and dotted lines in, e.g., \cref{fig:shuffled_ps_main}). Sample variance becomes more significant for high-mass haloes that contribute strongly to the power spectrum due to their higher bias and luminosities, but whose statistics are relatively poorly sampled. The combination of low number statistics and large individual contributions of such massive haloes amplifies the impact of sample variance on the
power spectrum. The large variance in the amplitude of the power spectrum across randomly shuffled realisations may hide any clustering bias. The sample variance also makes it difficult to identify scale-dependent bias.

Observational surveys typically map volumes that are significantly larger than the $L\sim 205$ \lenunit{} of \tng{}, and hence suffer less from sample variance.
This motivates the use of alternative methods such as dark matter-only simulations
combined with a model to assign \sfr{} to haloes. However, not accounting for 
secondary bias will bias the comparison between model and data. Hydrodynamical simulations can provide valuable information about the nature of secondary galaxy bias.

\citet{Mao_2018} find strong correlations between the clustering bias of haloes
of a given mass and the presence of substructure inside them. They
used the $1\, \mathrm{Gpc^3 h^{-3}}$ MultiDark Planck 2 $N$-body simulation described by \cite{Klypin_2016}. The large size of the box means that they can more reliably sample secondary halo bias for massive haloes ($\log M_h$ $\gtrsim$ \mhalounit{14}). While we have found that satellite mass, for example, does reduce the relative bias, the large variance makes it difficult to disentangle bias from random variation.

The large variance seen in the power spectra in the exercises done in this work highlights the need for larger boxes not only to reduce the variance due to the dark matter but also the variance due to the scatter in the $L-M$ relation.

\subsubsection{Other limitations of this work}
In this paper, we assumed that the H$\alpha$ flux is directly proportional to the \sfr{}. 
We assumed a constant dust attenuation, such that the direct proportionality remains unaffected, though dust attenuation may not affect all galaxies equally.
While \sfr{} is a primary driver of \ha{} emission, other factors — including gas metallicity and \agn{} activity — can affect the relation between luminosity and \sfr{}. 
For example, \citet{Jimenez_2021} show that the secondary bias signals of [\ion{O}{iii}] and [\ion{O}{ii}] emitters differ from those of \sfr{}-selected galaxies in galaxy surveys. Similar tests can be applied to assess whether the secondary bias signal is affected by the dependence of luminosities on properties other than \sfr{}.

We have examined a snapshot at a single redshift and have not accounted for contamination from interlopers, or uncertainties resulting from the location of the continuum, which are significant challenges in \lim{} (see e.g. \citealt{Bernal_2025}). Even if contamination errors exceed the secondary bias signal, the systematic error introduced by secondary bias should not be overlooked.

The effect of redshift space distortions (\rsd{}s), which arise from galaxy motion along the line of sight, has also not been included. See e.g. \citet{Hamilton_1998} for a general review of \rsd{}s and e.g. \citet{Bernal_2019}
for their impact on \lim{}. Additionally, we have only considered the 3D power spectrum, but for surveys such as SPHEREx, which have low spectral resolution, the 2D power spectrum can sometimes be more appropriate.

This work focuses on $z \sim 1.5$, but future studies can explore redshift dependence. For instance, \citet{Contreras_2019} use a semi-analytic model to show that, while at $z=0$ clustering is correlated with both formation time and concentration, at $z=3$, the correlation reverses for concentration and diminishes for formation time. The redshift dependence of secondary biases can propagate to to the \lim{} power spectrum systematically.

While \illus{} provides valuable insights, as with other simulations, there may be systematic differences between its predictions and reality. Therefore, it may be informative to examine predictions of other simulations as well.

\ifSubfilesClassLoaded{%
  \bibliography{bibliography}%
}{}

\end{document}

\section{Summary and Conclusions}\label{sec:summary}

Line-intensity mapping (\lim{}) surveys have the potential to map large volumes whilst probing faint galaxies, enabling us to obtain accurate clustering information from large to small scales \citep[e.g.][]{Kovetz_2017}. The power spectrum measured in such surveys is weighted by the flux (or luminosity at a given redshift) of an emission line, which, for many lines, is closely tied to the star formation rate (\sfr{}) of the galaxy. The power spectrum provides a statistical description of the spatial distribution of these sources, enabling the quantification of the spatial distribution of galaxies, which reflects the clustering of the underlying dark matter. As the observed emission depends on how galaxies form and evolve within dark matter haloes, interpreting the power spectrum provides insight into the physics of galaxy formation.
At the same time, the distribution of dark matter haloes is sensitive to the initial conditions of the Universe and the cosmological model, making the power spectrum a powerful probe of cosmology. 

To infer the distribution of dark matter from the observed emission-line power spectrum requires a model that connects the observed luminosity to the underlying matter density. Hydrodynamical simulations offer a physically motivated approach by directly modelling galaxy formation through the equations of gravity and hydrodynamics, with subgrid models capturing the essence of star formation and their associated feedback processes. However, such simulations are computationally expensive and limited in the volume that can be modelled, and in the spatial and mass resolution that can be sampled. A commonly used alternative to simulations is to relate the total luminosity of galaxies within a halo to the halo mass using a model. However, we find that accounting for halo mass alone causes an underestimation of the power spectrum measured in the \illus{} hydrodynamical simulation \citep{Pillepich_2018}.

It is well established that the amplitude of the clustering of haloes 
depends on mass. However, at given mass, it can also depend on secondary properties, a phenomenon known as secondary halo bias. A well-known example of secondary bias is assembly bias, which refers to the dependence of halo bias on formation history (see e.g. \citealt{Mao_2018}). 
If galaxy properties linked to observables are correlated with halo bias at fixed halo mass, detectable signatures may be imprinted in observations.
We refer to such correlations under the broader term secondary bias, encompassing correlations of bias with both galaxy and secondary dark matter properties.
We show mathematically and using toy models that, in the absence of secondary bias, the large-scale power spectrum (excluding shot noise) and the linear matter power spectrum
are related by the mean luminosity-halo mass relation, independent of any random scatter. However, this relation changes if the scatter is not random but correlated with the bias of the host haloes (see \cref{sec:assignment_effect}).

We specifically investigate the impact of secondary bias on the \sfr{}-weighted power spectrum using the TNG300-1 (\tng{}) run of the \illus{} hydrodynamical simulation suite at $z \sim 1.5$ (linear extent $205~\lenunit$). This is particularly relevant for emission lines measured in \lim{} that are closely related to the \sfr{}. We investigate how the \sfr{} of central and satellite galaxies correlates with secondary properties (in particular, concentration parametrised by $\tilde{c} = V_\mathrm{max}/(10\,H\,R_\mathrm{max})$ and total satellite mass), and how these correlations impact the \lim{} power spectrum. To do so, we shuffle galaxies in bins of halo mass and in bins of the secondary parameter, and then compare the original power spectrum to that of the shuffled case.

Our main findings are as follows.
\begin{enumerate}
    \item Secondary bias causes a systematic enhancement of the two-halo term
    of the \sfr{}-weighted power spectrum of up to 5 per cent, on scales $\log_{10}( k\ [\kunit])\gtrsim -0.2$ (\cref{fig:shuffled_ps_main}).
    \item Although satellite galaxies only contribute 30 per cent of the star formation rate density and power spectrum, they contribute more to the secondary bias than central galaxies (\cref{fig:cent_vs_sat_shuffled}).
    \item Correlation between the \sfr{} of central and satellite galaxies (one-halo galactic conformity) results in a 10 per cent increase in the galaxy power spectrum on scales  $\log_{10}( k\ [\kunit])\gtrsim 0.0$, where the power spectrum is dominated by the one-halo term (\cref{fig:cent_vs_sat_shuffled}).
    \item Using satellite subhalo mass as a secondary parameter accounts for a large fraction of the secondary bias, in particular for higher mass haloes
    ($\log_{10} (M_h\ [\massunit]) \gtrsim 12$). Concentration only accounts for some of the secondary bias, which is due to central galaxies in lower mass haloes ($\log_{10} (M_h\ [\massunit]) \lesssim 12$, \cref{fig:msat_conc_effect_logMs}).
\end{enumerate}

In conclusion, secondary bias limits the accuracy with which cosmological and astrophysical parameters can be inferred from the \lim{} power spectrum, because it biases the power spectrum. Incorporating secondary parameters into mock catalogues can reduce the discrepancy, but care must be taken as using inappropriate parameters can introduce additional errors. We find that concentration and satellite subhalo mass are effective secondary parameters, though some residual discrepancy remains. Further work is needed to identify additional parameters -- that are potentially halo-mass dependent or even scale dependent -- for more accurate mock catalogue generation.

\section*{Acknowledgements}

RLJ has been supported by The University of Tokyo Fellowship. KM acknowledges JSPS KAKENHI Grant Number 23K03446, 23K20035, and 24H00004. SB is supported by the UK Research and Innovation (UKRI) Future Leaders Fellowship (grant number MR/V023381/1). This work is co-funded by the European Union (Widening Participation, ExGal-Twin, GA 101158446). Views and opinions expressed are however those of the author(s) only and do not necessarily reflect those of the European Union. Neither the European Union nor the granting authority can be held responsible for them.
This work used the DiRAC@Durham facility managed by the Institute for Computational Cosmology on behalf of the STFC DiRAC HPC Facility (www.dirac.ac.uk). The equipment was funded by BEIS capital funding via STFC capital grants ST/K00042X/1, ST/P002293/1, ST/R002371/1 and ST/S002502/1, Durham University and STFC operations grant ST/R000832/1. DiRAC is part of the National e-Infrastructure. 
Python packages used include \texttt{matplotlib} \citep{hunter_2007} and \texttt{numpy} \citep{numpy_2020}.

\section*{Data Availability}
The \illus\ data are publicly available at \url{https://www.tng-project.org/}.
The \eagle\ data are publicly available at \url{https://icc.dur.ac.uk/Eagle/}.
We will share the scripts used in this paper upon reasonable request.

\bibliography{bibliography}

\begin{thebibliography}{}
\makeatletter
\relax
\def\mn@urlcharsother{\let\do\@makeother \do\$\do\&\do\#\do\^\do\_\do\%\do\~}
\def\mn@doi{\begingroup\mn@urlcharsother \@ifnextchar [ {\mn@doi@} {\mn@doi@[]}}
\def\mn@doi@[#1]#2{\def\@tempa{#1}\ifx\@tempa\@empty \href {http://dx.doi.org/#2} {doi:#2}\else \href {http://dx.doi.org/#2} {#1}\fi \endgroup}
\def\mn@eprint#1#2{\mn@eprint@#1:#2::\@nil}
\def\mn@eprint@arXiv#1{\href {http://arxiv.org/abs/#1} {{\tt arXiv:#1}}}
\def\mn@eprint@dblp#1{\href {http://dblp.uni-trier.de/rec/bibtex/#1.xml} {dblp:#1}}
\def\mn@eprint@#1:#2:#3:#4\@nil{\def\@tempa {#1}\def\@tempb {#2}\def\@tempc {#3}\ifx \@tempc \@empty \let \@tempc \@tempb \let \@tempb \@tempa \fi \ifx \@tempb \@empty \def\@tempb {arXiv}\fi \@ifundefined {mn@eprint@\@tempb}{\@tempb:\@tempc}{\expandafter \expandafter \csname mn@eprint@\@tempb\endcsname \expandafter{\@tempc}}}

\bibitem[\protect\citeauthoryear{{Asgari}, {Mead}  \& {Heymans}}{{Asgari} et~al.}{2023}]{Asgari23}
{Asgari} M.,  {Mead} A.~J.,   {Heymans} C.,  2023, \mn@doi [The Open Journal of Astrophysics] {10.21105/astro.2303.08752}, \href {https://ui.adsabs.harvard.edu/abs/2023OJAp....6E..39A} {6, 39}

\bibitem[\protect\citeauthoryear{{Ayromlou}, {Kauffmann}, {Anand}  \& {White}}{{Ayromlou} et~al.}{2023}]{Ayromlou23}
{Ayromlou} M.,  {Kauffmann} G.,  {Anand} A.,   {White} S. D.~M.,  2023, \mn@doi [\mnras] {10.1093/mnras/stac3637}, \href {https://ui.adsabs.harvard.edu/abs/2023MNRAS.519.1913A} {519, 1913}

\bibitem[\protect\citeauthoryear{{Bah{\'e}} et~al.,}{{Bah{\'e}} et~al.}{2017}]{Bahe17}
{Bah{\'e}} Y.~M.,  et~al., 2017, \mn@doi [\mnras] {10.1093/mnras/stx1403}, \href {https://ui.adsabs.harvard.edu/abs/2017MNRAS.470.4186B} {470, 4186}

\bibitem[\protect\citeauthoryear{{Balaguera-Antol{\'\i}nez}, {Montero-Dorta}  \& {Favole}}{{Balaguera-Antol{\'\i}nez} et~al.}{2024}]{Balaguera_2024}
{Balaguera-Antol{\'\i}nez} A.,  {Montero-Dorta} A.~D.,   {Favole} G.,  2024, \mn@doi [\aap] {10.1051/0004-6361/202348694}, \href {https://ui.adsabs.harvard.edu/abs/2024A&A...685A..61B} {685, A61}

\bibitem[\protect\citeauthoryear{{Baldauf}, {Seljak}, {Smith}, {Hamaus}  \& {Desjacques}}{{Baldauf} et~al.}{2013}]{Baldauf_2013}
{Baldauf} T.,  {Seljak} U.,  {Smith} R.~E.,  {Hamaus} N.,   {Desjacques} V.,  2013, \mn@doi [\prd] {10.1103/PhysRevD.88.083507}, \href {https://ui.adsabs.harvard.edu/abs/2013PhRvD..88h3507B} {88, 083507}

\bibitem[\protect\citeauthoryear{{Ballardini} et~al.,}{{Ballardini} et~al.}{2024}]{Ballardini24}
{Ballardini} M.,  et~al., 2024, \mn@doi [\aap] {10.1051/0004-6361/202348162}, \href {https://ui.adsabs.harvard.edu/abs/2024A&A...683A.220B} {683, A220}

\bibitem[\protect\citeauthoryear{{Berlind} \& {Weinberg}}{{Berlind} \& {Weinberg}}{2002}]{Berlind02}
{Berlind} A.~A.,  {Weinberg} D.~H.,  2002, \mn@doi [\apj] {10.1086/341469}, \href {https://ui.adsabs.harvard.edu/abs/2002ApJ...575..587B} {575, 587}

\bibitem[\protect\citeauthoryear{{Bernal} \& {Baleato Lizancos}}{{Bernal} \& {Baleato Lizancos}}{2025}]{Bernal_2025}
{Bernal} J.~L.,  {Baleato Lizancos} A.,  2025, \mn@doi [\prd] {10.1103/PhysRevD.111.043539}, \href {https://ui.adsabs.harvard.edu/abs/2025PhRvD.111d3539B} {111, 043539}

\bibitem[\protect\citeauthoryear{{Bernal}, {Breysse}, {Gil-Mar{\'\i}n}  \& {Kovetz}}{{Bernal} et~al.}{2019}]{Bernal_2019}
{Bernal} J.~L.,  {Breysse} P.~C.,  {Gil-Mar{\'\i}n} H.,   {Kovetz} E.~D.,  2019, \mn@doi [\prd] {10.1103/PhysRevD.100.123522}, \href {https://ui.adsabs.harvard.edu/abs/2019PhRvD.100l3522B} {100, 123522}

\bibitem[\protect\citeauthoryear{{Bett}, {Eke}, {Frenk}, {Jenkins}, {Helly}  \& {Navarro}}{{Bett} et~al.}{2007}]{Bett_2007}
{Bett} P.,  {Eke} V.,  {Frenk} C.~S.,  {Jenkins} A.,  {Helly} J.,   {Navarro} J.,  2007, \mn@doi [\mnras] {10.1111/j.1365-2966.2007.11432.x}, \href {https://ui.adsabs.harvard.edu/abs/2007MNRAS.376..215B} {376, 215}

\bibitem[\protect\citeauthoryear{{Blanton} et~al.,}{{Blanton} et~al.}{2017}]{Blanton_2017}
{Blanton} M.~R.,  et~al., 2017, \mn@doi [\aj] {10.3847/1538-3881/aa7567}, \href {https://ui.adsabs.harvard.edu/abs/2017AJ....154...28B} {154, 28}

\bibitem[\protect\citeauthoryear{{Bose}, {Eisenstein}, {Hernquist}, {Pillepich}, {Nelson}, {Marinacci}, {Springel}  \& {Vogelsberger}}{{Bose} et~al.}{2019}]{Bose_2019}
{Bose} S.,  {Eisenstein} D.~J.,  {Hernquist} L.,  {Pillepich} A.,  {Nelson} D.,  {Marinacci} F.,  {Springel} V.,   {Vogelsberger} M.,  2019, \mn@doi [\mnras] {10.1093/mnras/stz2546}, \href {https://ui.adsabs.harvard.edu/abs/2019MNRAS.490.5693B} {490, 5693}

\bibitem[\protect\citeauthoryear{{Breysse}, {Kovetz}, {Behroozi}, {Dai}  \& {Kamionkowski}}{{Breysse} et~al.}{2017}]{Breysse_2017_pdf}
{Breysse} P.~C.,  {Kovetz} E.~D.,  {Behroozi} P.~S.,  {Dai} L.,   {Kamionkowski} M.,  2017, \mn@doi [\mnras] {10.1093/mnras/stx203}, \href {https://ui.adsabs.harvard.edu/abs/2017MNRAS.467.2996B} {467, 2996}

\bibitem[\protect\citeauthoryear{{CONCERTO Collaboration} et~al.,}{{CONCERTO Collaboration} et~al.}{2020}]{Concerto_2020}
{CONCERTO Collaboration} et~al., 2020, \mn@doi [\aap] {10.1051/0004-6361/202038456}, \href {https://ui.adsabs.harvard.edu/abs/2020A&A...642A..60C} {642, A60}

\bibitem[\protect\citeauthoryear{{Caplar} \& {Tacchella}}{{Caplar} \& {Tacchella}}{2019}]{Caplar_2019}
{Caplar} N.,  {Tacchella} S.,  2019, \mn@doi [\mnras] {10.1093/mnras/stz1449}, \href {https://ui.adsabs.harvard.edu/abs/2019MNRAS.487.3845C} {487, 3845}

\bibitem[\protect\citeauthoryear{{Chabrier}}{{Chabrier}}{2003}]{Chabrier2003}
{Chabrier} G.,  2003, \mn@doi [\pasp] {10.1086/376392}, \href {https://ui.adsabs.harvard.edu/abs/2003PASP..115..763C} {115, 763}

\bibitem[\protect\citeauthoryear{{Chaves-Montero}, {Angulo}, {Schaye}, {Schaller}, {Crain}, {Furlong}  \& {Theuns}}{{Chaves-Montero} et~al.}{2016}]{Chaves-Montero_2016}
{Chaves-Montero} J.,  {Angulo} R.~E.,  {Schaye} J.,  {Schaller} M.,  {Crain} R.~A.,  {Furlong} M.,   {Theuns} T.,  2016, \mn@doi [\mnras] {10.1093/mnras/stw1225}, \href {https://ui.adsabs.harvard.edu/abs/2016MNRAS.460.3100C} {460, 3100}

\bibitem[\protect\citeauthoryear{{Chittenden} \& {Tojeiro}}{{Chittenden} \& {Tojeiro}}{2023}]{Chittenden_2023}
{Chittenden} H.~G.,  {Tojeiro} R.,  2023, \mn@doi [\mnras] {10.1093/mnras/stac3498}, \href {https://ui.adsabs.harvard.edu/abs/2023MNRAS.518.5670C} {518, 5670}

\bibitem[\protect\citeauthoryear{{Cole} \& {Kaiser}}{{Cole} \& {Kaiser}}{1989}]{Cole89}
{Cole} S.,  {Kaiser} N.,  1989, \mn@doi [\mnras] {10.1093/mnras/237.4.1127}, \href {https://ui.adsabs.harvard.edu/abs/1989MNRAS.237.1127C} {237, 1127}

\bibitem[\protect\citeauthoryear{{Colless} et~al.,}{{Colless} et~al.}{2001}]{Colless_2001}
{Colless} M.,  et~al., 2001, \mn@doi [\mnras] {10.1046/j.1365-8711.2001.04902.x}, \href {https://ui.adsabs.harvard.edu/abs/2001MNRAS.328.1039C} {328, 1039}

\bibitem[\protect\citeauthoryear{{Conroy} \& {Wechsler}}{{Conroy} \& {Wechsler}}{2009}]{Conroy_2009}
{Conroy} C.,  {Wechsler} R.~H.,  2009, \mn@doi [\apj] {10.1088/0004-637X/696/1/620}, \href {https://ui.adsabs.harvard.edu/abs/2009ApJ...696..620C} {696, 620}

\bibitem[\protect\citeauthoryear{{Contreras}, {Zehavi}, {Padilla}, {Baugh}, {Jim{\'e}nez}  \& {Lacerna}}{{Contreras} et~al.}{2019}]{Contreras_2019}
{Contreras} S.,  {Zehavi} I.,  {Padilla} N.,  {Baugh} C.~M.,  {Jim{\'e}nez} E.,   {Lacerna} I.,  2019, \mn@doi [\mnras] {10.1093/mnras/stz018}, \href {https://ui.adsabs.harvard.edu/abs/2019MNRAS.484.1133C} {484, 1133}

\bibitem[\protect\citeauthoryear{{Cooray} \& {Sheth}}{{Cooray} \& {Sheth}}{2002}]{Cooray02}
{Cooray} A.,  {Sheth} R.,  2002, \mn@doi [\physrep] {10.1016/S0370-1573(02)00276-4}, \href {https://ui.adsabs.harvard.edu/abs/2002PhR...372....1C} {372, 1}

\bibitem[\protect\citeauthoryear{{Cortese}, {Catinella}  \& {Smith}}{{Cortese} et~al.}{2021}]{Dawes_Review9}
{Cortese} L.,  {Catinella} B.,   {Smith} R.,  2021, \mn@doi [\pasa] {10.1017/pasa.2021.18}, \href {https://ui.adsabs.harvard.edu/abs/2021PASA...38...35C} {38, e035}

\bibitem[\protect\citeauthoryear{{Crain} \& {van de Voort}}{{Crain} \& {van de Voort}}{2023}]{Crain23}
{Crain} R.~A.,  {van de Voort} F.,  2023, \mn@doi [\araa] {10.1146/annurev-astro-041923-043618}, \href {https://ui.adsabs.harvard.edu/abs/2023ARA&A..61..473C} {61, 473}

\bibitem[\protect\citeauthoryear{{Crain} et~al.,}{{Crain} et~al.}{2015}]{Crain15}
{Crain} R.~A.,  et~al., 2015, \mn@doi [\mnras] {10.1093/mnras/stv725}, \href {https://ui.adsabs.harvard.edu/abs/2015MNRAS.450.1937C} {450, 1937}

\bibitem[\protect\citeauthoryear{{Croton}, {Gao}  \& {White}}{{Croton} et~al.}{2007}]{Croton_2007}
{Croton} D.~J.,  {Gao} L.,   {White} S. D.~M.,  2007, \mn@doi [\mnras] {10.1111/j.1365-2966.2006.11230.x}, \href {https://ui.adsabs.harvard.edu/abs/2007MNRAS.374.1303C} {374, 1303}

\bibitem[\protect\citeauthoryear{{DESI Collaboration} et~al.,}{{DESI Collaboration} et~al.}{2016}]{Desi_2016}
{DESI Collaboration} et~al., 2016, \mn@doi [arXiv e-prints] {10.48550/arXiv.1611.00036}, \href {https://ui.adsabs.harvard.edu/abs/2016arXiv161100036D} {p. arXiv:1611.00036}

\bibitem[\protect\citeauthoryear{{DESI Collaboration} et~al.,}{{DESI Collaboration} et~al.}{2024}]{desi_2024}
{DESI Collaboration} et~al., 2024, \mn@doi [\aj] {10.3847/1538-3881/ad0b08}, \href {https://ui.adsabs.harvard.edu/abs/2024AJ....167...62D} {167, 62}

\bibitem[\protect\citeauthoryear{{DESI Collaboration} et~al.,}{{DESI Collaboration} et~al.}{2025}]{Desi25}
{DESI Collaboration} et~al., 2025, \mn@doi [arXiv e-prints] {10.48550/arXiv.2503.14745}, \href {https://ui.adsabs.harvard.edu/abs/2025arXiv250314745D} {p. arXiv:2503.14745}

\bibitem[\protect\citeauthoryear{{Davis}, {Efstathiou}, {Frenk}  \& {White}}{{Davis} et~al.}{1985}]{FOF}
{Davis} M.,  {Efstathiou} G.,  {Frenk} C.~S.,   {White} S.~D.~M.,  1985, \mn@doi [\apj] {10.1086/163168}, \href {https://ui.adsabs.harvard.edu/abs/1985ApJ...292..371D} {292, 371}

\bibitem[\protect\citeauthoryear{{Donnari} et~al.,}{{Donnari} et~al.}{2019}]{Donnari_2019_erratum}
{Donnari} M.,  et~al., 2019, \mn@doi [\mnras] {10.1093/mnras/stz2395}, \href {https://ui.adsabs.harvard.edu/abs/2019MNRAS.489.3036D} {489, 3036}

\bibitem[\protect\citeauthoryear{{Dor{\'e}} et~al.,}{{Dor{\'e}} et~al.}{2018}]{SPHEREx_2018}
{Dor{\'e}} O.,  et~al., 2018, \mn@doi [arXiv e-prints] {10.48550/arXiv.1805.05489}, \href {https://ui.adsabs.harvard.edu/abs/2018arXiv180505489D} {p. arXiv:1805.05489}

\bibitem[\protect\citeauthoryear{{Fonseca}, {Silva}, {Santos}  \& {Cooray}}{{Fonseca} et~al.}{2017}]{Fonseca_2017}
{Fonseca} J.,  {Silva} M.~B.,  {Santos} M.~G.,   {Cooray} A.,  2017, \mn@doi [\mnras] {10.1093/mnras/stw2470}, \href {https://ui.adsabs.harvard.edu/abs/2017MNRAS.464.1948F} {464, 1948}

\bibitem[\protect\citeauthoryear{{Gao} \& {White}}{{Gao} \& {White}}{2007}]{Gao_2007}
{Gao} L.,  {White} S. D.~M.,  2007, \mn@doi [\mnras] {10.1111/j.1745-3933.2007.00292.x}, \href {https://ui.adsabs.harvard.edu/abs/2007MNRAS.377L...5G} {377, L5}

\bibitem[\protect\citeauthoryear{{Gao}, {White}, {Jenkins}, {Stoehr}  \& {Springel}}{{Gao} et~al.}{2004}]{Gao_2004}
{Gao} L.,  {White} S.~D.~M.,  {Jenkins} A.,  {Stoehr} F.,   {Springel} V.,  2004, \mn@doi [\mnras] {10.1111/j.1365-2966.2004.08360.x}, \href {https://ui.adsabs.harvard.edu/abs/2004MNRAS.355..819G} {355, 819}

\bibitem[\protect\citeauthoryear{{Gao}, {Springel}  \& {White}}{{Gao} et~al.}{2005}]{Gao_2005}
{Gao} L.,  {Springel} V.,   {White} S. D.~M.,  2005, \mn@doi [\mnras] {10.1111/j.1745-3933.2005.00084.x}, \href {https://ui.adsabs.harvard.edu/abs/2005MNRAS.363L..66G} {363, L66}

\bibitem[\protect\citeauthoryear{{Gao} et~al.,}{{Gao} et~al.}{2024}]{Gao_2024}
{Gao} H.,  et~al., 2024, \mn@doi [\apj] {10.3847/1538-4357/ad09d6}, \href {https://ui.adsabs.harvard.edu/abs/2024ApJ...961...74G} {961, 74}

\bibitem[\protect\citeauthoryear{{Garn} et~al.,}{{Garn} et~al.}{2010}]{Garn10}
{Garn} T.,  et~al., 2010, \mn@doi [\mnras] {10.1111/j.1365-2966.2009.16042.x}, \href {https://ui.adsabs.harvard.edu/abs/2010MNRAS.402.2017G} {402, 2017}

\bibitem[\protect\citeauthoryear{{Geller} \& {Huchra}}{{Geller} \& {Huchra}}{1989}]{Geller89}
{Geller} M.~J.,  {Huchra} J.~P.,  1989, \mn@doi [Science] {10.1126/science.246.4932.897}, \href {https://ui.adsabs.harvard.edu/abs/1989Sci...246..897G} {246, 897}

\bibitem[\protect\citeauthoryear{{Gong}, {Cooray}, {Silva}, {Zemcov}, {Feng}, {Santos}, {Dore}  \& {Chen}}{{Gong} et~al.}{2017}]{Gong_2017}
{Gong} Y.,  {Cooray} A.,  {Silva} M.~B.,  {Zemcov} M.,  {Feng} C.,  {Santos} M.~G.,  {Dore} O.,   {Chen} X.,  2017, \mn@doi [\apj] {10.3847/1538-4357/835/2/273}, \href {https://ui.adsabs.harvard.edu/abs/2017ApJ...835..273G} {835, 273}

\bibitem[\protect\citeauthoryear{{Gonzalez-Perez} et~al.,}{{Gonzalez-Perez} et~al.}{2020}]{Gonzalez_2020}
{Gonzalez-Perez} V.,  et~al., 2020, \mn@doi [\mnras] {10.1093/mnras/staa2504}, \href {https://ui.adsabs.harvard.edu/abs/2020MNRAS.498.1852G} {498, 1852}

\bibitem[\protect\citeauthoryear{{Hadzhiyska}, {Tacchella}, {Bose}  \& {Eisenstein}}{{Hadzhiyska} et~al.}{2021}]{Hadzhiyska_2021}
{Hadzhiyska} B.,  {Tacchella} S.,  {Bose} S.,   {Eisenstein} D.~J.,  2021, \mn@doi [\mnras] {10.1093/mnras/stab243}, \href {https://ui.adsabs.harvard.edu/abs/2021MNRAS.502.3599H} {502, 3599}

\bibitem[\protect\citeauthoryear{{Hadzhiyska} et~al.,}{{Hadzhiyska} et~al.}{2023}]{Hadzhiyska_2023}
{Hadzhiyska} B.,  et~al., 2023, \mn@doi [\mnras] {10.1093/mnras/stad731}, \href {https://ui.adsabs.harvard.edu/abs/2023MNRAS.524.2507H} {524, 2507}

\bibitem[\protect\citeauthoryear{{Hamilton}}{{Hamilton}}{1998}]{Hamilton_1998}
{Hamilton} A.~J.~S.,  1998, in {Hamilton} D.,  ed.,  Astrophysics and Space Science Library Vol. 231, The Evolving Universe. p.~185 (\mn@eprint {arXiv} {astro-ph/9708102}), \mn@doi{10.1007/978-94-011-4960-0_17}

\bibitem[\protect\citeauthoryear{Harris et~al.,}{Harris et~al.}{2020}]{numpy_2020}
Harris C.~R.,  et~al., 2020, \mn@doi [Nature] {10.1038/s41586-020-2649-2}, 585, 357

\bibitem[\protect\citeauthoryear{{Hartley}, {Conselice}, {Mortlock}, {Foucaud}  \& {Simpson}}{{Hartley} et~al.}{2015}]{Hartley_2015}
{Hartley} W.~G.,  {Conselice} C.~J.,  {Mortlock} A.,  {Foucaud} S.,   {Simpson} C.,  2015, \mn@doi [\mnras] {10.1093/mnras/stv972}, \href {https://ui.adsabs.harvard.edu/abs/2015MNRAS.451.1613H} {451, 1613}

\bibitem[\protect\citeauthoryear{{Hasan} et~al.,}{{Hasan} et~al.}{2023}]{Hasan_2023}
{Hasan} F.,  et~al., 2023, \mn@doi [\apj] {10.3847/1538-4357/acd11c}, \href {https://ui.adsabs.harvard.edu/abs/2023ApJ...950..114H} {950, 114}

\bibitem[\protect\citeauthoryear{{Hearin}, {Watson}  \& {van den Bosch}}{{Hearin} et~al.}{2015}]{Hearin_2015}
{Hearin} A.~P.,  {Watson} D.~F.,   {van den Bosch} F.~C.,  2015, \mn@doi [\mnras] {10.1093/mnras/stv1358}, \href {https://ui.adsabs.harvard.edu/abs/2015MNRAS.452.1958H} {452, 1958}

\bibitem[\protect\citeauthoryear{{Hern{\'a}ndez}, {Gonz{\'a}lez}  \& {Padilla}}{{Hern{\'a}ndez} et~al.}{2023}]{Hernandez_2023}
{Hern{\'a}ndez} C.~A.,  {Gonz{\'a}lez} R.~E.,   {Padilla} N.~D.,  2023, \mn@doi [\mnras] {10.1093/mnras/stad2112}, \href {https://ui.adsabs.harvard.edu/abs/2023MNRAS.524.4653H} {524, 4653}

\bibitem[\protect\citeauthoryear{Hunter}{Hunter}{2007}]{hunter_2007}
Hunter J.~D.,  2007, \mn@doi [Computing in Science \& Engineering] {10.1109/MCSE.2007.55}, 9, 90

\bibitem[\protect\citeauthoryear{{Jespersen}, {Cranmer}, {Melchior}, {Ho}, {Somerville}  \& {Gabrielpillai}}{{Jespersen} et~al.}{2022}]{Jespersen_2022}
{Jespersen} C.~K.,  {Cranmer} M.,  {Melchior} P.,  {Ho} S.,  {Somerville} R.~S.,   {Gabrielpillai} A.,  2022, \mn@doi [\apj] {10.3847/1538-4357/ac9b18}, \href {https://ui.adsabs.harvard.edu/abs/2022ApJ...941....7J} {941, 7}

\bibitem[\protect\citeauthoryear{{Jim{\'e}nez}, {Padilla}, {Contreras}, {Zehavi}, {Baugh}  \& {Orsi}}{{Jim{\'e}nez} et~al.}{2021}]{Jimenez_2021}
{Jim{\'e}nez} E.,  {Padilla} N.,  {Contreras} S.,  {Zehavi} I.,  {Baugh} C.~M.,   {Orsi} {\'A}.,  2021, \mn@doi [\mnras] {10.1093/mnras/stab1819}, \href {https://ui.adsabs.harvard.edu/abs/2021MNRAS.506.3155J} {506, 3155}

\bibitem[\protect\citeauthoryear{{Jo} \& {Kim}}{{Jo} \& {Kim}}{2019}]{Jo_2019}
{Jo} Y.,  {Kim} J.-h.,  2019, \mn@doi [\mnras] {10.1093/mnras/stz2304}, \href {https://ui.adsabs.harvard.edu/abs/2019MNRAS.489.3565J} {489, 3565}

\bibitem[\protect\citeauthoryear{{Jun}, {Theuns}, {Moriwaki}  \& {Bose}}{{Jun} et~al.}{2025}]{Jun_2024}
{Jun} R.~L.,  {Theuns} T.,  {Moriwaki} K.,   {Bose} S.,  2025, \mn@doi [\mnras] {10.1093/mnras/staf647}, \href {https://ui.adsabs.harvard.edu/abs/2025MNRAS.540..433J} {540, 433}

\bibitem[\protect\citeauthoryear{{Kaiser}}{{Kaiser}}{1984}]{Kaiser_1984}
{Kaiser} N.,  1984, \mn@doi [\apjl] {10.1086/184341}, \href {https://ui.adsabs.harvard.edu/abs/1984ApJ...284L...9K} {284, L9}

\bibitem[\protect\citeauthoryear{{Kennicutt}}{{Kennicutt}}{1998}]{Kennicutt98}
{Kennicutt} Robert~C. J.,  1998, \mn@doi [\araa] {10.1146/annurev.astro.36.1.189}, \href {https://ui.adsabs.harvard.edu/abs/1998ARA&A..36..189K} {36, 189}

\bibitem[\protect\citeauthoryear{{Klypin}, {Yepes}, {Gottl{\"o}ber}, {Prada}  \& {He{\ss}}}{{Klypin} et~al.}{2016}]{Klypin_2016}
{Klypin} A.,  {Yepes} G.,  {Gottl{\"o}ber} S.,  {Prada} F.,   {He{\ss}} S.,  2016, \mn@doi [\mnras] {10.1093/mnras/stw248}, \href {https://ui.adsabs.harvard.edu/abs/2016MNRAS.457.4340K} {457, 4340}

\bibitem[\protect\citeauthoryear{{Kovetz} et~al.,}{{Kovetz} et~al.}{2017}]{Kovetz_2017}
{Kovetz} E.~D.,  et~al., 2017, \mn@doi [arXiv e-prints] {10.48550/arXiv.1709.09066}, \href {https://ui.adsabs.harvard.edu/abs/2017arXiv170909066K} {p. arXiv:1709.09066}

\bibitem[\protect\citeauthoryear{{Lacerna} \& {Padilla}}{{Lacerna} \& {Padilla}}{2012}]{Lacerna_2012}
{Lacerna} I.,  {Padilla} N.,  2012, \mn@doi [\mnras] {10.1111/j.1745-3933.2012.01316.x}, \href {https://ui.adsabs.harvard.edu/abs/2012MNRAS.426L..26L} {426, L26}

\bibitem[\protect\citeauthoryear{{Li}, {Wechsler}, {Devaraj}  \& {Church}}{{Li} et~al.}{2016}]{Li16}
{Li} T.~Y.,  {Wechsler} R.~H.,  {Devaraj} K.,   {Church} S.~E.,  2016, \mn@doi [\apj] {10.3847/0004-637X/817/2/169}, \href {https://ui.adsabs.harvard.edu/abs/2016ApJ...817..169L} {817, 169}

\bibitem[\protect\citeauthoryear{{Liu}, {Sun}, {Chang}, {Furlanetto}  \& {Bradford}}{{Liu} et~al.}{2024}]{Liu_2024}
{Liu} L.-J.,  {Sun} G.,  {Chang} T.-C.,  {Furlanetto} S.~R.,   {Bradford} C.~M.,  2024, \mn@doi [\apj] {10.3847/1538-4357/ad73d5}, \href {https://ui.adsabs.harvard.edu/abs/2024ApJ...974..175L} {974, 175}

\bibitem[\protect\citeauthoryear{{Madau} \& {Dickinson}}{{Madau} \& {Dickinson}}{2014}]{Madau_2014}
{Madau} P.,  {Dickinson} M.,  2014, \mn@doi [\araa] {10.1146/annurev-astro-081811-125615}, \href {https://ui.adsabs.harvard.edu/abs/2014ARA&A..52..415M} {52, 415}

\bibitem[\protect\citeauthoryear{{Mao}, {Zentner}  \& {Wechsler}}{{Mao} et~al.}{2018}]{Mao_2018}
{Mao} Y.-Y.,  {Zentner} A.~R.,   {Wechsler} R.~H.,  2018, \mn@doi [\mnras] {10.1093/mnras/stx3111}, \href {https://ui.adsabs.harvard.edu/abs/2018MNRAS.474.5143M} {474, 5143}

\bibitem[\protect\citeauthoryear{{Marinacci} et~al.,}{{Marinacci} et~al.}{2018}]{IllustrisTNG_Marinacci_2018}
{Marinacci} F.,  et~al., 2018, \mn@doi [\mnras] {10.1093/mnras/sty2206}, \href {https://ui.adsabs.harvard.edu/abs/2018MNRAS.480.5113M} {480, 5113}

\bibitem[\protect\citeauthoryear{{Matthee} \& {Schaye}}{{Matthee} \& {Schaye}}{2019}]{Matthee_2019}
{Matthee} J.,  {Schaye} J.,  2019, \mn@doi [\mnras] {10.1093/mnras/stz030}, \href {https://ui.adsabs.harvard.edu/abs/2019MNRAS.484..915M} {484, 915}

\bibitem[\protect\citeauthoryear{{Matthee}, {Schaye}, {Crain}, {Schaller}, {Bower}  \& {Theuns}}{{Matthee} et~al.}{2017}]{Matthee17}
{Matthee} J.,  {Schaye} J.,  {Crain} R.~A.,  {Schaller} M.,  {Bower} R.,   {Theuns} T.,  2017, \mn@doi [\mnras] {10.1093/mnras/stw2884}, \href {https://ui.adsabs.harvard.edu/abs/2017MNRAS.465.2381M} {465, 2381}

\bibitem[\protect\citeauthoryear{{Montero-Dorta} \& {Rodriguez}}{{Montero-Dorta} \& {Rodriguez}}{2024}]{Montero_2024}
{Montero-Dorta} A.~D.,  {Rodriguez} F.,  2024, \mn@doi [\mnras] {10.1093/mnras/stae796}, \href {https://ui.adsabs.harvard.edu/abs/2024MNRAS.531..290M} {531, 290}

\bibitem[\protect\citeauthoryear{{Montero-Dorta} et~al.,}{{Montero-Dorta} et~al.}{2020}]{Montero_Dorta_2020}
{Montero-Dorta} A.~D.,  et~al., 2020, \mn@doi [\mnras] {10.1093/mnras/staa1624}, \href {https://ui.adsabs.harvard.edu/abs/2020MNRAS.496.1182M} {496, 1182}

\bibitem[\protect\citeauthoryear{{Murmu} et~al.,}{{Murmu} et~al.}{2023}]{Murmu_2023}
{Murmu} C.~S.,  et~al., 2023, \mn@doi [\mnras] {10.1093/mnras/stac3304}, \href {https://ui.adsabs.harvard.edu/abs/2023MNRAS.518.3074M} {518, 3074}

\bibitem[\protect\citeauthoryear{{Naiman} et~al.,}{{Naiman} et~al.}{2018}]{IllustrisTNG_Naiman_2018}
{Naiman} J.~P.,  et~al., 2018, \mn@doi [\mnras] {10.1093/mnras/sty618}, \href {https://ui.adsabs.harvard.edu/abs/2018MNRAS.477.1206N} {477, 1206}

\bibitem[\protect\citeauthoryear{{Nelson} et~al.,}{{Nelson} et~al.}{2018}]{Illustris_TNG_Nelson_2018}
{Nelson} D.,  et~al., 2018, \mn@doi [\mnras] {10.1093/mnras/stx3040}, \href {https://ui.adsabs.harvard.edu/abs/2018MNRAS.475..624N} {475, 624}

\bibitem[\protect\citeauthoryear{{Nelson} et~al.,}{{Nelson} et~al.}{2019}]{IllustrisTNG_release}
{Nelson} D.,  et~al., 2019, \mn@doi [Computational Astrophysics and Cosmology] {10.1186/s40668-019-0028-x}, \href {https://ui.adsabs.harvard.edu/abs/2019ComAC...6....2N} {6, 2}

\bibitem[\protect\citeauthoryear{{Peebles}}{{Peebles}}{1980}]{Peebles80}
{Peebles} P.~J.~E.,  1980, {The large-scale structure of the universe}.
Princeton Univ. Press, Princeton, NJ

\bibitem[\protect\citeauthoryear{{Pillepich} et~al.,}{{Pillepich} et~al.}{2018a}]{Pillepich_2018}
{Pillepich} A.,  et~al., 2018a, \mn@doi [\mnras] {10.1093/mnras/stx2656}, \href {https://ui.adsabs.harvard.edu/abs/2018MNRAS.473.4077P} {473, 4077}

\bibitem[\protect\citeauthoryear{{Pillepich} et~al.,}{{Pillepich} et~al.}{2018b}]{IllustrisTNG_Pillepich_2018}
{Pillepich} A.,  et~al., 2018b, \mn@doi [\mnras] {10.1093/mnras/stx3112}, \href {https://ui.adsabs.harvard.edu/abs/2018MNRAS.475..648P} {475, 648}

\bibitem[\protect\citeauthoryear{{Planck Collaboration} et~al.,}{{Planck Collaboration} et~al.}{2016}]{Planck15}
{Planck Collaboration} et~al., 2016, \mn@doi [\aap] {10.1051/0004-6361/201525830}, \href {https://ui.adsabs.harvard.edu/abs/2016A&A...594A..13P} {594, A13}

\bibitem[\protect\citeauthoryear{{Reyes-Peraza}, {Avila}, {Gonzalez-Perez}, {Lopez-Cano}, {Knebe}, {Ramakrishnan}  \& {Yepes}}{{Reyes-Peraza} et~al.}{2024}]{Reyes-Peraza_2024}
{Reyes-Peraza} G.,  {Avila} S.,  {Gonzalez-Perez} V.,  {Lopez-Cano} D.,  {Knebe} A.,  {Ramakrishnan} S.,   {Yepes} G.,  2024, \mn@doi [\mnras] {10.1093/mnras/stae623}, \href {https://ui.adsabs.harvard.edu/abs/2024MNRAS.529.3877R} {529, 3877}

\bibitem[\protect\citeauthoryear{{Salcedo} et~al.,}{{Salcedo} et~al.}{2022}]{Salcedo2022}
{Salcedo} A.~N.,  et~al., 2022, \mn@doi [Science China Physics, Mechanics, and Astronomy] {10.1007/s11433-022-1955-7}, \href {https://ui.adsabs.harvard.edu/abs/2022SCPMA..6509811S} {65, 109811}

\bibitem[\protect\citeauthoryear{{Sato-Polito}, {Kokron}  \& {Bernal}}{{Sato-Polito} et~al.}{2023}]{Sato-Polito_2023}
{Sato-Polito} G.,  {Kokron} N.,   {Bernal} J.~L.,  2023, \mn@doi [\mnras] {10.1093/mnras/stad2498}, \href {https://ui.adsabs.harvard.edu/abs/2023MNRAS.526.5883S} {526, 5883}

\bibitem[\protect\citeauthoryear{{Schaan} \& {White}}{{Schaan} \& {White}}{2021a}]{Schaan+21-galaxy}
{Schaan} E.,  {White} M.,  2021a, \mn@doi [\jcap] {10.1088/1475-7516/2021/05/067}, \href {https://ui.adsabs.harvard.edu/abs/2021JCAP...05..067S} {2021, 067}

\bibitem[\protect\citeauthoryear{{Schaan} \& {White}}{{Schaan} \& {White}}{2021b}]{Schaan+21-multi}
{Schaan} E.,  {White} M.,  2021b, \mn@doi [\jcap] {10.1088/1475-7516/2021/05/068}, \href {https://ui.adsabs.harvard.edu/abs/2021JCAP...05..068S} {2021, 068}

\bibitem[\protect\citeauthoryear{{Schaye} et~al.,}{{Schaye} et~al.}{2015}]{Schaye15}
{Schaye} J.,  et~al., 2015, \mn@doi [\mnras] {10.1093/mnras/stu2058}, \href {https://ui.adsabs.harvard.edu/abs/2015MNRAS.446..521S} {446, 521}

\bibitem[\protect\citeauthoryear{{Scholtz} et~al.,}{{Scholtz} et~al.}{2018}]{Scholtz18}
{Scholtz} J.,  et~al., 2018, \mn@doi [\mnras] {10.1093/mnras/stx3177}, \href {https://ui.adsabs.harvard.edu/abs/2018MNRAS.475.1288S} {475, 1288}

\bibitem[\protect\citeauthoryear{{Sharma} \& {Theuns}}{{Sharma} \& {Theuns}}{2020}]{Sharma20}
{Sharma} M.,  {Theuns} T.,  2020, \mn@doi [\mnras] {10.1093/mnras/stz2909}, \href {https://ui.adsabs.harvard.edu/abs/2020MNRAS.492.2418S} {492, 2418}

\bibitem[\protect\citeauthoryear{{Silva}, {Zaroubi}, {Kooistra}  \& {Cooray}}{{Silva} et~al.}{2018}]{Silva_2018}
{Silva} B.~M.,  {Zaroubi} S.,  {Kooistra} R.,   {Cooray} A.,  2018, \mn@doi [\mnras] {10.1093/mnras/stx3265}, \href {https://ui.adsabs.harvard.edu/abs/2018MNRAS.475.1587S} {475, 1587}

\bibitem[\protect\citeauthoryear{{Sobral}, {Smail}, {Best}, {Geach}, {Matsuda}, {Stott}, {Cirasuolo}  \& {Kurk}}{{Sobral} et~al.}{2013}]{Sobral+13}
{Sobral} D.,  {Smail} I.,  {Best} P.~N.,  {Geach} J.~E.,  {Matsuda} Y.,  {Stott} J.~P.,  {Cirasuolo} M.,   {Kurk} J.,  2013, \mn@doi [\mnras] {10.1093/mnras/sts096}, \href {https://ui.adsabs.harvard.edu/abs/2013MNRAS.428.1128S} {428, 1128}

\bibitem[\protect\citeauthoryear{{Springel}}{{Springel}}{2010}]{Springel_2010}
{Springel} V.,  2010, \mn@doi [\mnras] {10.1111/j.1365-2966.2009.15715.x}, \href {https://ui.adsabs.harvard.edu/abs/2010MNRAS.401..791S} {401, 791}

\bibitem[\protect\citeauthoryear{{Springel}, {White}, {Tormen}  \& {Kauffmann}}{{Springel} et~al.}{2001}]{Subfind}
{Springel} V.,  {White} S. D.~M.,  {Tormen} G.,   {Kauffmann} G.,  2001, \mn@doi [\mnras] {10.1046/j.1365-8711.2001.04912.x}, \href {https://ui.adsabs.harvard.edu/abs/2001MNRAS.328..726S} {328, 726}

\bibitem[\protect\citeauthoryear{{Springel}, {Frenk}  \& {White}}{{Springel} et~al.}{2006}]{LSS_2006}
{Springel} V.,  {Frenk} C.~S.,   {White} S. D.~M.,  2006, \mn@doi [\nat] {10.1038/nature04805}, \href {https://ui.adsabs.harvard.edu/abs/2006Natur.440.1137S} {440, 1137}

\bibitem[\protect\citeauthoryear{{Springel} et~al.,}{{Springel} et~al.}{2018}]{IllustrisTNG_Springel_2018}
{Springel} V.,  et~al., 2018, \mn@doi [\mnras] {10.1093/mnras/stx3304}, \href {https://ui.adsabs.harvard.edu/abs/2018MNRAS.475..676S} {475, 676}

\bibitem[\protect\citeauthoryear{{Sun}, {Hensley}, {Chang}, {Dor{\'e}}  \& {Serra}}{{Sun} et~al.}{2019}]{Sun_2019}
{Sun} G.,  {Hensley} B.~S.,  {Chang} T.-C.,  {Dor{\'e}} O.,   {Serra} P.,  2019, \mn@doi [\apj] {10.3847/1538-4357/ab55df}, \href {https://ui.adsabs.harvard.edu/abs/2019ApJ...887..142S} {887, 142}

\bibitem[\protect\citeauthoryear{{Tinker}, {Robertson}, {Kravtsov}, {Klypin}, {Warren}, {Yepes}  \& {Gottl{\"o}ber}}{{Tinker} et~al.}{2010}]{Tinker10}
{Tinker} J.~L.,  {Robertson} B.~E.,  {Kravtsov} A.~V.,  {Klypin} A.,  {Warren} M.~S.,  {Yepes} G.,   {Gottl{\"o}ber} S.,  2010, \mn@doi [\apj] {10.1088/0004-637X/724/2/878}, \href {https://ui.adsabs.harvard.edu/abs/2010ApJ...724..878T} {724, 878}

\bibitem[\protect\citeauthoryear{{Tucci}, {Montero-Dorta}, {Abramo}, {Sato-Polito}  \& {Artale}}{{Tucci} et~al.}{2021}]{Tucci_2021}
{Tucci} B.,  {Montero-Dorta} A.~D.,  {Abramo} L.~R.,  {Sato-Polito} G.,   {Artale} M.~C.,  2021, \mn@doi [\mnras] {10.1093/mnras/staa3319}, \href {https://ui.adsabs.harvard.edu/abs/2021MNRAS.500.2777T} {500, 2777}

\bibitem[\protect\citeauthoryear{{Vakili} \& {Hahn}}{{Vakili} \& {Hahn}}{2019}]{Vakili2019}
{Vakili} M.,  {Hahn} C.,  2019, \mn@doi [\apj] {10.3847/1538-4357/aaf1a1}, \href {https://ui.adsabs.harvard.edu/abs/2019ApJ...872..115V} {872, 115}

\bibitem[\protect\citeauthoryear{{Vale} \& {Ostriker}}{{Vale} \& {Ostriker}}{2004}]{Vale04}
{Vale} A.,  {Ostriker} J.~P.,  2004, \mn@doi [\mnras] {10.1111/j.1365-2966.2004.08059.x}, \href {https://ui.adsabs.harvard.edu/abs/2004MNRAS.353..189V} {353, 189}

\bibitem[\protect\citeauthoryear{{Vogelsberger}, {Marinacci}, {Torrey}  \& {Puchwein}}{{Vogelsberger} et~al.}{2020}]{Vogelsberger20}
{Vogelsberger} M.,  {Marinacci} F.,  {Torrey} P.,   {Puchwein} E.,  2020, \mn@doi [Nature Reviews Physics] {10.1038/s42254-019-0127-2}, \href {https://ui.adsabs.harvard.edu/abs/2020NatRP...2...42V} {2, 42}

\bibitem[\protect\citeauthoryear{{Walsh} \& {Tinker}}{{Walsh} \& {Tinker}}{2019}]{Walsh2019}
{Walsh} K.,  {Tinker} J.,  2019, \mn@doi [\mnras] {10.1093/mnras/stz1351}, \href {https://ui.adsabs.harvard.edu/abs/2019MNRAS.488..470W} {488, 470}

\bibitem[\protect\citeauthoryear{{Wechsler} \& {Tinker}}{{Wechsler} \& {Tinker}}{2018}]{Wechsler_2018}
{Wechsler} R.~H.,  {Tinker} J.~L.,  2018, \mn@doi [\araa] {10.1146/annurev-astro-081817-051756}, \href {https://ui.adsabs.harvard.edu/abs/2018ARA&A..56..435W} {56, 435}

\bibitem[\protect\citeauthoryear{{Wechsler}, {Zentner}, {Bullock}, {Kravtsov}  \& {Allgood}}{{Wechsler} et~al.}{2006}]{Wechsler_2006}
{Wechsler} R.~H.,  {Zentner} A.~R.,  {Bullock} J.~S.,  {Kravtsov} A.~V.,   {Allgood} B.,  2006, \mn@doi [\apj] {10.1086/507120}, \href {https://ui.adsabs.harvard.edu/abs/2006ApJ...652...71W} {652, 71}

\bibitem[\protect\citeauthoryear{{Weinberger} et~al.,}{{Weinberger} et~al.}{2017}]{Weinberger_2017}
{Weinberger} R.,  et~al., 2017, \mn@doi [\mnras] {10.1093/mnras/stw2944}, \href {https://ui.adsabs.harvard.edu/abs/2017MNRAS.465.3291W} {465, 3291}

\bibitem[\protect\citeauthoryear{{Weinmann}, {van den Bosch}, {Yang}  \& {Mo}}{{Weinmann} et~al.}{2006}]{Weinmann06}
{Weinmann} S.~M.,  {van den Bosch} F.~C.,  {Yang} X.,   {Mo} H.~J.,  2006, \mn@doi [\mnras] {10.1111/j.1365-2966.2005.09865.x}, \href {https://ui.adsabs.harvard.edu/abs/2006MNRAS.366....2W} {366, 2}

\bibitem[\protect\citeauthoryear{{Xu} \& {Zheng}}{{Xu} \& {Zheng}}{2020}]{Xu_2020}
{Xu} X.,  {Zheng} Z.,  2020, \mn@doi [\mnras] {10.1093/mnras/staa009}, \href {https://ui.adsabs.harvard.edu/abs/2020MNRAS.492.2739X} {492, 2739}

\bibitem[\protect\citeauthoryear{{Xu}, {Zehavi}  \& {Contreras}}{{Xu} et~al.}{2021}]{Xu_2021}
{Xu} X.,  {Zehavi} I.,   {Contreras} S.,  2021, \mn@doi [\mnras] {10.1093/mnras/stab100}, \href {https://ui.adsabs.harvard.edu/abs/2021MNRAS.502.3242X} {502, 3242}

\bibitem[\protect\citeauthoryear{{Yuan}, {Hadzhiyska}, {Bose}, {Eisenstein}  \& {Guo}}{{Yuan} et~al.}{2021}]{Yuan2021}
{Yuan} S.,  {Hadzhiyska} B.,  {Bose} S.,  {Eisenstein} D.~J.,   {Guo} H.,  2021, \mn@doi [\mnras] {10.1093/mnras/stab235}, \href {https://ui.adsabs.harvard.edu/abs/2021MNRAS.502.3582Y} {502, 3582}

\bibitem[\protect\citeauthoryear{{Yuan}, {Hadzhiyska}, {Bose}  \& {Eisenstein}}{{Yuan} et~al.}{2022}]{Yuan_2022}
{Yuan} S.,  {Hadzhiyska} B.,  {Bose} S.,   {Eisenstein} D.~J.,  2022, \mn@doi [\mnras] {10.1093/mnras/stac830}, \href {https://ui.adsabs.harvard.edu/abs/2022MNRAS.512.5793Y} {512, 5793}

\bibitem[\protect\citeauthoryear{{Yuan} et~al.,}{{Yuan} et~al.}{2025}]{Yuan_2025}
{Yuan} S.,  et~al., 2025, \mn@doi [\mnras] {10.1093/mnras/staf368}, \href {https://ui.adsabs.harvard.edu/abs/2025MNRAS.538.1216Y} {538, 1216}

\bibitem[\protect\citeauthoryear{{Zehavi}, {Contreras}, {Padilla}, {Smith}, {Baugh}  \& {Norberg}}{{Zehavi} et~al.}{2018}]{Zehavi_2018}
{Zehavi} I.,  {Contreras} S.,  {Padilla} N.,  {Smith} N.~J.,  {Baugh} C.~M.,   {Norberg} P.,  2018, \mn@doi [\apj] {10.3847/1538-4357/aaa54a}, \href {https://ui.adsabs.harvard.edu/abs/2018ApJ...853...84Z} {853, 84}

\bibitem[\protect\citeauthoryear{{Zentner}, {Hearin}  \& {van den Bosch}}{{Zentner} et~al.}{2014}]{Zentner_2014}
{Zentner} A.~R.,  {Hearin} A.~P.,   {van den Bosch} F.~C.,  2014, \mn@doi [\mnras] {10.1093/mnras/stu1383}, \href {https://ui.adsabs.harvard.edu/abs/2014MNRAS.443.3044Z} {443, 3044}

\bibitem[\protect\citeauthoryear{{Zhai} et~al.,}{{Zhai} et~al.}{2023}]{Zhai2023}
{Zhai} Z.,  et~al., 2023, \mn@doi [\apj] {10.3847/1538-4357/acc65b}, \href {https://ui.adsabs.harvard.edu/abs/2023ApJ...948...99Z} {948, 99}

\bibitem[\protect\citeauthoryear{{van den Bosch}, {Tormen}  \& {Giocoli}}{{van den Bosch} et~al.}{2005}]{Bosch_2005}
{van den Bosch} F.~C.,  {Tormen} G.,   {Giocoli} C.,  2005, \mn@doi [\mnras] {10.1111/j.1365-2966.2005.08964.x}, \href {https://ui.adsabs.harvard.edu/abs/2005MNRAS.359.1029V} {359, 1029}

\makeatother
\end{thebibliography}

\appendix

\crefalias{section}{appendix}

\section{Modelling the luminosity}
\subsection{IllustrisTNG}\label{app:tng}

Most of the numerical analysis in this paper uses the TNG300-1 hydrodynamical simulation of the \illus{} project \citep{Illustris_TNG_Nelson_2018,IllustrisTNG_Pillepich_2018,IllustrisTNG_Springel_2018,IllustrisTNG_Marinacci_2018,IllustrisTNG_Naiman_2018}. 
The linear extent of the cubic simulation volume is 205~\lenunit, with dark matter particle mass of $m_{\mathrm{DM}} \sim 3.98 \times$ \mhalounit{7} and gas particle mass of $m_{\mathrm{gas}} \sim 7.44 \times$ \mhalounit{6}.
The \illus{} simulations use the \arepo{} moving-mesh code described by \citet{Springel_2010} to solve for gravity and magneto-hydrodynamics. The \illus{} simulations include detailed subgrid modelling of the physics of galaxy formation, including radiative and collisional gas cooling, a model for star formation, supernova feedback and black hole seeding, accretion, merging and the feedback associated with accretion \citep{Weinberger_2017, Pillepich_2018}. 
Compared to its predecessor, Illustris, the stellar and {\sc agn} feedback model in \illus{} is more effective at suppressing star formation. 
The \illus\ simulations assume a cosmology consistent with the \emph{Planck 2015} values \citep{Planck15}. We analyse the simulation at redshift $z = 1.5$.

Dark matter haloes in \illus{} are identified with the usual friends-of-friends ({\sc fof}) algorithm with standard linking length in units of the mean dark matter particle separation of $b=0.2$ \citep{FOF}, while {\sc subfind} identifies substructures \citep{Subfind}. Each {\sc fof} halo may contain a number of subhaloes, each of which may contain zero or one galaxy containing stars, gas and black holes. 

The halo mass we quote throughout is the virial mass, \mvir\ (given by the \texttt{Group\_M\_TopHat200} field in the \illus{} database\footnote{\href{https://www.illustris-project.org/data/}{https://www.illustris-project.org/data}}; \citealt{IllustrisTNG_release}), such that haloes of the same \mvir\ have by contruction the same virial radius, \rvir. The \sfr{} quoted is the sum of the \sfr{}s of all gas cells associated with subhaloes (galaxy \sfr{}, \texttt{SubhaloSFR}) or haloes (halo \sfr{}, \texttt{GroupSFR}). We use these values directly without applying aperture corrections and include all subhaloes regardless of the value of the \texttt{SubhaloFlag} (introduced to signal that a structure may be associated with substructure in a given galaxy rather than being an independent structure). We do so because this most closely mimics what a \lim{} observation would associate with the \sfr{} in a voxel (\citepaperone).

The subhalo at the bottom of the gravitational potential of the {\sc fof} halo is labelled as the central subhalo (identified using the \texttt{GroupFirstSub} field), while all remaining subhaloes are classified as satellite subhaloes, regardless of whether or not they lie within the virial radius of the main halo. Central subhaloes host central galaxies and satellite subhaloes host satellite galaxies. As a halo grows, the identification of which galaxy is the central galaxy may occasionally flip between a pair of galaxies, especially during mergers, but such misidentifications do not affect the statistical properties of the samples discussed here.

Some of the results shown in this paper may be dependent on the galaxy formation model. To test 
how general our conclusions are, we also compare to results of the \eagle{} simulations \citep{Schaye15, Crain15} in \cref{app:sfr-halo_mass_eagle_vs_tng}.

\subsection{The SFR-luminosity relation}\label{app:sfr-luminosity}

The line luminosity of many emission lines is closely related to the galaxy's \sfr{}. For example, the \ha\ emission line originates from recombining gas in \ion{H}{II} regions where the gas is photo-ionised by
short-lived massive stars. The ionising flux of these stars, and hence the \ha\ flux of the \ion{H}{II} regions, are therefore closely related to the total amount of massive stars formed over the past several \unit{Myr}'s and hence the recent \sfr{} of these regions. 

The relation between this time-averaged \sfr{} and the instantaneous emitted \ha\ luminosity is given by
\citet{Kennicutt98} as 
\begin{align}\label{eq:L-sfr}
    L_{\mathrm{H\alpha}}\, (\unit{erg\ s^{-1}}) &= 2.0 \times 10^{41} \mathrm{SFR}\, (\unit{M_\odot\ yr^{-1}})\,,
\end{align}
after applying a $\sim 0.63$ factor for converting from the Salpeter stellar initial mass function ({\sc imf}) used by \cite{Kennicutt98} to a Chabrier \citep{Chabrier2003} {\sc imf}, following \citet{Madau_2014}.
The \ha\ luminosity will be attenuated by dust, which can be modelled for \ha\ using
\begin{align}\label{eq:dust}
    L_{\mathrm{H\alpha}}^{\text{dust}} &= 10^{-A_{\mathrm{H\alpha}}/2.5} L_{\mathrm{H\alpha}}^{\text{no dust}},
\end{align}
where $A_{\mathrm{H\alpha}}$ = 1 \citep{Garn10,Sobral+13}.

We use these same relations to model \ha{} emission for the simulated galaxies, but the \sfr{} we use from the simulation corresponds to the instantaneous rate. The timescale of star formation measured by \ha\ is of the order $10^7$~yr, making it relatively instantaneous. However, other emission lines tend to correspond to longer timescales of star formation (see \citealt{Donnari_2019_erratum} for the effect of averaging over different timescales). Using the instantaneous rate rather than a time-averaged rate may affect the level of scatter, but should not affect the mean relation between \ha{} flux and \sfr{}. 

The luminosity of other emission lines, such as optical and infrared oxygen lines, nitrogen lines, [\ion{C}{II}] and CO roto-vibrational lines, are also closely related to the \sfr{}, but may also depend on other parameters of the interstellar medium of the galaxy - for example, its metallicity 
\citep[e.g.][]{Fonseca_2017}. These effects should be accounted for in \lim{} mocks.

\section{Comparing EAGLE SFR-halo mass relation to TNG}\label{app:sfr-halo_mass_eagle_vs_tng}

\begin{figure*}
\centering
\begin{subfigure}[t]{.5\textwidth}
    \centering
    \includegraphics[width=\linewidth]{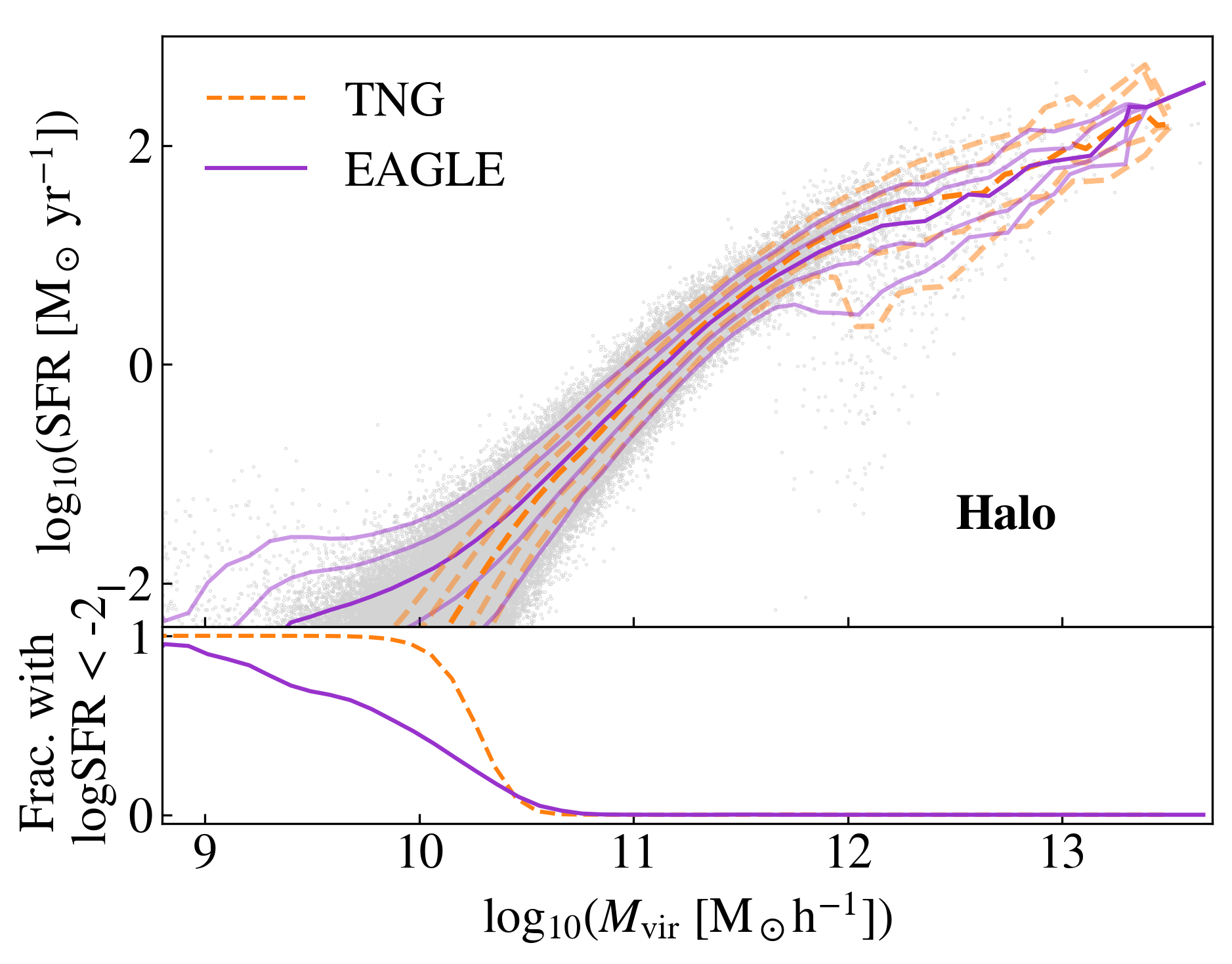}
\end{subfigure}%
\begin{subfigure}[t]{.5\textwidth}
    \centering
    \includegraphics[width=\linewidth]{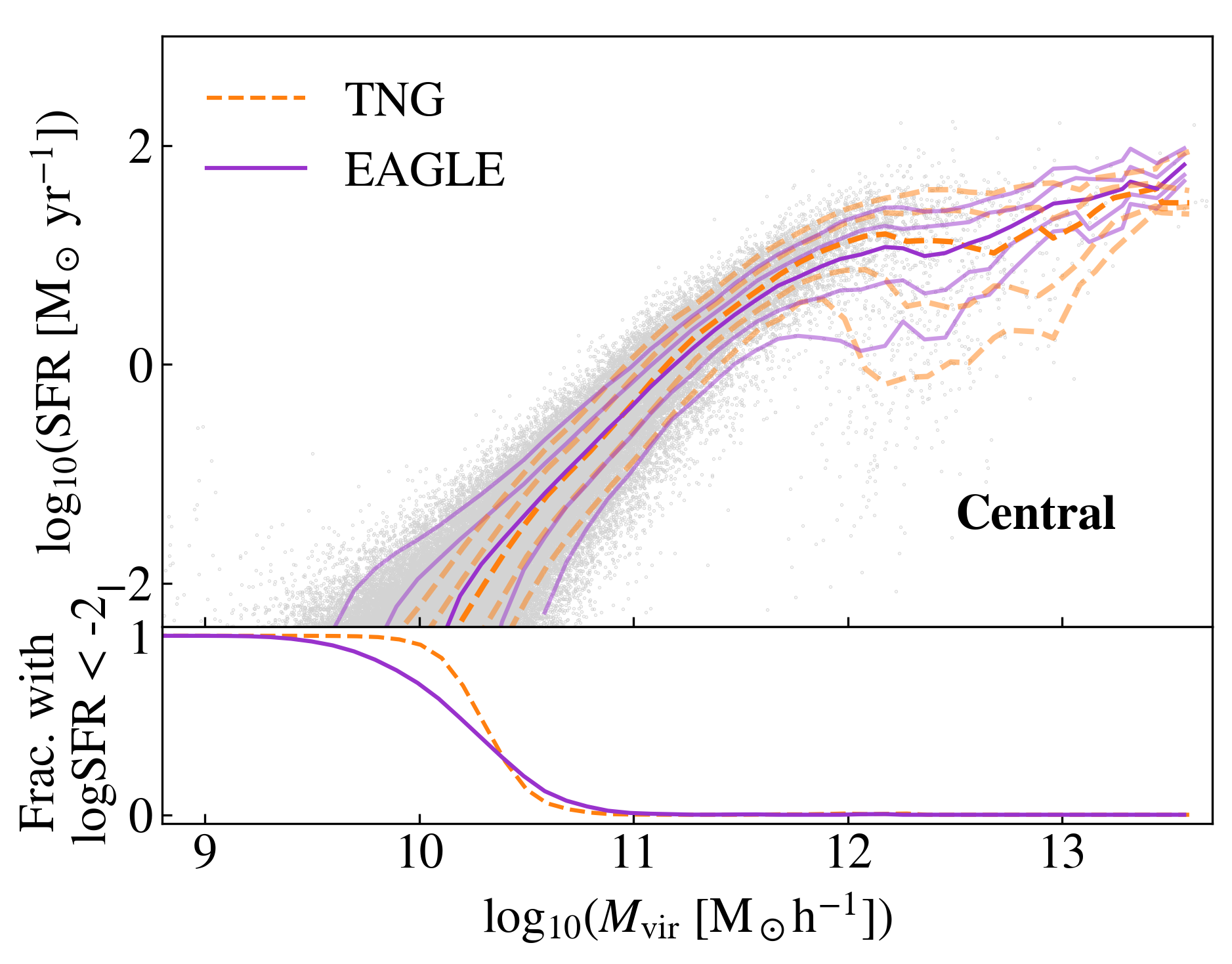}
\end{subfigure}
\caption{Comparing TNG100-1 and \eagle{} \sfr{}-virial mass relation at $z\sim1.5$.
\textbf{\textit{Left Panel}}: The \emph{grey dots} represent individual halo \sfr{}s from TNG100-1.
The \emph{dashed orange lines} represent the 10th, 25th, 50th, 75th and 90th percentiles in \sfr{} from bottom to top.
The \emph{solid purple lines} represent the equivalent for \eagle{}. The bottom panel shows the fraction of haloes with $\log\mathrm{SFR} < -2$ for TNG100-1 (\emph{orange}) and \eagle{} (\emph{purple}).
\textbf{\textit{Right Panel}}: Same as left panel, but for central galaxies rather than haloes. 
\textbf{\textit{Summary}}: At halo mass around \mhalounit{12}, TNG100-1 has more quenched central galaxies than in \eagle{}, but the total halo \sfr{}-halo mass relation is very similar.
}
\label{fig:sfr-mass-eagle}
\end{figure*}

\Cref{fig:sfr-mass-eagle} compares the \sfr{}-halo mass relation for the TNG100-1 (75 \lenunit{}) simulation and \eagle{} Ref-L0100N1504 (67.77 \lenunit{}) simulation. The left panel shows the results for the halo \sfr{} and the right panel shows that for the central galaxy \sfr{}. 
\agn{} feedback is stronger in \illus{} than in \eagle{}, causing a greater decrease in the \sfr{} of central galaxies at $\log M_h \sim 12$ (right panel). However, when we sum up all the \sfr{} in the haloes, including that of satellite galaxies, then the \sfr{}-halo mass relations are very similar between the two simulations, despite their different galaxy formation models.

\section{The average bias is dependent only on the mean luminosity-halo mass relation} \label{app:scatter_maths}

In this appendix, we demonstrate mathematically that the bias, and therefore two-halo term, averaged over realisations of scatter, is dependent only on the mean luminosity-halo mass relation.

Let $\mu_b^j = \sum_q b_{j,q} /N_j$ be the unweighted mean of the biases of haloes of a given mass.
For the case where all haloes of a given mass have luminosity $\bar{L}_j$, the weighted mean bias for a given mass is equivalent to the unweighted mean since all the weights are equal.

Now assume that the luminosities of haloes, $L_j$, are  randomly distributed with mean $\mu_L^j = \sum_q L_{j,q}/N_j$. 
We pull from this distribution $N_j$ times, where $N_j$ is the number of haloes of that mass. This gives us a set of haloes with luminosities in the set $\{L_{j,1},..., L_{j,q}, ..., L_{j,{N_j}}\}$. 
If we repeat this process multiple times, we may get different $\bar{b}(M_j)$ each time.
The weighted mean $\bar{b}(M_j)$ is now a random variable with some mean and variance. We will now show that the expected value of $\bar{b}(M_j)$ is equal to $\mu_b^j$ in general, making no assumptions about the shape or variance of the distribution of $L_j$.

We pull from this distribution $N_j$ times, where $N_j$ is the number of haloes of that mass. This gives us a set of haloes with luminosities in the set $\{L_{j,1},..., L_{j,q}, ..., L_{j,{N_j}}\}$. This set will have a mean bias $\bar{b}(M_j)$, given by \cref{eq:b-bar_M}. If we repeat this process again, we may get a different $\bar{b}(M_j)$. 

The expected value for the weighted mean bias for a given mass can be written as
\begin{align}
    E(\bar{b}(M_j)) = E\left(\frac{\sum_{q=1}^{N_j}{b_{j,q} L_{j,q}}}{\sum_{r=1}^{N_j}{L_{j,r}}}\right) = \sum_{q=1}^{N_j} E\left(b_{j,q} \frac{L_{j,q}}{\sum_{r=1}^{N_j}{L_{j,r}}}\right),
\end{align}
by linearity of expectation.
Since we assume $L_{j,q}$ to be drawn randomly and independently of $b_{j,q}$:
\begin{align}
    \sum_{q=1}^{N_j} E\left(b_{j,q} \frac{L_{j,q}}{\sum_{r=1}^{N_j}{L_{j,r}}}\right) = \sum_{q=1}^{N_j} E(b_{j,q}) E\left(\frac{L_{j,q}} {\sum_{r=1}^{N_j}{L_{j,r}}}\right).
\end{align}
The expectation $E(b_{j,q})$ is equivalent to the unweighted mean $\mu_b^j$. Since this is a constant, it is independent of $q$, and can be pulled out of the sum. Then, again by linearity of expectation, we have
\begin{align}
     \mu_b^j E\left(\frac{\sum_{q=1}^{N_j}L_{j,q}} {\sum_{r=1}^{N_j}{L_{j,r}}}\right) = \mu_b^j E(1) = \mu_b^j. 
\end{align}
Therefore, the expected value of $\bar{b}(M_j)$ is equal to the unweighted mean bias, 
\begin{align}
     E(\bar{b}(M_j)) = \sum_q b_{j,q} /N_j,
\end{align}
for any random distribution of $L_j$.

This same analysis can be extended to the mean bias of the whole sample. The contribution of $\bar{b}(M_j)$ to the whole sample depends on the total luminosity contributed by haloes of that mass $M_j$:
\begin{align}\label{eq:bias_whole_sample2}
    \bar{b} = \sum_j{\bar{b}(M_j) \cdot \frac{\sum_q{L_{j,q}}}{\sum_r(\sum_q{L^{M_r}_q})}}.
\end{align}
A similar procedure can be applied to compute the mean of the bias of all the haloes to arrive at
\begin{align}
    E(\bar{b}) = \mu_B,
\end{align}
where $\mu_B$ is the bias obtained when all haloes have luminosities $\bar{L}(M)$, where $\bar{L}(M)$ is a one-to-one relation with no scatter.
In other words, the mean weighted bias of all the haloes is independent of the variance of the luminosities, as long as the mean $\bar{L}(M)$ is preserved.

As derived in eq.13 of \citepaperone, we see that the shot noise is dependent on the variance of L:
\begin{align}
     P_{\mathrm{shot}} &=\frac{1}{\bar{n}}\left(\frac{{\rm Var}(L) }{\langle L \rangle ^2} + 1\right).\label{eq:shotnoise_var}
\end{align}
Here $\bar{n}$ is the number density of galaxies, which is independent of $L$.
If we apply some scatter then the variance of $L$ will change, therefore the shot noise is affected. If scatter is applied at each halo mass, the total variance will also increase. For the same mean, the larger the variance, the larger the shot noise.

\subsection{The two-halo term is independent of the shape of the scatter}\label{app:tng_gaussian}

\begin{figure*}
        \centering
        \begin{subfigure}[b]{0.49\textwidth}
            \centering
            \includegraphics[width=\textwidth]{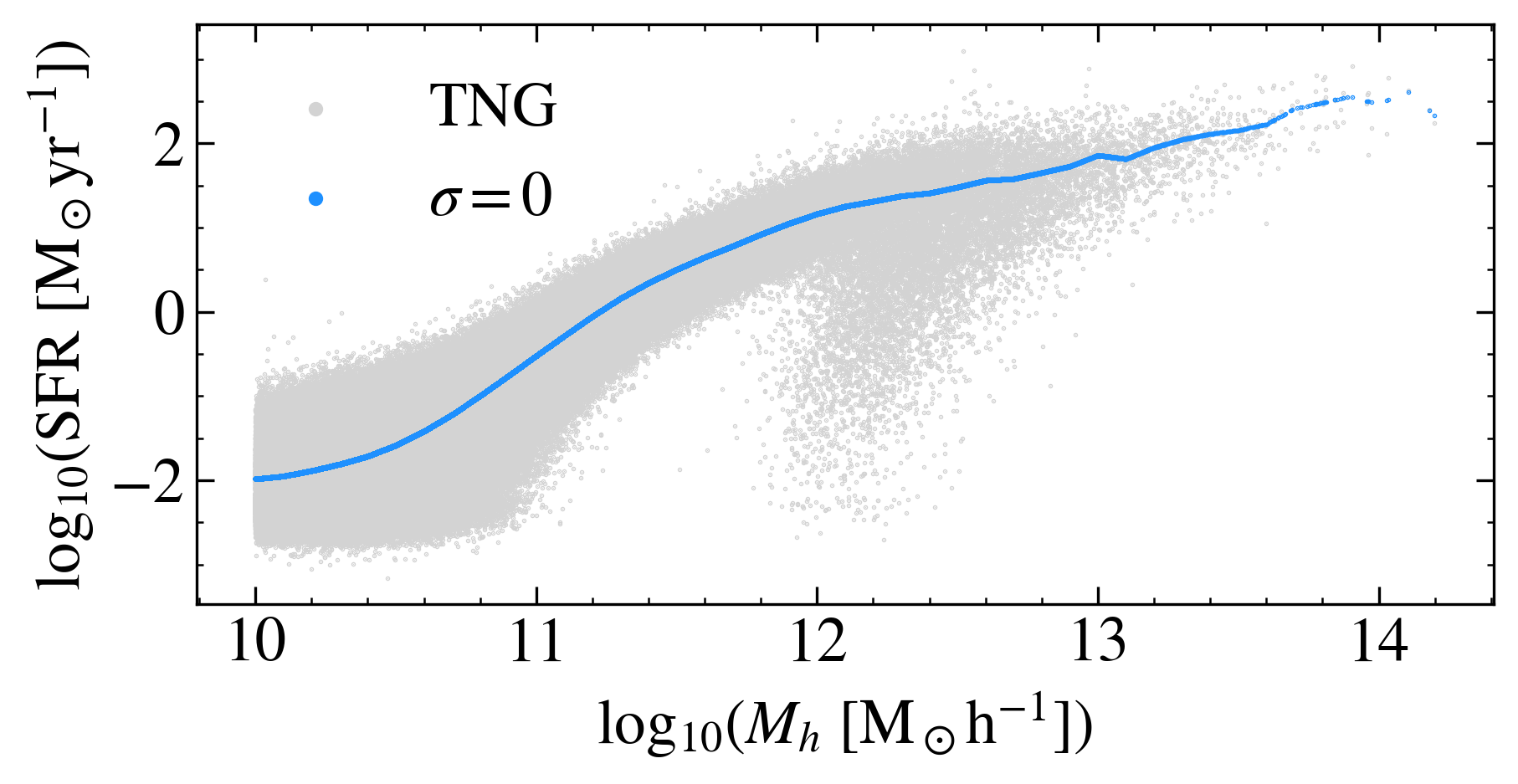}
        \end{subfigure}
        \hfill
        \begin{subfigure}[b]{0.49\textwidth}  
            \centering 
            \includegraphics[width=\textwidth]{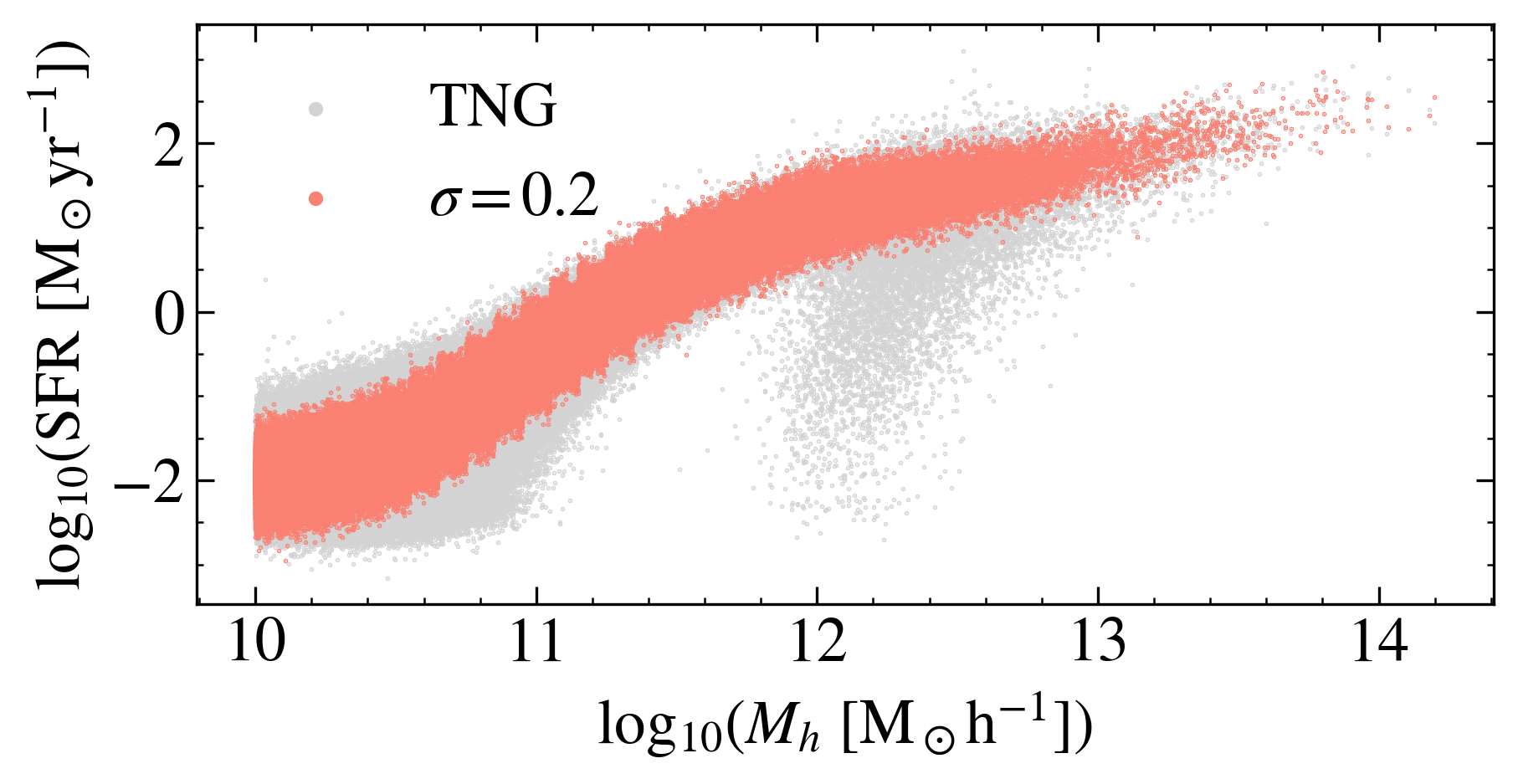}
        \end{subfigure}
        \vskip\baselineskip
        \begin{subfigure}[b]{0.49\textwidth}   
            \centering 
            \includegraphics[width=\textwidth]{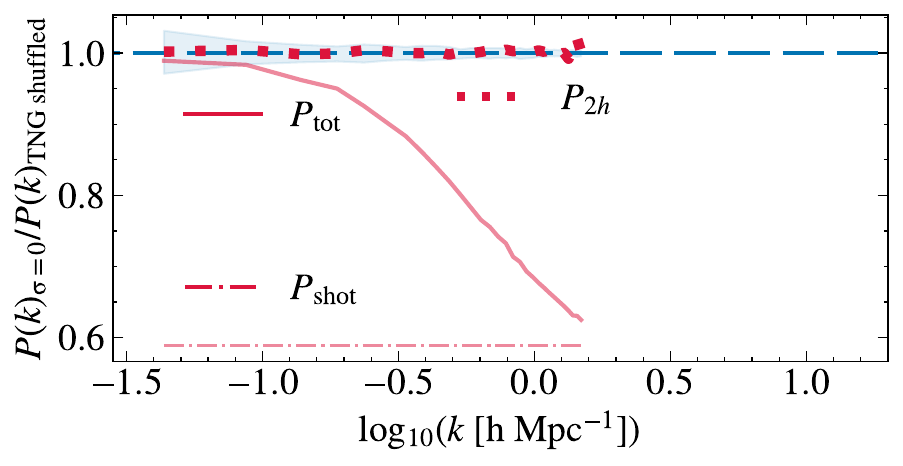}
        \end{subfigure}
        \hfill
        \begin{subfigure}[b]{0.49\textwidth}   
            \centering 
            \includegraphics[width=\textwidth]{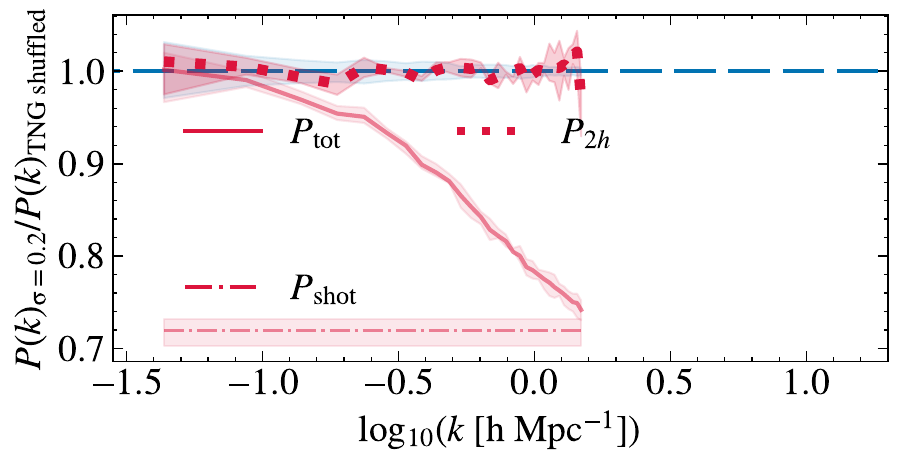}
        \end{subfigure}
        \caption
        {\textbf{\textit{Upper left panel}}: The \emph{solid blue line} shows the linear mean \sfr{}-halo mass relation for \tng{}. The \emph{grey dots} show \sfr{} against virial mass for \tng{} haloes.
        \textbf{\textit{Lower left panel}}: The \emph{red lines} show the ratio of the power spectra computed for the case where \sfr{}s are assigned to halos according to the linear mean \sfr{}-halo mass relation of \tng{} with no scatter, compared to the case where the \sfr{}s of \tng{} halos have been shuffled in bins of halo mass. The \emph{blue shaded region} shows the 25th-75th percentile of power spectra generated by shuffling with 100 different random seeds. 
        \textbf{\textit{Upper right panel}}: The \emph{red dots} show the \sfr{}s where a lognormal scatter with $\sigma=0.2$ dex is applied such that the linear mean is approximately the same as \tng{}. The \emph{grey dots} show \sfr{} against virial mass for \tng{} haloes. 
\textbf{\textit{Lower right panel}}: The \emph{red lines} show the ratio of the power spectra computed for the sample with  $\sigma=0.2$ dex lognormal scatter (mean over 5 realisations) relative to the shuffled \tng{} power spectrum.
\textbf{\textit{Summary}}: The mean two-halo term depends only on the linear mean luminosity-halo mass relation, regardless of the shape of the scatter (provided the scatter is random).} 
        \label{fig:tng_gaussian}
\end{figure*}

To check if the same holds for the case when we have non-Gaussian scatter, we use the probability distribution of \sfr{}s in \tng{}. 
To ensure that for a given halo mass, the \sfr{} is randomly distributed, we randomly shuffle the \sfr{}s amongst haloes within the same halo mass bin of  width d$\log M_h = 0.1$ dex, and we refer to this as TNG$_{\mathrm{shuffled}}$.

We first consider the case where \sfr{}s are assigned according to the linear mean \sfr{}-halo mass relation with no scatter and compare this to TNG$_{\mathrm{shuffled}}$ (upper left panel of \cref{fig:tng_gaussian}). The ratio of the power spectra are shown in the lower left panel of  \cref{fig:tng_gaussian}. The two-halo term is reproduced, showing that the mean two-halo term is dependent only on the linear mean luminosity-halo mass relation regardless of the shape of the scatter. 

Next, we consider the case where we apply a 0.2 dex lognormal scatter such that the linear mean \sfr{}-halo mass relation is preserved (upper right panel of \cref{fig:tng_gaussian}). Again, the average two-halo term is approximately the same as for TNG$_{\mathrm{shuffled}}$ (lower left panel of \cref{fig:tng_gaussian}). 
However, adding scatter introduces random variation around the mean. Here, we have only computed the power spectrum for 5 realisations -- taking the average of more realisations should cause the mean power spectra to converge to each other.

If we are not concerned with reproducing the shot noise component or the luminosity function, then using the linear mean relation without scatter (left panels) could reduce the uncertainty due to scatter. One method that is adopted is to apply some scatter and then rank order to match the luminosity function. However, the fact that this changes the linear mean relation, and therefore the two-halo term, should be taken into account.

\section{Dependence of SFR on concentration and satellite mass}
\label{app:sfr-c-msat-corr}

\begin{figure*}
\centering
\begin{subfigure}[b]{.49\textwidth}
    \centering
    \includegraphics[width=.99\textwidth]{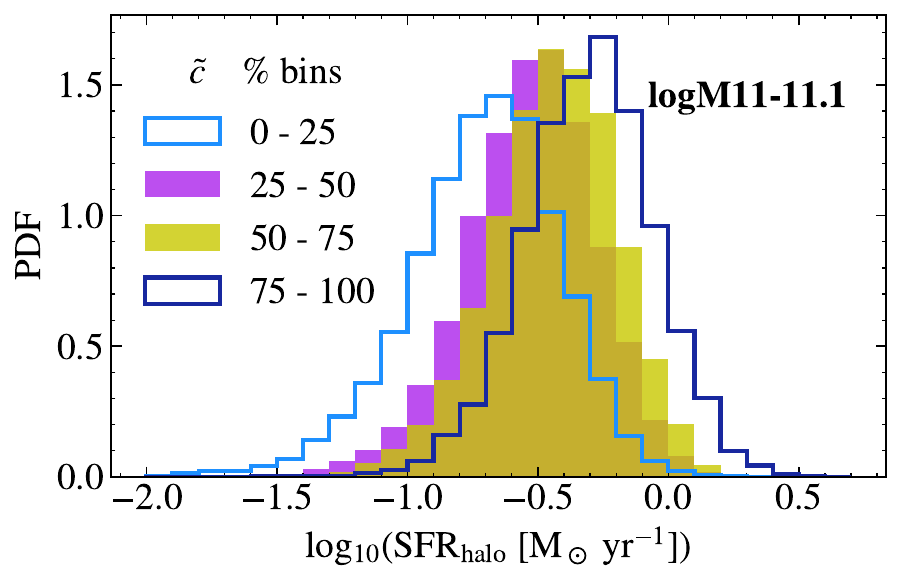}
\end{subfigure}%
\begin{subfigure}[b]{.49\textwidth}
    \centering
    \includegraphics[width=.99\textwidth]{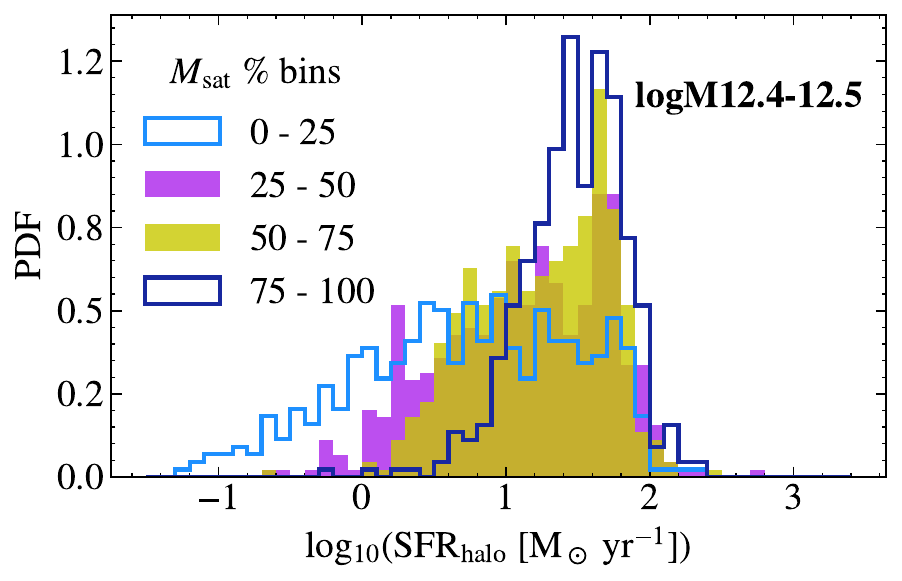}
\end{subfigure}
\caption{Histograms showing the effect of dividing haloes in a given mass bin into further bins according to a secondary property. \textbf{\textit{Left panel}}: Haloes in the mass bin $\log M_h \in (11, 11.1)$ are divided into quartiles based on \vmax. The variance of the \sfr{} in each quartile is smaller than if we include all haloes in the mass bin. \textbf{\textit{Right panel}}: 
Haloes in the mass bin $\log M_h \in (12.4, 12.5)$ are divided into quartiles based on total satellite mass. The variance is smaller than including all haloes, in particular for the higher $M_{\mathrm{sat}}$ quartiles.}
\label{fig:hists}
\end{figure*}

\begin{figure*}
\centering
\begin{subfigure}[b]{.49\textwidth}
    \centering
    \includegraphics[width=.99\textwidth]{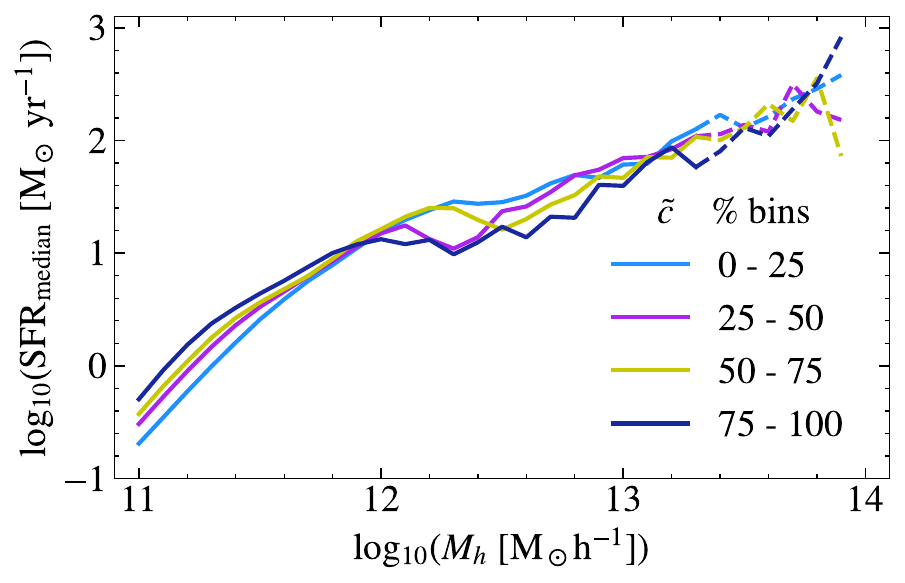}
\end{subfigure}%
\begin{subfigure}[b]{.49\textwidth}
    \centering
    \includegraphics[width=.99\textwidth]{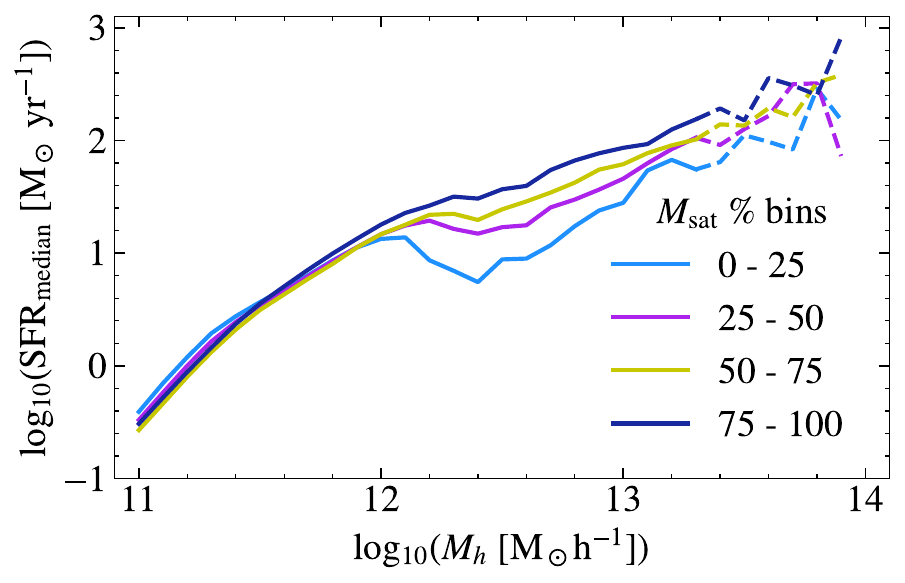}
\end{subfigure}
\caption{Effect on the median \sfr{} within each quartile when dividing mass bins (d$\log M_h=0.1$ dex) into \vmax\ (left) or satellite mass (right) quartiles. The \emph{dashed lines} indicate where there are fewer than 100 haloes within the mass bin. \textbf{\textit{Left panel}}: For $\log M_h \lesssim 12$, the median \sfr{} is higher for higher \vmax. There is no clear trend for $\log M_h \gtrsim 12$. 
\textbf{\textit{Right panel}}: For $\log M_h \lesssim 12$, satellite mass is zero for many haloes. For $\log M_h \gtrsim 12$, the median \sfr{} is higher for higher satellite mass.}
\label{fig:median_property2}
\end{figure*}

\begin{figure*}
\centering
\begin{subfigure}[b]{.49\textwidth}
    \centering
    \includegraphics[width=.99\textwidth]{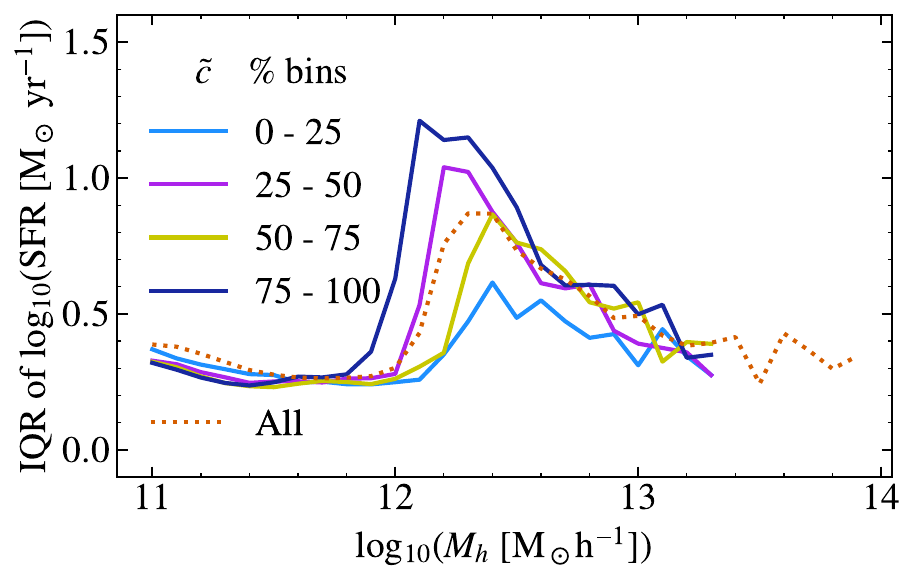}
\end{subfigure}%
\begin{subfigure}[b]{.49\textwidth}
    \centering
    \includegraphics[width=.99\textwidth]{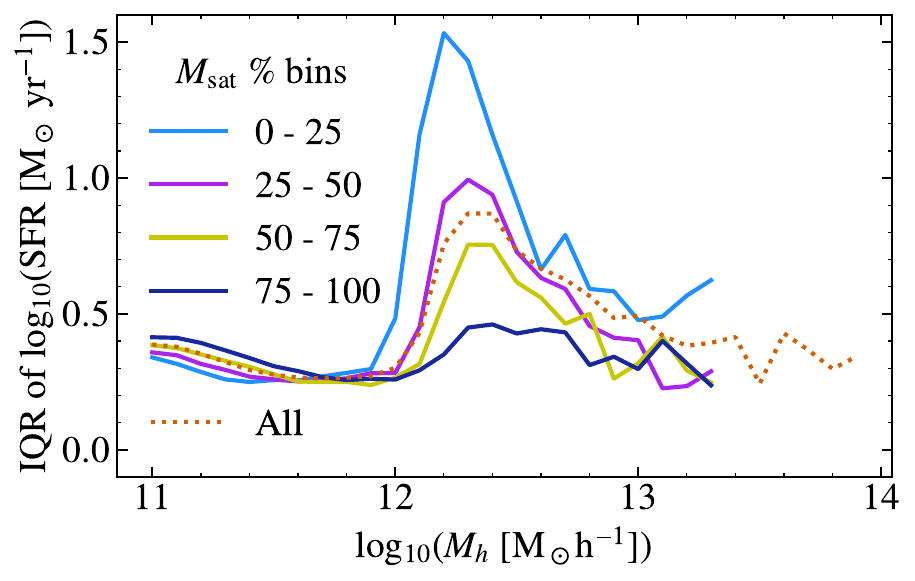}
\end{subfigure}
\caption{Effect on the interquartile range when introducing a secondary property. The \emph{dotted orange line} shows the interquartile range (\iqr{}) when including all haloes in the mass bin. In each mass bin (d$\log M_h =0.1$ dex), haloes are further divided into quartiles based on \vmax{} (left) and satellite mass (right), and the \emph{solid lines} indicate the \iqr{}s for each quartile. \iqr{}s for the quartiles are only shown for mass bins containing at least 100 haloes.
\textbf{\textit{Left panel}}: For $\log M_h \lesssim 11.5$, the variance of the \sfr{} in each quartile is smaller than if we include all halos in the mass bin. \textbf{\textit{Right panel}}: For $\log M_h \gtrsim 12$, The higher satellite mass bins have smaller variance than including all haloes.}
\label{fig:iqr_property2}
\end{figure*}

To take a closer look at the correlation between \sfr{} and \vmax{} that we find in \cref{sec:corr_coef}, we divide haloes into quartiles split by \vmax. The left panel of \cref{fig:hists} shows the histogram of halo \sfr{} 
for mass in the range $\log M_h\in (11,11.1)$ as an example, and the left panel of \cref{fig:median_property2} shows the median \sfr{} for each \vmax\ quartile bin as a function of halo mass.
The left panel of \cref{fig:iqr_property2} shows the interquartile ranges of those quartiles, as a function of halo mass.

We find that at $\log M_h \lesssim 12$, the haloes in the lower \vmax\ quartiles have on average lower \sfr{}s and the variance of the \sfr{} in each quartile is smaller than if we included all the haloes in the mass bin.
For $\log M_h \lesssim 11.5$, the interquartile ranges are smaller when divided into bins of \vmax{} (the solid lines are below the dashed line in the left panel of \cref{fig:iqr_property2}), with no clear effect at higher masses. 

This means that for more accurate modelling of the galaxy-halo connection, one could assign lower \sfr{}s to lower \vmax\ haloes, with some scatter, for this mass range. However, the correlation of \sfr{} with \vmax\ is only strong for log$M_h \lesssim 12$ (top left panel of \cref{fig:corr_coefs}), therefore including \vmax\ as a secondary parameter for higher halo masses may not improve the modelling. 

The right panel of \cref{fig:hists} shows that dividing haloes in the mass bin $\log M_h \in (12.4, 12.5)$ into quartiles of $M_\mathrm{sat}$ reduces the variance of \sfr{}s, especially for higher satellite masses. In the lowest $M_\mathrm{sat}$ quartile, the range of \sfr{}s is as large as for all haloes combined. However, we see that if $\log \mathrm{SFR} \lesssim 0$, then it likely means that $M_\mathrm{sat}$ is also low. Similarly, if $M_\mathrm{sat}$ is high, then it is unlikely that $\log \mathrm{SFR} < 0.5$. The right panel of \cref{fig:iqr_property2} shows that other halo mass bins for $\log M_h > 12$ also follow this trend. 
The interquartile range is reduced relative to the whole mass bin for the 50-75 percentile and 75-100 percentile ranges for $\log M_h \gtrsim 12$. 
This suggests that if the satellite mass is high then it should have a high \sfr{}. However, for the lowest quartile, the interquartile range remains large.

The right panel of \cref{fig:median_property2} shows the median \sfr{} for each $M_\mathrm{sat}$ quartile bin as a function of $M_h$. For $\log M_h \lesssim 12$, few haloes have satellites, so $M_\mathrm{sat}$ is not a good indicator of the \sfr{}. However, for $\log M_h \gtrsim 12$, the haloes with \rev{higher} $M_\mathrm{sat}$ do have a higher median \sfr{}, on average.

\begin{figure*}
\centering
\begin{subfigure}[b]{.49\textwidth}
    \centering
    \includegraphics[width=.99\textwidth]{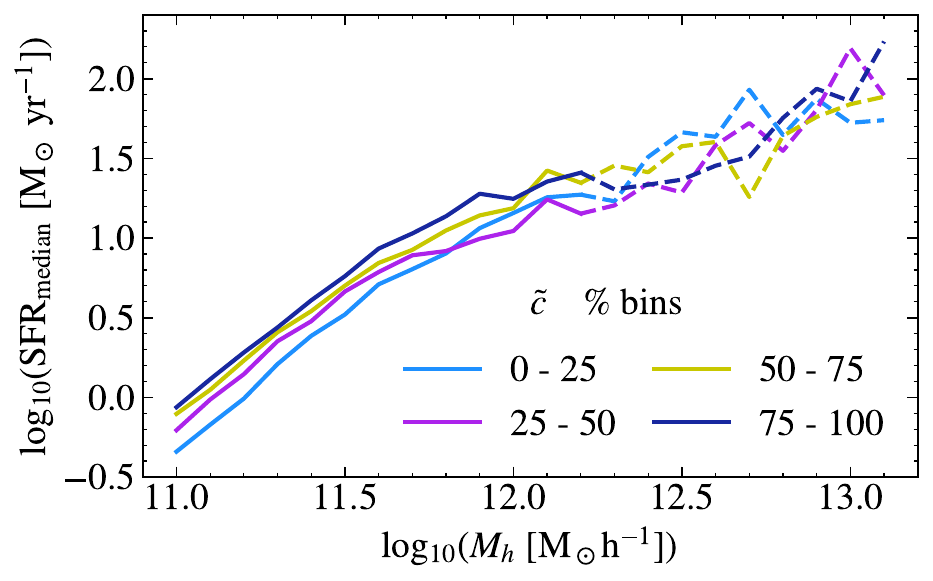}
\end{subfigure}%
\begin{subfigure}[b]{.49\textwidth}
    \centering
    \includegraphics[width=.99\textwidth]{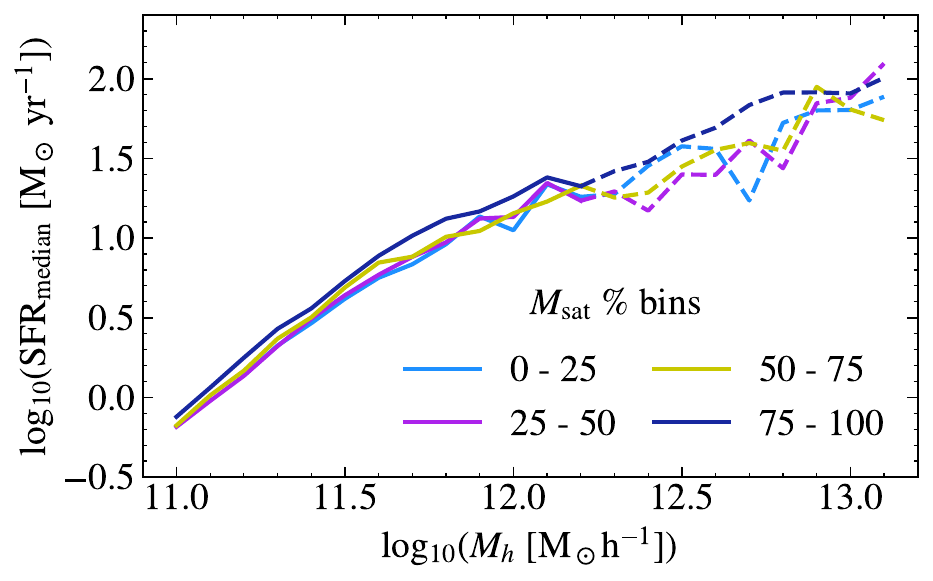}
\end{subfigure}
\caption{\eagle{} equivalent of \cref{fig:median_property2} (note that the $x$ and $y$ axis ranges differ from \cref{fig:median_property2}). 
Effect on the median \sfr{} within each quartile when dividing mass bins (d$\log M_h =0.1$ dex) into \vmax\ (left) or satellite mass (right) quartiles.
The \emph{dashed lines} indicate where there are fewer than 100 haloes within the mass bin. \textbf{\textit{Left panel}}: For $\log M_h \lesssim 12$, the median \sfr{} is higher for higher \vmax. There is no clear trend for $\log M_h \gtrsim 12$. 
\textbf{\textit{Right panel}}: No clear trend is seen for satellite mass, as low mass haloes have few or zero satellites, and there is a lack of high mass haloes in due to the smaller box size.}
\label{fig:median_property2_eagle}
\end{figure*}

\Cref{fig:median_property2_eagle} shows for \eagle{} the median \sfr{} in each quartile when haloes within mass bins of d$\log M_h = 0.1$ dex are divided into quartiles in \vmax\ (left) and satellite mass (right). 
Similar to the trend seen in \tng{} (\cref{fig:median_property2}), higher \vmax\ corresponds to higher \sfr{} for $\log M_h \lesssim 12$ and there is no clear trend for $\log M_h \gtrsim 12$. While a clear trend with satellite mass is seen for \tng{}, no clear trend is seen for \eagle{}.
This is likely due to the smaller volume of the \eagle{} simulation, which results in a limited number of haloes with satellites in each mass bin.

\section{Reducing the bias in the satellite power spectrum}\label{app:sat_ps}

\begin{figure}
    \centering
    \includegraphics[width=\linewidth]{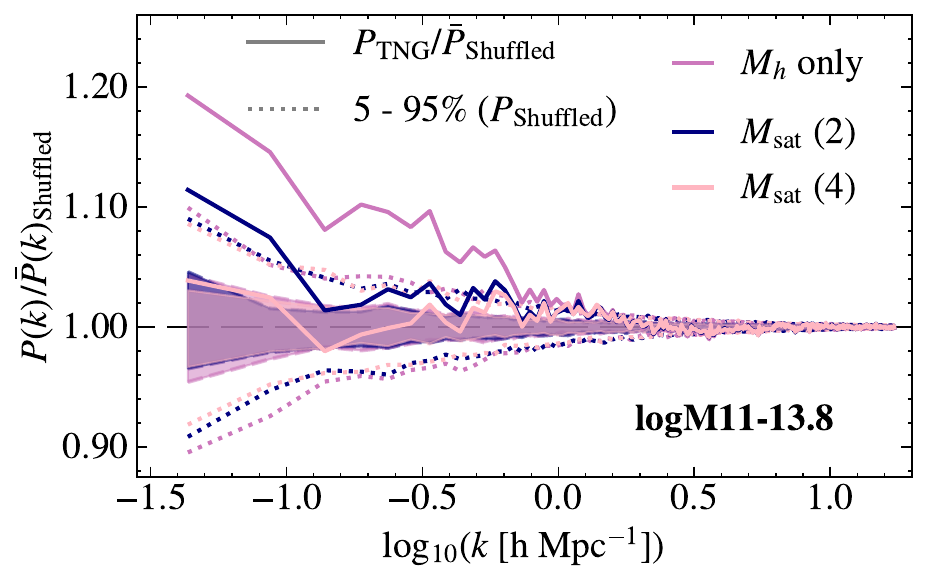}
    \caption{The ratio of the \tng{} satellite power spectrum to the shuffled satellite power spectrum when including satellite mass as a secondary property. The linestyles are the same as in \cref{fig:msat_conc_effect_logMs}, except that here only satellite galaxies have been shuffled and their power spectra computed. Dividing each halo mass bin into 4 bins of satellite mass can remove almost all of the bias discrepancy due to satellite galaxies.
    }  \label{fig:msat_sat_ps}
\end{figure}

\Cref{fig:msat_sat_ps} shows that when we only consider satellite galaxies, then the offset between \tng{} and the shuffled case can be almost completely removed when we use satellite mass as a secondary parameter. This is because the satellite mass is strongly correlated with the satellite \sfr{} for all mass ranges (bottom left panel of \cref{fig:corr_coefs}). 
From this, we can conclude that the small offset remaining in \cref{fig:property2_all} is caused by the central galaxies.
Neither \vmax\ nor satellite mass can completely remove the offset for central galaxies. 
This suggests that additional parameters should be considered to resolve the offset for the central galaxies, in order to reduce the overall offset. Nevertheless, \cref{fig:cent_vs_sat_shuffled} shows that satellite galaxies contribute a lot more to the bias discrepancy than central galaxies, and \cref{fig:property2_all} shows that using satellite mass alone can remove a significant amount of the bias discrepancy. The bias of central galaxies needs to be investigated further in order to remove the remaining discrepancy in the bias.

\end{document}